\documentclass[aps,preprint,floatfix,showpacs,superscriptaddress]{revtex4}

\usepackage{graphicx}
\usepackage{amsmath}
\usepackage{bm}

\newcommand{\be}{\begin{equation}}
\newcommand{\ee}{\end{equation}}
\newcommand{\bel}[1]{\be\label{#1}}
\newcommand{\re}[1]{Eq.~(\ref{#1})}
\newcommand{\sst}{\scriptstyle}

\newcommand{\ds}{\displaystyle}
\newcommand{\ov}[1]{\overline{#1}}
\newcommand{\hsp}{\hspace*{1pt}}

\newcommand{\Bb}{\ov{B}}
\newcommand{\BbN}{\Bb N}
\newcommand{\Nb}{\ov{N}}
\newcommand{\Lb}{\ov{\Lambda}}
\newcommand{\NbN}{\Nb N}

\newcommand{\Oap}{$^{16}\hspace*{-5pt}$\raisebox{-4pt}
           {$\sst\ov{p}$}\hsp O}
\newcommand{\Pbap}{$^{208}\hspace*{-8pt}$\raisebox{-4pt}
           {$\sst\ov{p}$} Pb}
\newcommand{\Oal}{$^{16}\hspace*{-6pt}$\raisebox{-5pt}
           {$\sst\ov{\Lambda}$}\hsp O}
\newcommand{\Pbal}{$^{208}\hspace*{-6pt}$\raisebox{-5pt}
           {$\sst\ov{\Lambda}$}\hsp Pb}

\begin{document}

\title
{Antibaryons bound in nuclei}

\author{I.N. Mishustin}

\affiliation{The Kurchatov Institute, Russian Research Center,
123182 Moscow, Russia}

\affiliation{Institut~f\"{u}r Theoretische Physik,
J.W. Goethe Universit\"{a}t,\\
D--60054 Frankfurt am Main, Germany}

\affiliation{The Niels Bohr Institute, DK--2100 Copenhagen {\O}, Denmark}

\author{L.M. Satarov}

\affiliation{The Kurchatov Institute, Russian Research Center,
123182 Moscow, Russia}

\affiliation{Institut~f\"{u}r Theoretische Physik,
J.W. Goethe Universit\"{a}t,\\
D--60054 Frankfurt am Main, Germany}

\author{T.J. B\"{u}rvenich}

\affiliation{Institut~f\"{u}r Theoretische Physik,
J.W. Goethe Universit\"{a}t,\\
D--60054 Frankfurt am Main, Germany}

\affiliation{Theoretical Division, Los Alamos National Laboratory,
87545 New Mexico, USA}

\author{H. St\"ocker}

\affiliation{Institut~f\"{u}r Theoretische Physik,
J.W. Goethe Universit\"{a}t,\\
D--60054 Frankfurt am Main, Germany}

\author{W. Greiner}

\affiliation{Institut~f\"{u}r Theoretische Physik,
J.W. Goethe Universit\"{a}t,\\
D--60054 Frankfurt am Main, Germany}

\begin{abstract}
We study the possibility of producing a new kind of nuclear
systems which in addition to ordinary nucleons contain a few
antibaryons ($\ov{B}=\ov{p}$, $\ov{\Lambda}$, etc.). The properties of
such systems are described within the relativistic mean--field
model by employing G--parity transformed interactions for
antibaryons. Calculations are first done for infinite systems and
then for finite nuclei from~$^4$He to $^{208}$Pb. It is
demonstrated that the presence of a real antibaryon leads to a
strong rearrangement of a target nucleus resulting in a
significant increase of its binding energy and local compression.
Noticeable effects remain even after the antibaryon coupling constants
are reduced by factor $3-4$ compared to G--parity motivated values.
We have performed detailed calculations of the antibaryon
annihilation rates in the nuclear environment by applying a kinetic
approach. It is shown that due to significant reduction of the
reaction Q--values, the in--medium annihilation rates should be
strongly suppressed leading to relatively long--lived
antibaryon--nucleus systems. Multi--nucleon annihilation channels
are analyzed too. We have also estimated formation probabilities
of bound $\ov{B}+A$ systems in $\ov{p}A$ reactions and have found
that their observation will be feasible at the future GSI antiproton
facility. Several observable signatures are proposed. The possibility
of producing multi--quark--antiquark clusters is discussed.
\end{abstract}

\pacs{25.43.+t, 21.10.-k, 21.30.Fe, 21.80.+a}

\maketitle


\section{Introduction}

Antibaryons are extremely interesting particles for nuclear
physics. They are building blocks of antimatter which can be
produced in the laboratory. When in close contact baryons and
antibaryons promptly annihilate each other producing
mesons in the final state. These reactions have been actively
studied in 80's and 90's, most extensively at LEAR (see reviews
\cite{Wal89,Ams91}). In particular, many attempts have been made
to find $\ov{N}N$ bound states with mass close to the
threshold \cite{Kle02}, but up to now no clear evidence was found
(see, however, the recent paper \cite{BES03}).

In contrast to elementary $\ov{B}B$ interactions, much less is
known about antibaryon-nuclear interactions. Main information
in this case is provided by antiprotonic atoms and scattering data.
However, due to strong absorption this information is limited to
a far periphery of the nuclear density distribution. As follows from
the analysis of Refs. \cite{Won84,Bat97}, the real part of the
antiproton optical potential in nuclei might be as large
as~\mbox{$-(200\div 300)$~MeV} with uncertainty reaching 100\%
in the deep interior.
The imaginary part is also quite uncertain in this region.

On the other hand, many interesting predictions concerning the
antibaryon behavior in nuclear medium have been made.  In particular,
appearance of antinucleon bound states in nuclei is one of the most
popular conjectures. This possibility was first studied in early 80's
\cite{Aue81,Won84,Hei83,Bal85} using antinucleon optical potentials
consistent with the $\ov{p}$--atomic data. However, much deeper
antinucleon potentials have been obtained \cite{Bog81,Bou82} on the
basis of relativistic nuclear models \cite{Due56,Wal74}. They predict a
large number of deeply bound antinucleon states in nuclei
\cite{Aue86,Mao99}. Several observable signatures of such states have
been discussed in Refs.~\cite{Aue86,Mis90,Mis93,Hof99}. Besides, it
was demonstrated in Ref.~\cite{Sch91} that in-medium reduction of the
antibaryon masses may lead to enhanced yields of baryon-antibaryon
pairs in relativistic heavy-ion collisions. This mechanism was also
used in Refs.~\cite{Tei94,Sib98} to explain subthreshold antibaryon
production in pA and AA collisions.

The Relativistic Mean--Field (RMF) models \cite{Ser85,Rei89} are widely
used now for describing nuclear matter and finite nuclei. Within this
approach nucleons are described by the Dirac equation coupled to scalar
and vector meson fields. Scalar $S$ and vector $V$ potentials generated
by these fields modify the spectrum of the Dirac equation in
homogeneous isospin--symmetric nuclear matter as follows
\begin{equation} \label{pmnuc}
E^{\pm}(\bm{p})=V\pm\sqrt{(m_N-S)^2+{\bf p}^2}~,
\end{equation}
where $m_N$ and ${\bf p}$ are the vacuum mass and 3-momentum of a nucleon,
respectively. The $+$ sign corresponds to nucleons with positive energy
\bel{nuce}
E_N(\bm{p})=E^{+}(\bm{p})=V+\sqrt{(m_N-S)^2+{\bf p}^2}~,
\ee
and the $-$ sign corresponds to antinucleons with energy
\bel{anuce}
E_{\Nb}({\bf p})=-E^{-}(-{\bf p})=-V+\sqrt{(m_N-S)^2+{\bf p}^2}.
\ee
Changing sign of the vector potential for antinucleons is exactly what
is expected from the G--parity transformation of the nucleon potential.
As follows from \re{pmnuc}, the spectrum of single--particle states of
the Dirac equation in nuclear environment is modified in two ways.
First, the mass gap between positive-- and negative--energy states,
$2\hsp (m_N-S)$, is reduced due to the scalar potential and, second,
all states are shifted upwards due to the vector potential.

It is well known from nuclear phenomenology that a good description of
nuclear ground state is achieved with $S\simeq 350$\,MeV and
$V\simeq 300$\,MeV so that the net potential for slow nucleons is
$V-S\simeq -50$\,MeV. Using the same values one obtains for
antinucleons a very deep potential, $-V-S\simeq -650$\,MeV.  Such a
potential would produce many strongly bound states in the Dirac sea.
However, when these states are occupied they are hidden from the direct
observation.  Only creating a hole in this sea, i.e.  inserting a real
antibaryon into the nucleus, produces an observable effect.  If this
picture is correct one can expect the existence of strongly bound
states of antinucleons in nuclei.  Qualitatively the same conclusions
can be made about antihyperons
($\ov{\Lambda}, \ov{\Sigma}, \ov{\Xi},\ldots$) which are coupled to
the meson fields generated by nucleons.

Of course, these bound antibaryon--nuclear systems will have limited
life times because of unavoidable annihilation of antibaryons. The
naive estimates based on the vacuum annihilation cross sections would
predict life times of the order of 1 fm/c. However, due to strong
reduction of phase space available for annihilation in medium, these
life times can be much longer. When calculating the structure of bound
antibaryon--nuclear systems we first consider them as stationary
objects. Afterwards their life times are estimated on the basis of
kinetic approach  taking into account in--medium effects.

In our previous paper \cite{Bue02} we have performed self-consistent
calculations of bound antibaryon--nuclear systems which take into
account rearrangement of nuclear structure due to the presence of a
real antibaryon.  We have found not only a significant increase in the
binding energy but also a strong local compression of nuclei
induced by the antibaryon. The calculations were mainly done for
antiprotons with potentials obtained by G-parity transformation. In the
present work we extend our study in several directions. We report on
new results for antiprotons and antilambdas for wider range of target
nuclei.  Calculations for reduced couplings of antibaryons to the meson
fields are done too in order to simulate effects which are missing in
a simple mean--field approximation.  We have also performed a detailed
study of the annihilation processes which determine the life time of
the antibaryon--nuclear systems.
The possibility of a delayed annihilation due to the reduction of the
available phase space is demonstrated by direct calculations.
Different scenarios of producing bound antibaryon--nuclear
systems by using antiproton beams are considered and their observable
signatures are discussed. In particular, we point out the possibility
of inertial compression of nuclei and deconfinement phase transition
induced by antibaryons.

Our paper is organized as follows. In Sect II we describe the formalism
used in calculations of antibaryon--nuclear systems and briefly explain
the numerical procedure applied for solving the RMF equations.  The
limiting case of infinite matter with admixture of antibaryons is
studied in Sect.~III. The numerical results concerning bound
antibaryon--nuclear systems are presented in Sect.~IV. The problem of
antibaryon annihilation both in infinite and finite nuclear systems is
discussed in Sect.~V. Probabilities of creating nuclei with bound
antiprotons and antilambdas in $\ov{p}A$ collisions are estimated in
Sect.~VI. Possible observable signatures of such systems are considered
in Sect.~VII. Finally, in Sect. VIII we summarize our results,
formulate several open questions and tasks for future.

\section{Theoretical Framework}

To study antibaryon--nuclear bound states we use different
versions of the relativistic mean--field (RMF) model. This approach
has several advantages:
a) its effective degrees of
freedom are adequate for the description of nuclear many--body systems,
b) it describes nuclear ground--state
observables with high accuracy,
c) antinucleons are naturally included in the theory through
the nucleonic Dirac equation,
d) its generalization for antihyperons  is straightforward .

There remain, however,
uncertainties in its application to antibaryons.
We list only some of them:
a) the G-parity symmetry is not necessarily valid in a many--body system
(we shall discuss this issue later),
b) the coupling constants of the model as well as its
functional form are determined for nuclear ground states and extrapolation
to higher densities is somewhat uncertain,
c) it is believed that the RMF model can be considered as an effective
field theory only at low enough energies, $\Delta E\lesssim 1$\,GeV.
On the other hand, inserting an antibaryon into a nucleus delivers an
excitation energy of about 2 GeV, which might be beyond the applicability
domain of such models.

Despite of these uncertainties, the RMF approach allows an exploratory
study of new types of finite nuclear systems.  The RMF model (for
reviews, see \cite{Ser85,Rei89}) is formulated in terms of a covariant
Lagrangian density of nucleons and mesons. We modify the usual approach
by introducing real antibaryons in addition to protons and neutrons.
The Lagrangian density~${\cal{L}}$ reads ($\hbar=c=1$)
\begin{eqnarray}
{\cal{L}} &=& \sum_{j=N,\ov{B}}
\ov{\psi}_j (i\gamma^\mu\partial_\mu - m_j) \psi_j \nonumber \\
& + & \frac{1}{2}\hsp
\partial^{\hsp\mu}\sigma\partial_\mu\sigma - \frac{1}
{2}\hsp m_\sigma^2 \sigma^2 -\frac{b}{3}\hsp\sigma^3
- \frac{c}{4}\hsp \sigma^4 \nonumber \\
& - & \frac{1}{4}\hsp \omega^{\mu\nu}\omega_{\mu\nu} +
\frac{1}{2}\hsp m^2_\omega\omega^\mu\omega_\mu + \frac{d}{4}
(\omega^\mu\omega_\mu)^2 \nonumber \\
& - & \frac{1}{4}\hsp\vec{\rho}^{\hsp\hsp\mu\nu}\vec{\rho}_{\mu\nu} +
\frac{1}{2}\hsp m_\rho^2\vec{\rho}^{\hsp\hsp\mu}
\vec{\rho}_{\hsp\mu} \nonumber \\
& + & \sum_{j=N,\ov{B}} \ov{\psi}_j \left[ g_{\sigma j}\sigma -
g_{\omega j} \omega^\mu\gamma_\mu -
g_{\rho j}\vec{\rho}^{\hsp\hsp\mu} \gamma_\mu \vec{\tau}_j - e_j A^\mu
\frac{1+ \tau_3}{2}\gamma_\mu \right] \psi_j \nonumber \\
& - & \frac{1}{4} F_{\mu\nu}F^{\mu\nu}\,.
\label{Lagr}
\end{eqnarray}
Here the degrees of freedom are nucleons ($N$), antinucleons
($\Bb=\Nb$) or antihyperons
(\mbox{$\Bb=\ov{\Lambda},\ov{\Sigma},\ldots$}), the isoscalar $\sigma$
($J^\pi = 0^+$) and $\omega$ ($J^\pi =1^-$) mesons, the isovector
\mbox{$\vec{\rho}$ ($J^\pi = 1^-$)} meson as well as the Coulomb field
$A_{\mu}$. The field tensors $G_{\mu\nu}$ of the vector fields
($G=\omega, \vec{\rho}, A$) are defined as
\bel{ften}
G_{\mu\nu} = \partial_\mu G_\nu - \partial_\nu G_\mu\,.
\ee
Arrows refer to the isospin space. The quantities $m_j$ in~\re{Lagr}
denote the vacuum masses of nucleons and antibaryons, $m_\omega$ and
$m_\rho$ are, respectively, the vacuum masses of $\omega$ and $\rho$
mesons.The $\sigma$ meson mass $m_\sigma$ and the coupling constants
$g_{ij}$ are adjustable parameters which are found by fitting the
observed data on medium and heavy nuclei (see below).

Our approach employs two common approximations, the {\em mean--field}
and the {\em no--sea} approximations. The latter implies that we
consider explicitly only the valence Fermi and Dirac sea states.  On
the other hand, the filled Dirac sea states are included only
implicitly via nonlinear terms of scalar and vector potentials.
According to recent work \cite{Fur02}, the vacuum contribution and the
related counter terms have a structure similar to the meson field
terms.  Implementing a real antinucleon inside a target nucleus means
appearance of a hole in the nucleonic Dirac sea.  The mean-field
approximation consists in replacing the meson field operators and the
corresponding source currents by their expectation values. This leads
to classical mean fields, for example:
\be
\sigma \rightarrow \langle \sigma \rangle, ~~~~~~~~~~~
\omega^{\mu} \rightarrow \langle \omega^{\mu} \rangle\,.
\ee
Below we treat the RMF model in {\em Hartree} approximation i.e.
neglect all exchange terms arising from antisymmetrization
of single--particle states.

Equation of motion for all meson and fermion fields are obtained from
the Euler-Lagrange equations
\bel{eleq}
\frac{\partial}{\partial x^{\mu}} \left(
\frac{\partial{\cal{L}}}{\partial(\partial q_{i}/
\partial x^{\mu})}\right) - \frac{\partial{\cal{L}}}{\partial q_{i}} = 0,
\quad q_i = N,\ov{B},\sigma,\omega,\vec{\rho},A\,.
\ee
In the equilibrium (static) case this gives the stationary Dirac equations
for (anti)baryons
\bel{direq}
\epsilon_j^\alpha  \psi_j^\alpha = \left[ -i\hsp\bm{\alpha}\bm{\nabla}
+ \beta\hsp (m_j - S_j) + V_j \right] \psi_j^\alpha\,.
\ee
Here $j = N, \ov{B}$ and $\alpha$ denotes various
single--particle valence states with the wave functions~$\psi_j^\alpha$
and energies $\epsilon_j^\alpha$. The scalar and vector potentials
acting on (anti)baryons are defined as
\begin{eqnarray}
S_j & = & g_{\sigma j}\hsp\sigma\,, \label{spot} \\
V_j & = & g_{\omega j}\hsp\omega^0 + g_{\rho j}\hsp\rho^0_3 \tau_3 +
e_j A^0 \frac{1+\tau_3}{2}\,.\label{vpot}
\end{eqnarray}

Since we consider static systems with even numbers of neutrons and
protons, only time--like components of vector fields give nonzero
contributions in the mean--field approximation.  Disregarding the
mixing of proton and neutron states we retain only the~$\rho_3$
components of the $\rho$--meson field. We also assume that the
time--reversal invariance is valid even in the presence of an unpaired
particle like an antiproton.

Equations of motion for the meson fields read
\begin{eqnarray}
(- \Delta  + m_\sigma^2 + b\hsp\sigma + c\hsp\sigma^2)\hsp\sigma & = &
\sum\limits_j g_{\sigma j} \rho_{Sj}\,, \label{mess} \\
( - \Delta + m_{\omega}^{2} + d\hsp\omega_0^2)\hsp \omega_0 & = &
\sum\limits_j g_{\omega j} \rho_j\,, \label{mesv} \\
( - \Delta + m_{\rho}^{2})\hsp \rho_3^0 & = &
\sum\limits_j g_{\rho j} \rho_{Ij}\,, \label{mesr} \\
\Delta A_0 & = &\sum\limits_j e_j \rho_{Qj}
\label{mesa}
\end{eqnarray}
The scalar, vector, isovector and charge densities are defined as
\begin{eqnarray}
\rho_{Sj} & = & \langle \ov{\psi}_j \psi_j \rangle\,,\\
\rho_j & = & \langle \psi_j^\dagger \psi_j \rangle\,,\\
\rho_{Ij} & = & \langle \psi^\dagger_j \tau_3 \psi_j \rangle\,,\\
\rho_{Qj} & = & \frac{1}{2}\hsp\langle \psi^\dagger_j\hsp (1+\tau_3)\hsp
\psi_j \rangle\,,
\end{eqnarray}
where angular brackets mean averaging over the ground state wave
function.

\begin{table}[thb!]
\caption{\label{tab:parn}
The parameter sets of the RMF models used in this paper.}
\vspace*{3mm}
\begin{ruledtabular}
\begin{tabular}{l|r|r|r|r}
{\em }                   & {NLZ}    & {NLZ2}       &  {NL3}  & TM1\\
\hline
$m_N\,({\rm MeV})$       & 938.9    &  938.9       &  939.0    &938.0\\
$m_{\sigma}\,({\rm MeV})$& 488.67   &  493.150     &\,508.194  &511.198\\
$m_{\omega}\,({\rm MeV})$& 780.0    &  780.0       &  782.501  &783.0 \\
$m_{\rho}\,({\rm MeV})$  & 763.0    &  763.0       &  763.0    &763.0\\
$g_{\sigma N}$           & 10.0553  &  10.1369     &  10.217   &10.0289\\
$g_{\omega N}$           & 12.9086  &  12.9084     &  12.868   &\,12.6139\\
$g_{\rho N}$             & 4.84944  &  4.55627     &  4.4740   &4.6322\\
$b\,({\rm fm}^{-1})$     &\,13.5072 &\,13.7561     & 10.431    & 7.2325\\
$c$                      &\,$-40.2243$&\,$-41.4013$& $-28.885$   &0.6183\\
$d$                      & 0        &  0           &  0        &71.305\\
\end{tabular}
\end{ruledtabular}
\end{table}
To investigate sensitivity of the results to the model parameters,
we have considered several versions of the RMF model, namely
the NL3~\cite{Lal97}, NLZ~\cite{Ruf88}, NLZ2~\cite{Ben99} and
TM1~\cite{Sug94} models. The corresponding parameter sets are listed
in Table~\ref{tab:parn}. They were found by fitting binding energies and
radii of spherical nuclei from $^{16}{\rm O}$ (not included
in the TM1 fit) to Pb isotopes~\footnote{
Note that masses $m_\rho$
and $m_\omega$ in the NLZ and NLZ2 models were not fitted, but
fixed to the values indicated in Table~\ref{tab:parn}.
}.

Regarding the antibaryon couplings, there is no reliable information
suitable for high density nuclear matter. In this situation, as a
starting point, we choose antibaryon--meson coupling constants
motivated by the G--parity transformation. It is analogous to the ordinary
parity transformation in the configurational space, which inverts the
directions of 3--vectors.
The G--parity transformation is defined as the combination of the
charge conjugation and the $180^{\circ}$ rotation around the second axis
of the isospin space~\cite{Gre01}. As known, $\sigma$ and $\rho$
mesons and the Coulomb field have positive G--parity, while $\omega$
meson has negative G--parity.  Therefore, applying the G--parity
transformation to the baryon potentials (\ref{spot})--(\ref{vpot}) one
obtains the corresponding potentials for antibaryons. The results of this
transformation can be formally expressed by the following relations
between the coupling constants:
\bel{gpar}
g_{\sigma\Nb} =
g_{\sigma N}\,, \quad g_{\omega\Nb} = - g_{\omega N}\,, \quad g_{\rho\Nb}
= g_{\rho N}\,.
\ee
If these relations are valid, one can make two conclusions. First,
equal scalar potentials lead to equal effective masses for nucleons and
antinucleons.  Second, the vector potentials have opposite signs for
these particles.  This is in agreement with the conclusion made by
considering positive-- and negative--energy solutions of the Dirac
equation (see Eqs.~\mbox{(\ref{nuce})--(\ref{anuce})}).

The simple consideration based on the G--parity transformation of mean
meson fields is certainly an idealization. There are several effects in
many--body systems which can, in principle, significantly distort this
picture.  Here we mention only two of them. First, inclusion of
exchange (Fock) terms leads generally to smaller magnitudes of scalar
and vector potentials for antinucleons in dense baryon--rich matter as
compared to nucleons~\cite{Sou90}. Second, the expressions
(\ref{spot})--(\ref{vpot}) correspond to the tadpole (single bubble)
diagrams associated with the $\sigma, \omega$ and $\rho$ meson
exchanges. However, for antibaryons one should consider also the
contribution of more complicated multi--meson diagrams originating from
annihilation channels in the intermediate state.  By using dispersion
relations, the corresponding contribution to the real part of the
antibaryon self energy can be expressed through the~$\BbN$ annihilation
cross section. The calculations of Refs.~\cite{Tei94,Sib98} show that
such a contribution can reach $100-150$\, MeV for slow antinucleons at
normal nuclear density.

Taking into account all these uncertainties, in our calculations we consider
not only the G--parity motivated couplings~(\ref{gpar}), but also
the reduced antibaryon couplings
\bel{gpar1}
g_{\sigma\Nb} = \xi g_{\sigma N}\,,
\quad  g_{\omega\Nb} = - \xi g_{\omega N}\,,
\quad  g_{\rho\Nb} = \xi g_{\rho N}\,
\ee
with the same reduction parameter $\xi$ for interactions with
$\sigma,~\omega$ and $\rho$ meson fields. By choosing $\xi$ in the
interval $0\le\xi\le 1$ we can investigate all possibilities from
maximally strong antibaryon couplings to noninteracting antibaryons.
In calculations of antihyperon--nuclear system we use the coupling
constants motivated by the SU(3) flavor symmetry, namely assume that
$g_{\sigma\Lb}=\frac{2}{3}g_{\sigma\Nb},~g_{\omega\Lb}=\frac{2}
{3}g_{\omega\Nb}$\,.

By using~\re{gpar1} we can rewrite the source terms in
Eqs.~(\ref{mess})--(\ref{mesv}) in a more transparent form
\begin{eqnarray}
&&\sum\limits_j g_{\sigma j} \rho_{Sj}=
g_{\sigma N}\hsp (\rho_{SN}+\xi\rho_{S\Nb})\,,
\label{sder}\\
&&\sum\limits_j g_{\omega j} \rho_j=
g_{\omega N}\hsp (\rho_N-\xi\rho_{\Nb})\,.
\label{vder}
\end{eqnarray}
One can infer from these relations that presence of antibaryons in
nuclear matter leads to increase of the scalar (attractive) potential
and decrease of the vector (repulsive) potential acting on surrounding
nucleons. Unlike some previous works, we take into account the
rearrangement of nuclear structure due to the presence of a real
antibaryon This leads to enhanced binding and additional compression of
nuclei as was first demonstrated in~Ref.~\cite{Bue02}.

A few remarks about the numerical procedures are in place here.
Calculations are performed for nuclear systems obeying the axial and
reflection symmetries.
We replace the Dirac equations (\ref{direq}) for
single--particle wave functions $\psi_j^\alpha$ by the effective
Schr\"odinger equations for their upper components.
Introducing explicitly the upper ($\psi_+$) and lower~($\psi_-$)
spinor components of the wave function and omitting indices $j$ and $\alpha$\,,
we rewrite Eqs.~(\ref{direq}) in the form
\bel{direq1}
\epsilon \left(
\begin{array}{c} \psi_+ \\ \psi_-
\end{array} \right) = \left(
\begin{array}{cc}  m - S + V &  \bm{\sigma}\bm{p} \\
\bm{\sigma}\bm{p} & -m + S +V
\end{array}\right)
\left(
\begin{array}{c} \psi_+\\ \psi_-
\end{array}\right)\,.
\ee
Here $\bm{\sigma}$ denote the spin Pauli matrices and
$\bm{p}=-i\bm{\nabla}$\,. From the lower part of \re{direq1} one gets
\bel{lcom}
\psi_-=(\epsilon +m-S-V)^{-1}(\bm{\sigma}\bm{p})\hsp\psi_+\,.
\ee
Substituting this expression into the upper part
leads to the following equation for~$\psi_+$
\bel{shr1}
\epsilon\hsp\psi_+=h\hsp\psi_+\,,
\ee
where $h$ is the effective, energy--dependent Hamiltonian
\bel{shr2}
h=(\bm{\sigma}\bm{p})\hsp (\epsilon+m-S-V)^{-1}\hsp
(\bm{\sigma}\bm{p}) +m-S+V\,.
\ee

Eqs.~(\ref{shr1}) for nucleon and antibaryon wave functions
are solved by iterations, using the damped gradient
method suggested in Ref.~\cite{Rei82}. At each iteration step we
use the potentials $S,V$ determined by solving Eqs.~(\ref{mess})--(\ref{mesa})
for mean meson and Coulomb fields at the preceding step.
Numerical calculations are carried out on a
spatial grid with equidistant grid points. The Fourier transformation
of fermion and meson fields is used to evaluate spatial derivatives as
matrix multiplication in the momentum space. This allows us to use a
rougher grid with the cell size $\sim 0.7$\,fm, as compared to much finer
grids required by the finite difference schemes. Most calculations below
were carried out with the cell size of 0.3 fm which guarantees the
the numerical accuracy in binding energies and density profiles
better than 0.5\%. For nucleons we implement the paring correlations
using the BCS model with a $\delta$
pairing force and a smooth cut--off given by the Fermi distribution in
single--particle energies~\cite{Ben99}. The center of mass corrections
are taken into account in the same way as for normal
nuclei (for details see Ref.~\cite{Ben99}).

\section{Infinite matter with admixture of antibaryons}

To study qualitative effects due to the presence of antibaryons, let us
consider first homogeneous isospin--symmetric nucleonic matter with
admixture of antibaryons ($\BbN$~matter). As compared to the general
formalism developed in the preceding section, here we neglect the
isospin--asymmetry and Coulomb effects assuming vanishing $\rho$--meson
and electromagnetic fields. At zero temperature, disregarding medium
polarization (density rearrangement) effects, we calculate
thermodynamic functions of $\Bb N$ matter at fixed, spatially
homogeneous densities of nucleons~$\rho_N$ and antibaryons
$\rho_{\Bb}$\,. In this case one can omit the spatial derivatives in
Eqs.~(\ref{mess})--(\ref{mesa}) and solve the Dirac equations
(\ref{direq}) in the plane wave representation. For example, assuming
that the G-parity concept is valid on the mean--field level, one gets
the energy spectra of $N,\Nb$ states given by
Eqs.~(\ref{nuce})--(\ref{anuce}) where $S=g_{\sigma N}\sigma,
V=g_{\omega N}\omega_0$.

Within the mean field approximation the occupation numbers of
nucleons ($j=N$) and antibaryons ($j=\Bb$) at zero temperature
have the form of Fermi distributions \mbox{$\Theta(p_{Fj}-p)$}
where $\Theta(x)\equiv (1+{\rm sqn}\,x)/2$ and
$p_{Fj}=(6\pi^2\rho_j/\nu_j)^{1/3}$ is the Fermi momentum of
corresponding particles (here $\nu_j$ denote their spin-isospin
degeneracy factors). Applying the standard formalism~\cite{Ser85}
one can obtain the energy density $e=E/V=T^{00}$ directly from the
Lagrangian~(\ref{Lagr}) with the result
\bel{ende}
e=\sum\limits_{j=N,\Bb}(e_j^{\rm kin}+g_{\omega
j}\rho_j\hsp\omega_0)+ \frac{1}{2}m_\sigma^2 \sigma^2
+\frac{b}{3}\sigma^3+\frac{c}{4}\hsp\sigma^4 -\frac{1}{2}\hsp
m_\omega^2 \omega_0^2-\frac{d}{4}\hsp\omega_0^4\,.
\ee
Here $e^{\rm kin}_j$ is the ''kinetic'' part of the energy density
\bel{enkin}
e^{\rm kin}_j=\frac{\nu_j}{2\pi^2}\int\limits_0^{p_{Fj}}{\rm d}p\,p^2
\sqrt{m_j^{*\hsp 2}+p^2}\,,
\ee
where $m_j^*$ is the effective
mass of $j$--th particles, related to the scalar meson field as
\bel{meff}
m_j^*=m_j-g_{\sigma j}\sigma\,.
\ee
After calculating
the integral in~\re{enkin} we get the formula
\bel{ende1}
e^{\rm kin}_j=\frac{\nu_j p_{Fj}^{\hsp 4}}{8\pi^2}\hsp\Psi
\left(\frac{m_j^*}{p_{Fj}}\right)\,,
\ee
where $\Psi(z)$ is a
dimensionless function
\bel{psif}
\Psi(z)\equiv 4\int\limits_0^1
{\rm d}\hsp t\hsp t^2\sqrt{t^2+z^2}=
\sqrt{1+z^2}\left(1+\frac{z^2}{2}\right)-\frac{z^4}{4}\ln
\frac{\sqrt{1+z^2}+1}{\sqrt{1+z^2}-1}\,.
\ee

Equations for meson fields can be obtained by minimizing the energy
density with respect to $\sigma$\, and $\omega_0$\,. From
Eqs.~(\ref{ende})--(\ref{meff}) one gets
~\footnote{
The same equations
follows from (\ref{mess})--(\ref{mesv}) after omitting the derivative
terms.}
\begin{eqnarray}
&&(m_\sigma^2+b\sigma+c\sigma^2)\hsp\sigma=
\sum\limits_{j=N,\Bb}g_{\sigma j}\rho_{Sj}\,,
\label{mess1}\\
&&(m_\omega^2+d\omega_0^2)\hsp\omega_0=
\sum\limits_{j=N,\Bb}g_{\omega j}\rho_j\,,
\label{mesv1}
\end{eqnarray}
where $\rho_{Sj}\equiv \partial e/\partial m_j$ coincides
with  the scalar density
defined in preceding section. The explicit expression for $\rho_{Sj}$
has the form
\bel{scld}
\rho_{Sj}=\frac{\nu_j}{2\pi^2}\int\limits_0^{p_{Fj}}{\rm d}p\,p^2
\frac{m_j^*}{\sqrt{m_j^{*\hsp 2}+p^2}}=
\frac{\nu_j\hsp p_{Fj}^{\hsp 2}\hsp m_j^*}{4\pi^2}\hsp
\Phi\left(\frac{m_j^*}{p_{Fj}}\right)\,,
\ee
where
\bel{phif}
\Phi(z)\equiv \frac{\Psi^{\hsp\prime}(z)}{2z}=
\sqrt{1+z^2}-\frac{z^2}{2}\ln\frac{\sqrt{1+z^2}+1}{\sqrt{1+z^2}-1}\,.
\ee
Using (\ref{mesv1}) one can rewrite \re{ende} in a simpler form
\bel{ende2}
e=\sum\limits_{j=N,\Bb}e_j^{\rm kin}+
\frac{1}{2}m_\sigma^2 \sigma^2 +\frac{b}{3}\sigma^3+\frac{c}{4}\hsp\sigma^4
+\frac{1}{2}\hsp m_\omega^2 \omega_0^2+\frac{3}{4}\hsp d\omega_0^4\,.
\ee
Notice that the terms containing the vector field $\omega_0$
generate a repulsive contribution.

\begin{figure*}[htb!]
\vspace*{-2cm}
\hspace*{1cm}\includegraphics[width=13cm]{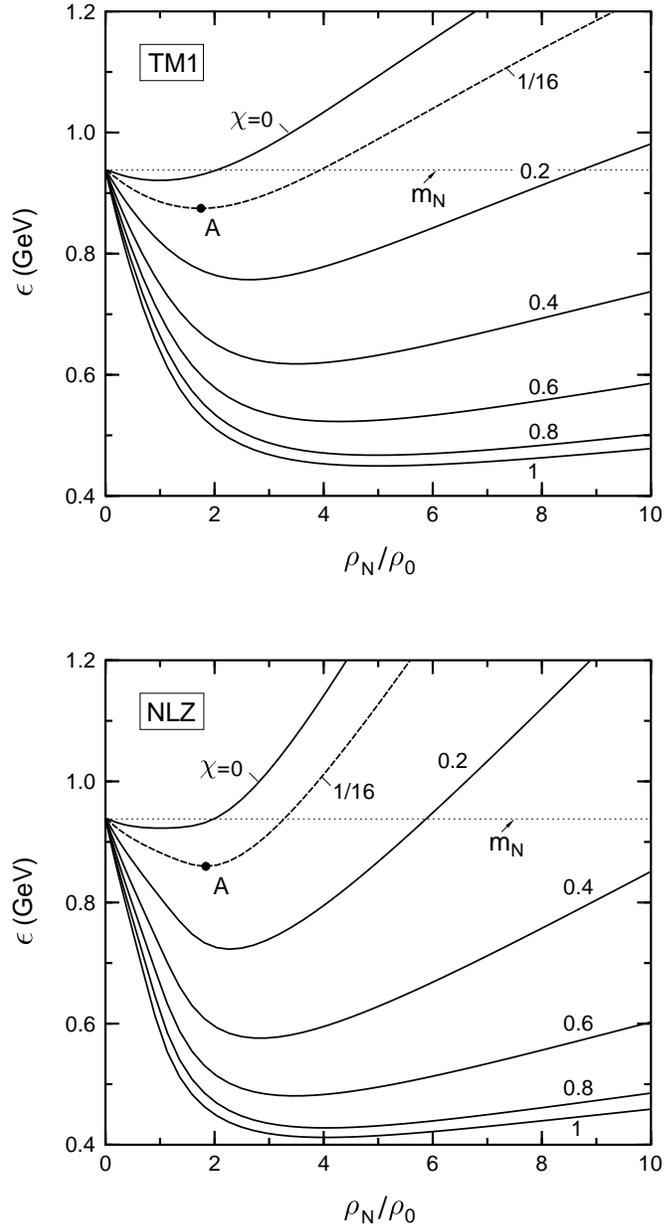}
\caption{Energy per particle of homogeneous $\NbN$ matter,
$\epsilon=E/(N_N+N_{\ov{N}})$\,, as a function of nucleon density
(in units of $\rho_0=0.15$\,fm$^{-3}$) at fixed
$\chi=\rho_{\Nb}/\rho_N$. Upper (lower) panel shows results of the
TM1 (NLZ) calculation. Points~A correspond to minima of $\epsilon$
at $\chi=1/16$\,.}
\label{fig1}
\end{figure*}

To find bound states of the $\BbN$\, matter we calculate
the energy per particle,
\bel{epp}
\epsilon\equiv\frac{E}{N_N+N_{\Bb}}=\frac{e}{\rho_N+\rho_{\Bb}}\,,
\ee
at different $\rho_N$\, and $\rho_{\Bb}$\,. It is useful
to introduce instead of $\rho_{\Bb}$ the relative concentration of
antibaryons
\bel{relc}
\chi=\frac{N_{\Bb}}{N_N}=\frac{\rho_{\Bb}}{\rho_N}\,.
\ee
Due to the charge--conjugation invariance of $\NbN$ matter,
its energy density must be symmetric
under the replacement $\rho_{\Nb}\leftrightarrow\rho_N$\,.
Therefore, it is sufficient
to consider only the values $\chi\leq 1$\,. At fixed $\chi$, using
Eqs.~(\ref{ende})--(\ref{phif}) one has in the low density limit
\bel{epp0}
\epsilon_0\equiv\lim\limits_{\rho_N\to 0}\epsilon=
\frac{m_N+m_{\Bb}\chi}{1+\chi}\,.
\ee
By definition, the total binding energy of the system is equal to
\bel{binde}
BE=(\epsilon_0-\epsilon)\cdot (N_N+N_{\Bb}).
\ee
In the case of $\NbN$ matter $\epsilon_0=m_N$\,. To
investigate qualitatively properties of bound \mbox{$\ov{B}+{}^{16}{\rm O}$}
systems (see Sect.~IV), we study separately the case of $\chi=1/16$\,.

\begin{figure*}[htb!]
\vspace*{-8cm}
\hspace*{1cm}\includegraphics[width=13cm]{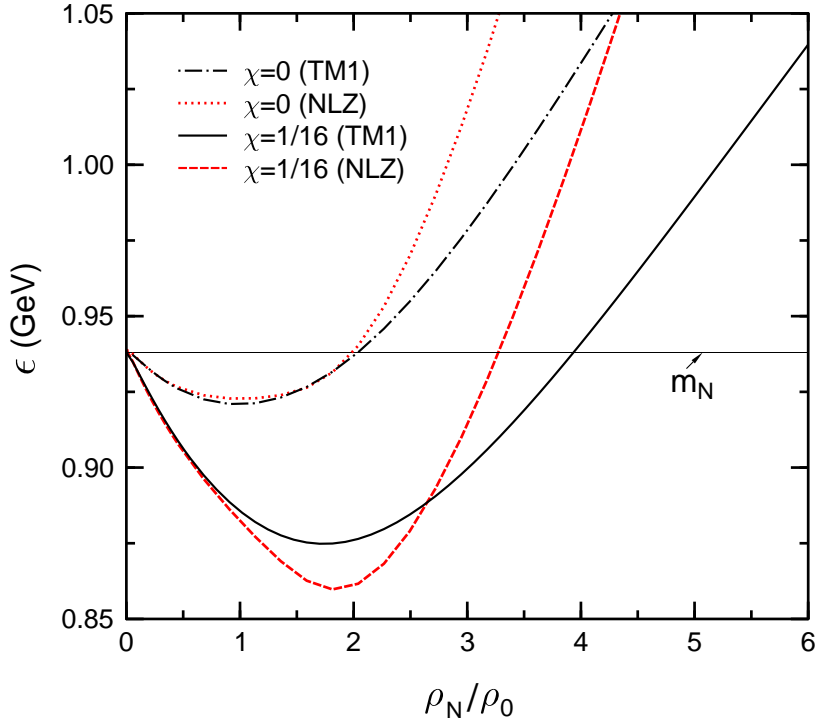}
\caption{Comparison of energies per particle calculated within
the TM1 and NLZ
models for $\chi=0$ and $1/16$.}
\label{fig2}
\end{figure*}

Figures \ref{fig1}, \ref{fig2} show the energies
per particle of $\NbN$ matter calculated within the TM1 and NLZ models.
The coupling constants of $\Nb$ interactions are fixed by the
G--parity transformation (see~\re{gpar}). Different curves corresponds
to different values of anti\-nucleon concentration $\chi$. Dashed lines
in Fig.~\ref{fig1} represent the results for the case $\chi=1/16$\,. It
is seen that nucleonic matter becomes more bound after inserting a
certain fraction of antibaryons. The maximal binding takes place at
\mbox{$\chi=1$} i.e. for the ''baryon--free'' matter with net baryon density
equal to zero. As one can see from Eqs.~(\ref{mesv1}), (\ref{ende2}),
the repulsive vector contribution to the energy density vanishes in
this case ($\omega_0=0$), therefore, the behavior of $\epsilon(\rho_N)$
is determined only by the counterbalance of scalar attraction and
effects of Fermi motion. It is interesting to note that such
baryon--free $\NbN$ matter is fully symmetric with respect to
interchanging nucleons and antinucleons.  Thus, there is absolutely no
reason that the $N$ and $\Nb$ coupling constants would violate the
G--parity symmetry.

To illustrate the model dependence of the results, in
Fig.~\ref{fig2} we compare predictions of the TM1 and NLZ
models for $\chi=0$\, and $\chi=1/16$.  One can see that the NLZ
parametrization predicts larger binding of the $\NbN$ matter at
densities \mbox{$\rho_N\sim 2\rho_0$}\,. The calculated parameters of
bound states for $\chi=0$\, (pure nucleonic matter), 1/16 and 1 are
presented in Table~\ref{tab:bspar}.
\begin{table}[ht]
\caption{Characteristics of bound states of pure nucleonic matter
($\chi=0$) as well as the $\NbN$ matter predicted by the TM1 and NLZ
models.}
\vspace*{3mm}
\label{tab:bspar}
\begin{ruledtabular}
\begin{tabular}{l|r|r|r|r|r|r}
Model &\multicolumn{3}{c|}{TM1}&\multicolumn{3}{c}{NLZ}\\
\cline{1-7}
$\chi$ & 0 & $1/16$ & 1 & 0 & $1/16$ & 1\\
\colrule
$\rho_N/\rho_0$& 0.99 & 1.75 & 4.99& 1.01 & 1.84 & 4.04\\
$\epsilon$\,(MeV)& 922 & 875 & 449& 923 & 860 & 412\\
$m_N^*$\,(MeV)& 593 & 393 & 90& 544 & 224 & 42\\
\end{tabular}
\end{ruledtabular}
\end{table}
By using~\re{binde} and the values of minimal energies per particle
given in this table (for $\chi=1/16$), one can estimate the
binding energy of a bound $\ov{p}\,+^{16}\hspace*{-2pt}{\rm O}$\, system.
This leads to the result
\bel{bineo}
BE\hsp (^{16}_{\hspace*{3pt}\ov{p}}{\rm O})
=\left\{\begin{array}{ll}1070\, {\rm MeV}~&(\mbox{\rm TM1})\,,\\
1330\, {\rm MeV}~&({\rm NLZ})\,.
\end{array}\right.
\ee
On the other hand, applying the formalism of
Sect.~II to a finite system \Oap~\cite{Bue02}
gives for its binding energy 1159 ~MeV~(TM1) and 828 MeV (NLZ2).
Comparison of these results shows influence of the finite size
(surface), rearrangement (polarization) and Coulomb effects which are
not taken into account in the infinite matter calculations.

\begin{figure*}[htb!]
\vspace*{-8.5cm}
\hspace*{1cm}\includegraphics[width=13cm]{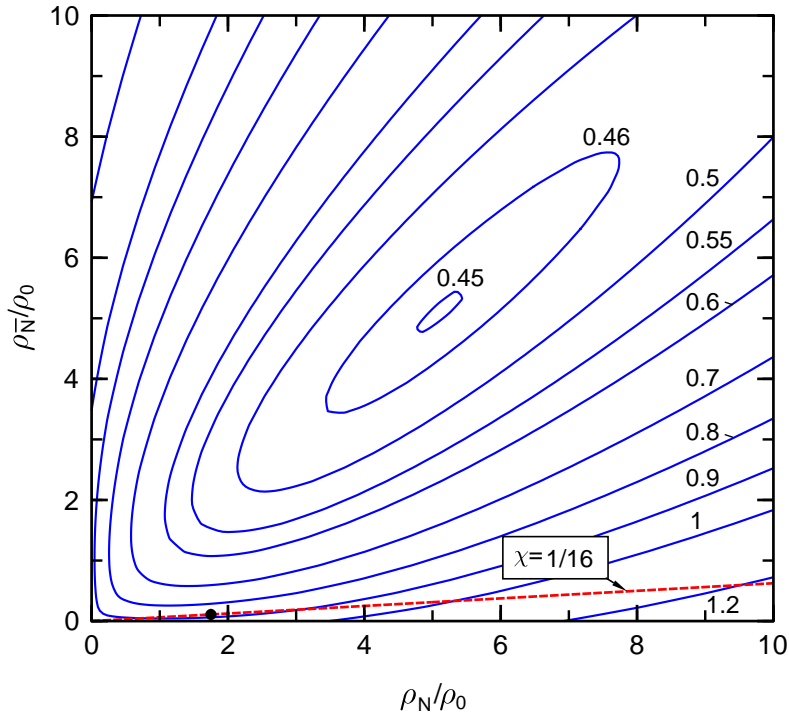}
\vspace*{-5mm}
\caption{Contours of energy per particle (shown in GeV near the
corresponding curves) on the $\rho_N-\rho_{\Nb}$\, plane,
calculated within the TM1 model. The dashed line corresponds to
states with $\chi=1/16$\,. Solid point shows position of minimal
$\epsilon$  on this line. }
\label{fig3}
\end{figure*}
More detailed results for the $\NbN$ matter are represented in
Fig.~\ref{fig3} in the form of contour plots of
$\epsilon\hsp (\rho_N,\rho_{\Nb})$ on the plane $\rho_N-\rho_{\Nb}$\,
calculated within the TM1 model. The states with fixed $\chi$ lie on
straight lines going from the origin of the plane. The maximal binding,
about~490 MeV per particle, is predicted  for symmetric systems with
$\rho_{\Nb}=\rho_N\simeq 5\hsp\rho_0$\,. Most likely, the hadronic
language is not valid at such high densities and the quark--antiquark
degrees of freedom are more appropriate in this case.

\begin{figure*}[htb!]
\vspace*{-12.5cm}
\hspace*{2cm}\includegraphics[width=15cm]{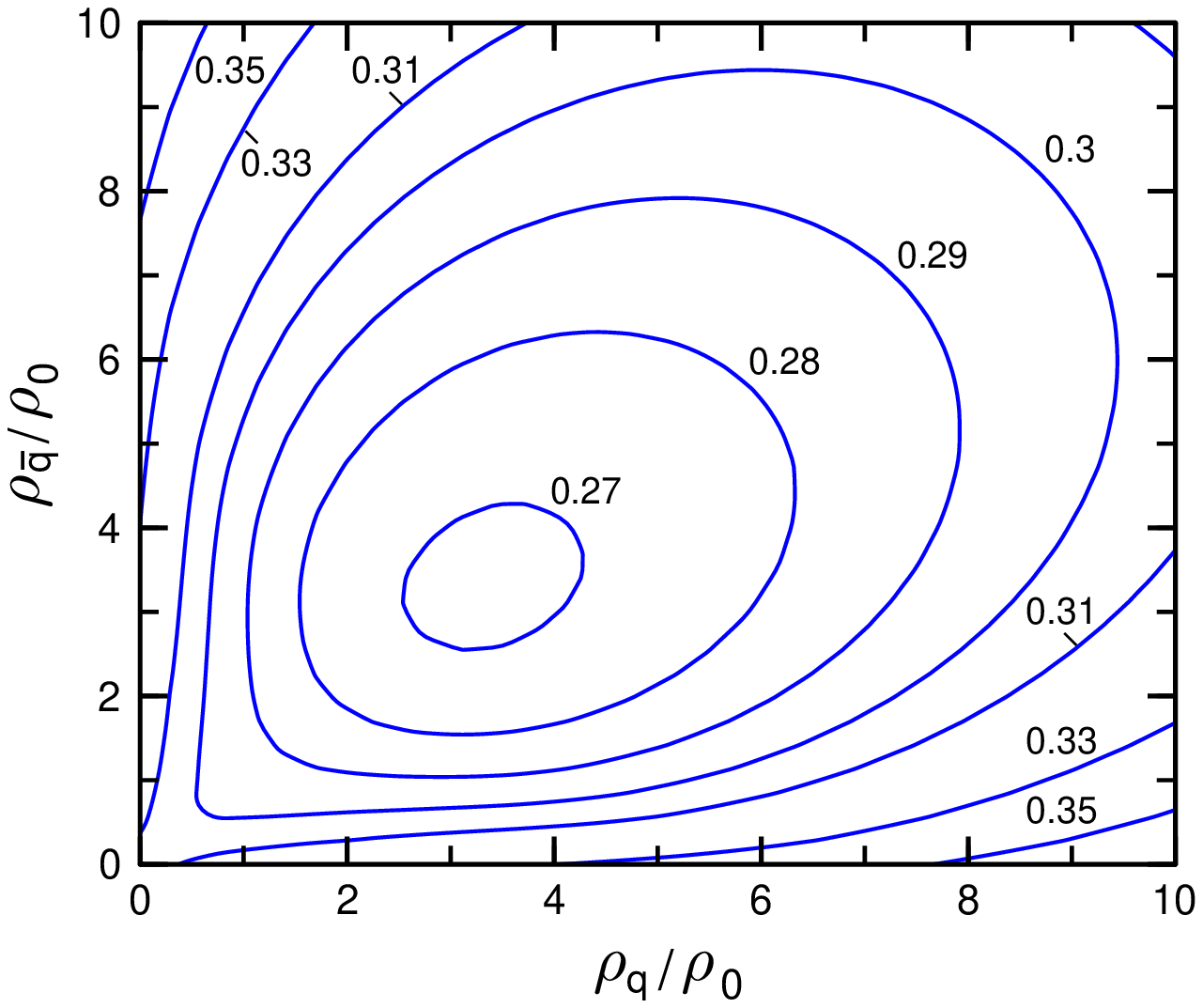}
\caption{Contours of energy per particle,
$\epsilon=E/(N_q+N_{\bar{q}})$\,, (shown in GeV near the
corresponding curves) of cold quark--antiquark matter calculated
within the SU(2) NJL model~\cite{Mis99}. The horizontal (vertical)
axis corresponds to the total density of $u+d$ quarks (antiquarks)
in units of $\rho_0=0.15$\,fm$^{-3}$\,. }
\label{fig4}
\end{figure*}
Strong binding effects can also be expected in a deconfined $q\ov{q}$
matter. In Refs.~\cite{Mis99, Mis00} we have performed calculations for
such matter within the generalized Nambu--Jona-Lasinio (NJL) model. As
an example, Fig.~\ref{fig4} shows energy per particle,
$\epsilon=E/(N_q+N_{\bar{q}})$, for nonstrange isospin--symmetric
$q\bar{q}$ matter at zero temperature. Here $N_q\,(N_{\bar{q}})$
denotes the total number of $u,d$ quarks (antiquarks). The figure
displays contour plots of $\epsilon\hsp (\rho_q, \rho_{\bar{q}})$
predicted by the SU(2) NJL model (for details, see Ref.~\cite{Mis99}).
The binding in this case is generated by an attractive interaction in
the scalar--pseudoscalar channel.  Again, one can see that the
strongest binding is predicted for the baryon--symmetric case
$\rho_q=\rho_{\bar{q}}$\,. The maximum binding energy per $q\bar{q}$ pair
is
\bel{qqbe}
BE\simeq (m_q^{\rm vac}-270\,{\rm MeV})\times 2\simeq 60\,{\rm MeV}\,,
\ee
where $m_q^{\rm vac}\simeq 300$\,MeV is the constituent mass of light
quarks in the vacuum. Approximately three times larger binding energies
have been found in Ref.~\cite{Mis00} for cold quark--antiquark matter
with admixture of strange $s, \bar{s}$ quarks.  On the basis of this
finding, in Refs.~\cite{Mis99, Mis00} we have predicted possible
existence of new metastable systems, mesoballs, consisting of many
quarks and antiquarks in a common ''bag''. Such systems are
characterized by a small (or zero) baryon number and their decay should
occur mainly via emitting pions from the surface. It is interesting to
note that multi--quark--antiquark clusters were also
predicted~\cite{Buc79} on the basis of the MIT bag model. It is
instructive to compare the value (\ref{qqbe}) with the binding energy
of baryon--symmetric $\NbN$ matter calculated within the RMF model.
Using Table~\ref{tab:bspar} for the case $\chi=1$\,, we get the binding
energy per quark--antiquark pair, $2\hsp (m_N-\epsilon)/3\simeq 326\,
(350)$\, MeV, within the TM1 (NLZ) model. Therefore, the hadronic
approach may overestimate real binding energies of cold and dense
baryon--free matter by a factor $\sim 5$\,.

\begin{figure*}[htb!]
\vspace*{-10cm}
\hspace*{1cm}\includegraphics[width=13cm]{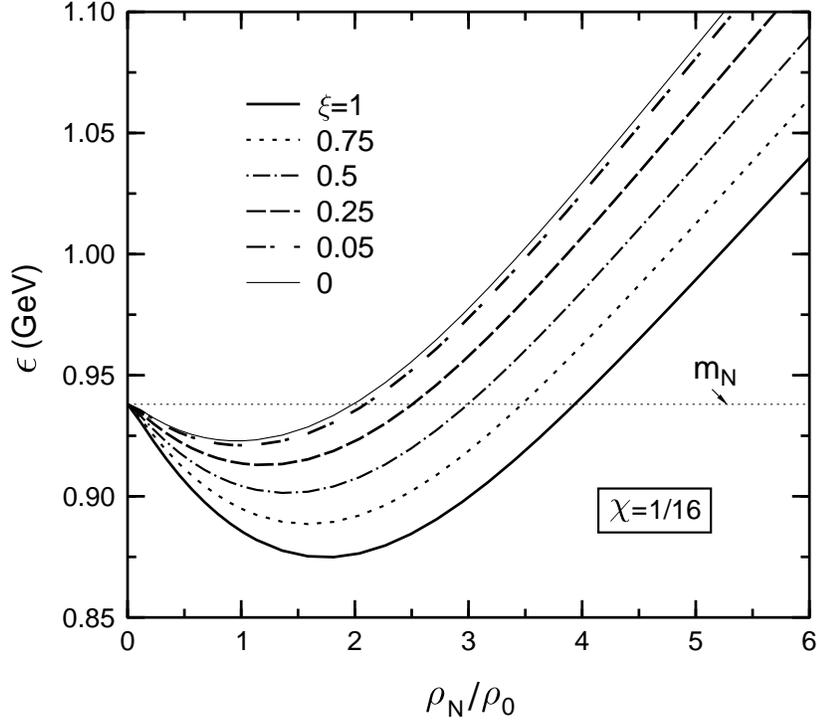}
\vspace*{-5mm}
\caption{Energy per particle of the $\NbN$ matter versus the nucleon
density at fixed $\chi=1/16$\,. Different curves correspond to the
TM1 calculation at different values of the parameter $\xi$
characterizing deviation of the $\Nb$ coupling constants from
the G--parity motivated values. }
\label{fig5}
\end{figure*}
As noted above, the G--parity symmetry may be violated in a dense
baryon--rich matter. As a consequence, the G--parity motivated
choice of antibaryon coupling constants, \re{gpar},  may overestimate
binding energies of
the $\BbN$ matter. To elaborate on this issue, we have performed an
analogous calculation, but with reduced couplings $g_{\sigma\Bb}\,,~
g_{\omega\Bb}$\, as defined in~\re{gpar1}. The results of
calculations within the TM1 model for different values of $\xi$ are
presented in Fig.~\ref{fig5}. It shows the energy per
particle of $\NbN$ matter with $\chi=1/16$. The lower curve
corresponds to the case of the exact G--parity symmetry. One can
see that a reduction of $\xi$ results in a smaller binding and
compression of the $\NbN$ matter. However, the admixture of
antinucleons becomes relatively unimportant only at very small
antinucleon couplings, corresponding to
$\xi\hsp < 0.25$\,. It is shown below that the same conclusion
follows from more refined calculations for the finite \Oap~system.

\begin{table}[htb]
\caption{The parameters of the TM1 and NLZ models in the hyperonic sector.
The remaining parameters are given in Table~\ref{tab:parn}.}
\label{tab:parl}
\vspace*{3mm}
\begin{ruledtabular}
\begin{tabular}{r|r|r|r|r|r|r|r|r}
 &$m_{\Lb}\,({\rm MeV})$ & $m_{\sigma^*}\,({\rm MeV})$
 & $m_{\phi}\,({\rm MeV})$&
 $g_{\sigma\Lb}$ &  $g_{\omega\Lb}$ & $g_{\rho\hsp\Lb}$
 & $g_{\sigma^*\Lb}$ &  $g_{\phi\Lb}$\\
\colrule
NLZ & 1116 & 975 & 1020 & 6.23 & $-8.61$ & 0 & 6.77 & 6.09\\
TM1 & 1116 & 975 & 1020 & 6.21 & $-8.41$ & 0 & 6.67 & 5.95\\
\end{tabular}
\end{ruledtabular}
\end{table}
We end this section by considering cold nucleonic matter
with admixture of antihyperons
$\ov{Y}=\ov{\Lambda},\,\ov{\Sigma}\ldots$ As proposed in
Ref.~\cite{Sch96}, the observed data on $\Lambda\Lambda$
interaction can be reproduced within the RMF model by
including additional scalar ($\sigma_*$) and vector ($\phi$) meson fields
coupled only to hyperons. By the same reason, to take into account
the interaction between antihyperons in the~$\ov{Y}N$ matter,
we generalize the Lagrangian~(\ref{Lagr}) by
introducing the additional terms~\footnote{ These terms are not
needed when only one antihyperon is trapped in the nucleus. }
\begin{eqnarray}
\delta{\cal{L}}_{YY}=&&\hspace*{-3pt}\frac{1}
{2}\left(\partial^\mu\sigma_*\partial_\mu\sigma_* -
m_\sigma^2 \sigma_*^2\right)
- \frac{1}{4} \phi^{\mu\nu}\phi_{\mu\nu} +
\frac{1}{2} m^2_\phi \phi^\mu\phi_\mu +\nonumber \\
&&+\ov{\psi}_{\ov{Y}}\left( g_{\sigma_* \ov{Y}}\hsp\sigma_* -
g_{\phi {\ov{Y}}}\hsp\phi^\mu\gamma_\mu\right)\psi_{\ov{Y}}\,.
\label{Lagry}
\end{eqnarray}
In our calculations of the $\Lb N$ matter and bound $\Lb$--nuclear
systems we use the values of parameters suggested in Ref.~\cite{Sch96}.
The $\Lb$--meson couplings were obtained from the $\Lambda$--meson
coupling constants by using the G--parity transformation.

\begin{figure*}[htb!]
\vspace*{-12.5cm}
\hspace*{2cm}\includegraphics[width=15cm]{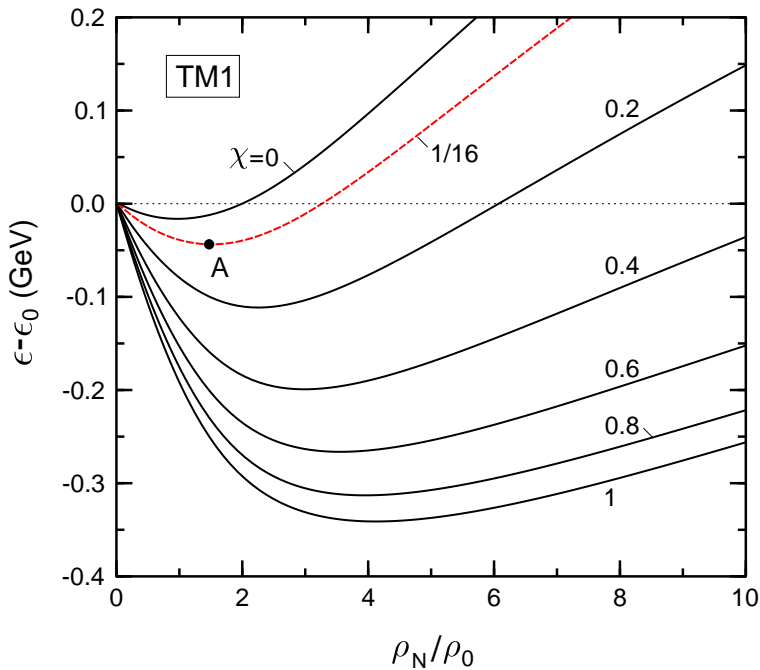}
\caption{Energy per particle of $\Lb N$ matter versus the nucleon
density calculated within the TM1 model. Different curves
correspond to different values of the $\Lb$ concentration
$\chi=\rho_{\Lb}/\rho_N$\,. The curves are shifted along the
vertical axis by a $\chi$--dependent value
$\epsilon_0=\lim\limits_{\rho_N\to 0}\epsilon$\,.
Point A marks the position of energy minimum for
$\chi=1/16$\,. }
\label{fig6}
\end{figure*}

\begin{table}[ht]
\caption{Characteristics of bound states of pure nucleonic matter
($\chi=0$) as well as the $\Lb N$--matter predicted by the TM1
models.}
\vspace*{3mm}
\label{tab:al_npar}
\begin{ruledtabular}
\begin{tabular}{l|r|r|r|r}
$\chi$ & 0 & $1/16$ & 1/4 & 1 \\
\colrule
$\rho_N/\rho_0$& 0.99 & 1.47 & 2.49& 4.08\\
$\epsilon_0-\epsilon$\,(MeV)& 16 & 44 & 135& 252\\
$m_N^*$\,(MeV)& 593 & 456 & 257& 44\\
$m_{\Lb}^*$\,(MeV)& 905 & 812 & 662 & 420\\
\end{tabular}
\end{ruledtabular}
\end{table}
It is easy to see that the hyperon-hyperon interactions
lead to the following effects. First, the gap equation for the antihyperon
effective mass is modified (as compared to~\re{meff}) due to the additional
$\sigma_*$ field:
\bel{alefm}
m_{\ov{Y}}^*=m_{\Lambda}-g_{\sigma\ov{Y}}\sigma -g_{\sigma_*\ov{Y}}\sigma_*\,.
\ee
Second, the energy density of the $\ov{Y} N$ matter
contains now the additional contribution of $\sigma_*$\, and $\phi$\,
mesons,
\bel{ende3}
\delta e_{YY}=\frac{1}{2}\left(m_{\sigma_*}^2\sigma_*^2+
m_{\phi}^2\phi_0^2\right)\,.
\ee

Figure \ref{fig6} shows the energy per particle of the $\Lb N$
matter, calculated within the TM1 model, generalized in accordance with
Eqs.~(\ref{alefm})--(\ref{ende3}), with parameters listed in
Tables~\ref{tab:parn},~\ref{tab:parl}. To compare results for different
$\chi$\,, the energy per particle is shifted by a corresponding vacuum
value $\epsilon_0$ defined in~\re{epp0}.  The calculated parameters of
bound states are given in Table~\ref{tab:al_npar}. Comparison with
results obtained earlier for the $\NbN$ matter shows that binding
energies and equilibrium densities are noticeably smaller in the~$\Lb
N$ case.  For example, the binding energy predicted for the
\Oal\, ''nucleus'' equals approximately $44\times 17\simeq 748$\, MeV
which is about 30\% smaller than for \Oap~bound state. This difference is
explained by a smaller scalar coupling of $\Lb$ particles as compared
to antinucleons~(\mbox{$g_{\sigma\Lb}/g_{\sigma\Nb}\simeq 2/3$})\,.

\section{Finite antibaryon--nuclear systems}

\subsection{Light nuclear systems with antiprotons}

\begin{figure*}[htb!]
\centerline{\includegraphics[width=16cm]{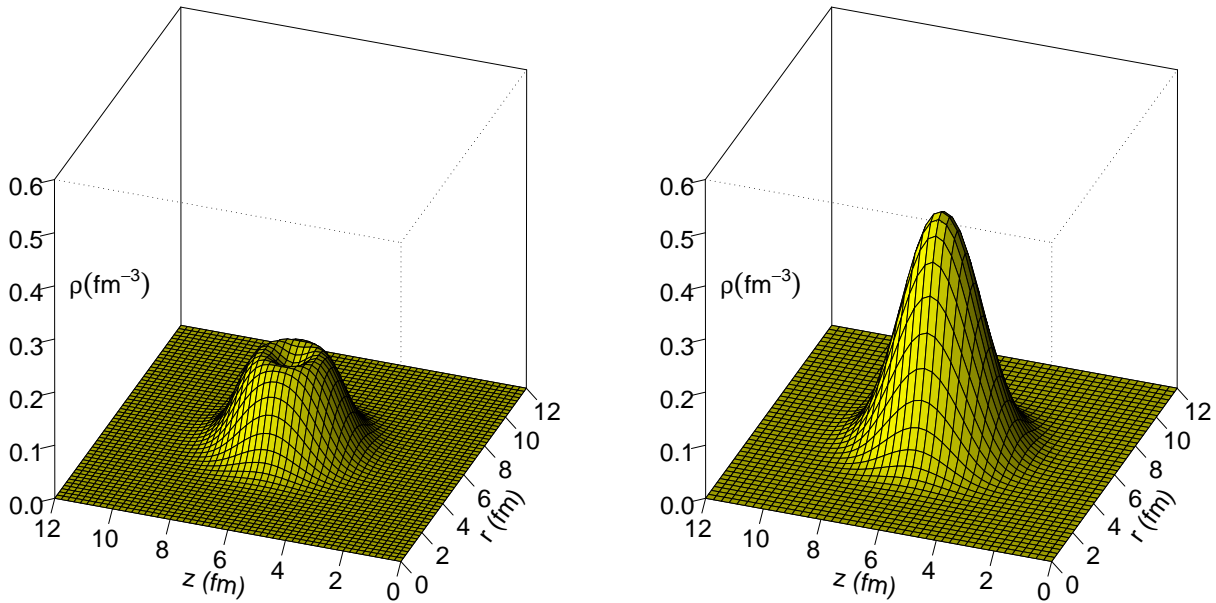}}
\caption{3D plots of nucleon density in the
$^{16}$O nucleus (left) and in the bound $\ov{p}\hsp\hsp +^{16}$O
system (right) calculated within the NL3 model.}
\label{fig7}
\end{figure*}

The nucleus $^{16}$O is the lightest nuclear system for which the RMF
approach is considered to be reliable. This nucleus is included into
the fit of the effective forces NL3 and NLZ2. Therefore, we choose this
nucleus as a basic system to study the antibaryon--nuclear bound
states. First, let us consider an antiproton bound in a $^{16}$O
nucleus. For clarity we use the notation \Oap~for
such a combined system. A priori it is unclear which quantum numbers has
the lowest bound state. We first assume that this is the
$\frac{1}{2}^+$ state and later on we shall check this assumption.
Results of our self-consistent calculations for both $^{16}$O and
\Oap~are presented in Fig.~\ref{fig7}.
It shows 3D plots of the nucleon densities for the G--parity motivated
antiproton couplings ($\xi$=1). One can see that inserting an
antiproton into the nucleus gives rise to a dramatic rearrangement of
nuclear structure. This effect has a simple origin. As explained above,
the antiproton contributes with the same sign as nucleons to the scalar
density (see~\re{sder}), but with the negative sign to the vector
density.  This leads to an overall increase of attraction and decrease
of repulsion for surrounding nucleons. To maximize attraction, protons
and neutrons move to the center of the nucleus, where the antiproton
has its largest occupation probability. This leads to a strong
compression of the nucleus.

\begin{figure*}[htb!]
\vspace*{5mm}
\includegraphics[width=16cm]{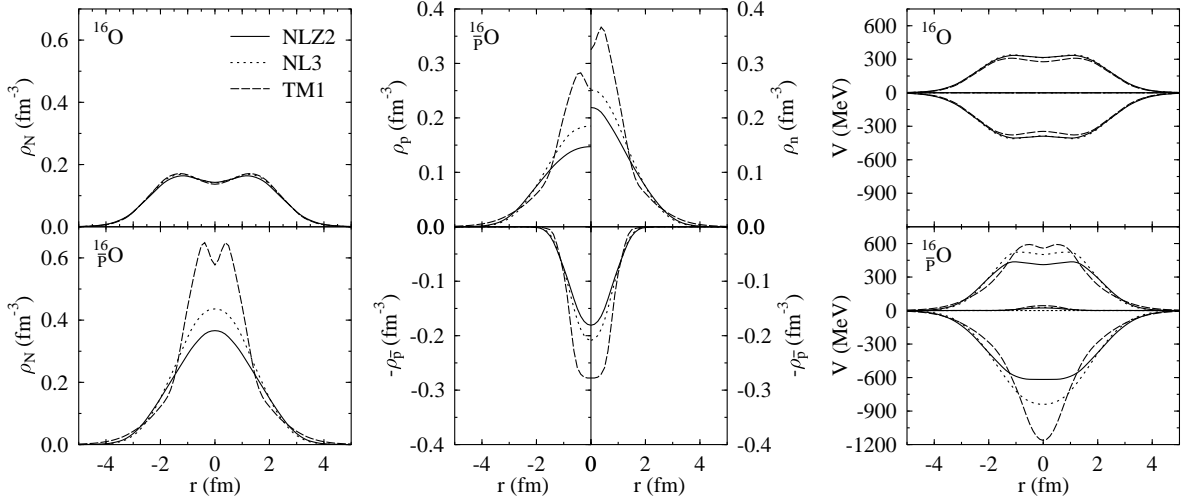}
\caption{
The left panel represents the sum of proton and neutron densities as
function of nuclear radius for $^{16}$O without (top) and with an
antiproton. The left and right parts of the upper
middle panel show separately the proton and neutron densities, the
lower part of this panel displays the antiproton density (with minus
sign). The right panel shows the scalar (negative)  and vector
(positive) parts of the nucleon potential. Small contributions shown in
the lower row correspond  to the isovector ($\rho$--meson) part.}
\label{fig8}
\end{figure*}
Figure~\ref{fig8} shows the densities and potentials for $^{16}$O with
and without the antiproton. For normal $^{16}$O all RMF
parametrizations considered produce very  similar results. The presence
of the antiproton changes drastically the structure of the nucleus.
Depending on the parametrization, the sum of proton and neutron
densities reaches a  maximum value of $(2-4)\,\rho_0$\,, where
$\rho_0\simeq 0.15$\,fm$^{-3}$ is the normal nuclear density. The
largest compression is predicted by the TM1 model. This follows from
the fact that this parametrization gives the softest equation of state
as compared to other forces considered here. According to our
calculations, the difference between proton and neutron  densities is
quite large, which leads to an increase in symmetry energy.  The reason
is that protons, though they feel additional Coulomb attraction to the
antiproton, repel each other. As a consequence, neutrons are
concentrated closer to the center than protons and the symmetry energy
increases.

\begin{figure}[ht]
\centerline{\includegraphics[width=16cm]{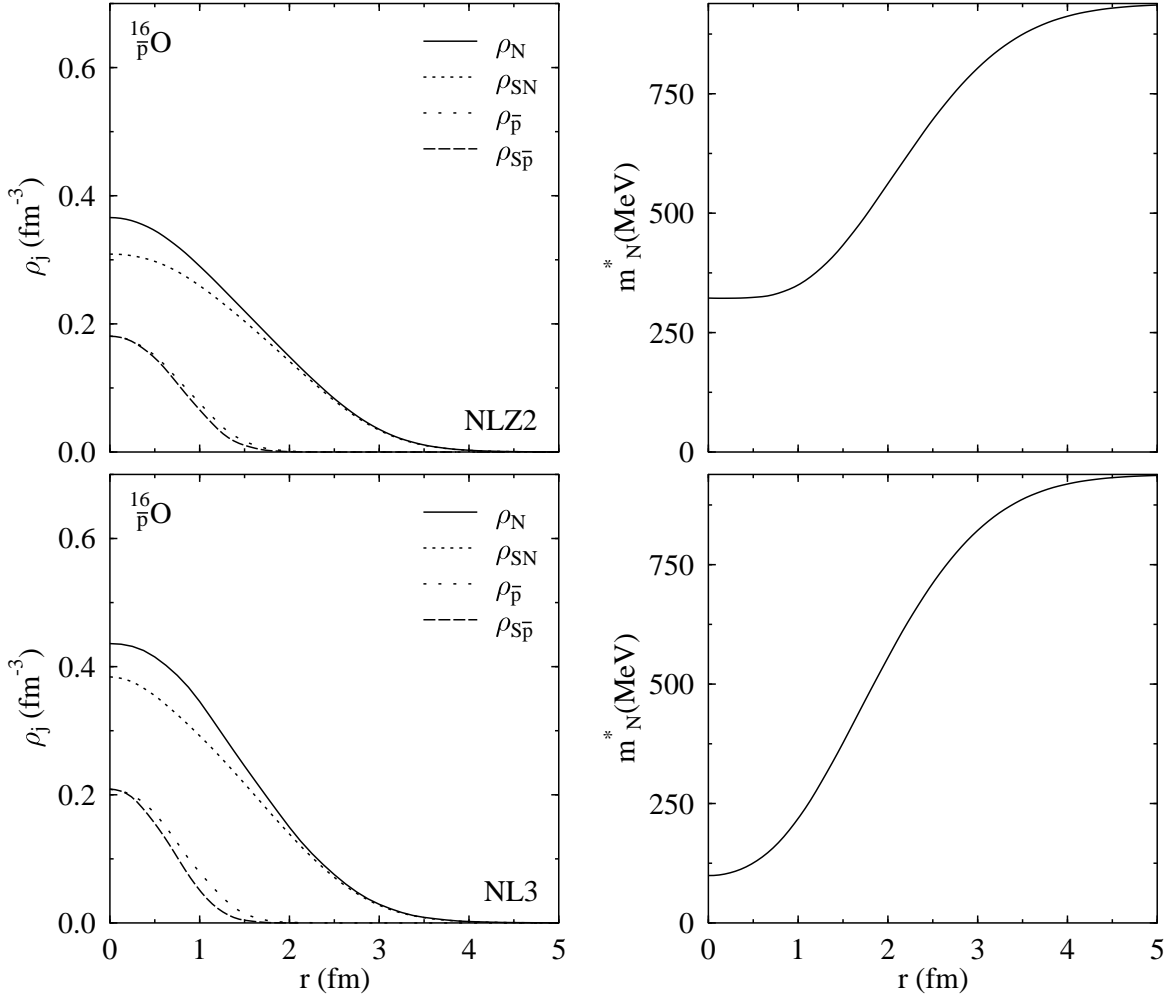}}
\caption{Left panels: radial profiles of the vector ($\rho_j$) and
scalar ($\rho_{Sj}$) densities of antiproton~($j=\ov{p}$) and nucleons
($j=N$) in the
ground state of $^{16}_{\hspace*{3pt}\ov{p}}$O as predicted by the
NLZ2 and NL3 calculations. Right panels show the corresponding
profiles of the effective nucleon mass.}
\label{fig9}
\end{figure}
Figure~\ref{fig9} shows radial profiles of scalar and vector
densities of nucleons and the antiproton in \Oap\hsp. As compared to
the normal $^{16}$O nucleus, absolute values of vector and scalar
densities increase  in the central region of the nucleus. This leads
to a strong  drop of the effective nucleon mass near the nuclear
center~\footnote{
The effective mass within the TM1 model even becomes
negative at \mbox{$r\lesssim 1$\,fm}.
},
which in turn suppresses the local annihilation rate of the antiproton
(see Sect.~V).

\begin{figure*}[htb!]
\vspace*{5mm}
\includegraphics[width=16cm]{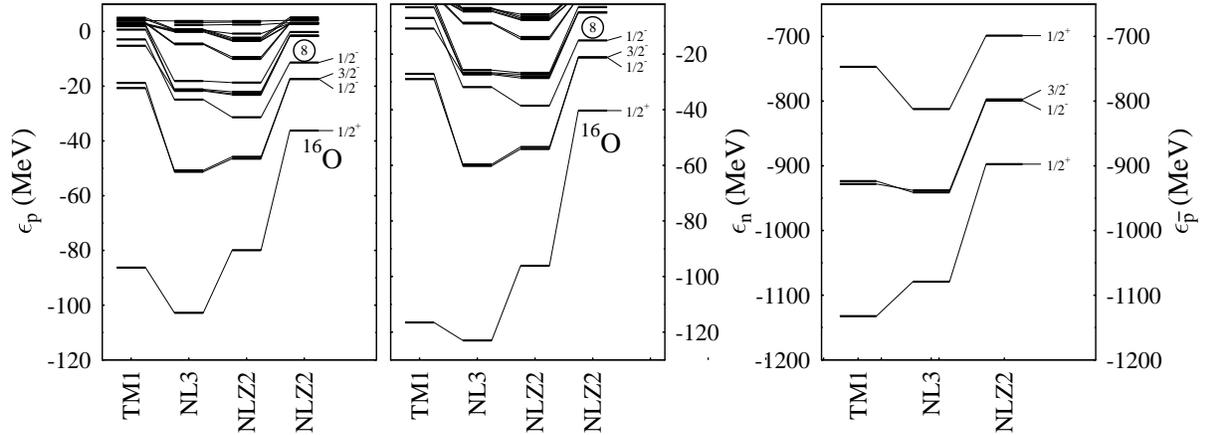}
\caption{
Proton (left), neutron (middle) and antiproton (right)
energy levels for the nucleus $^{16}$O with one antiproton and
without it (rightmost columns in the left and middle panels).
}
\label{fig10}
\end{figure*}
Since nucleons feel a deeper potential due to the presence of
the antiproton, their binding energy increases too. This can be
seen in Fig.~\ref{fig10}. The nucleon binding is largest within the
NL3 parametrization. In the TM1 case, the $s1/2^+$ state is
also deep, but higher levels are less bound as compared to the
NL3 and NLZ2 calculations. This is a consequence of the smaller
spatial extension of the potential in this case. The highest $s1/2^-$ level
in the TM1 calculation (see Fig.~\ref{fig10}) is even less bound than
for the system without an antiproton.

For the antiproton levels, the TM1 parametrization predicts the deepest
bound state with binding energy of about $1130$~MeV. The NL3
calculation gives nearly the same binding,  while in the NLZ2 case,
antiproton levels are more shallow and have smaller spacing. It should
be noted that the antiprotons are more strongly bound than was obtained
in Ref.~\cite{Mao99}. This follows from the fact that here we consider
both nucleons and the antinucleon self--consistently allowing the
target nucleus to change its shape and structure due to the presence of
the antiproton. The total binding energy of the~\Oap\, system is
predicted to be~828~MeV for NLZ2, 1051~MeV for NL3, and 1159~MeV for
TM1.  For comparison, the binding energy of the normal $^{16}$O nucleus
is 127.8, 128.7 and 130.3~MeV in NLZ2, NL3, and TM1, respectively.

\vspace*{5mm}
\begin{figure*}[htb!]
\vspace*{1cm}
\centerline{\includegraphics[width=16cm]{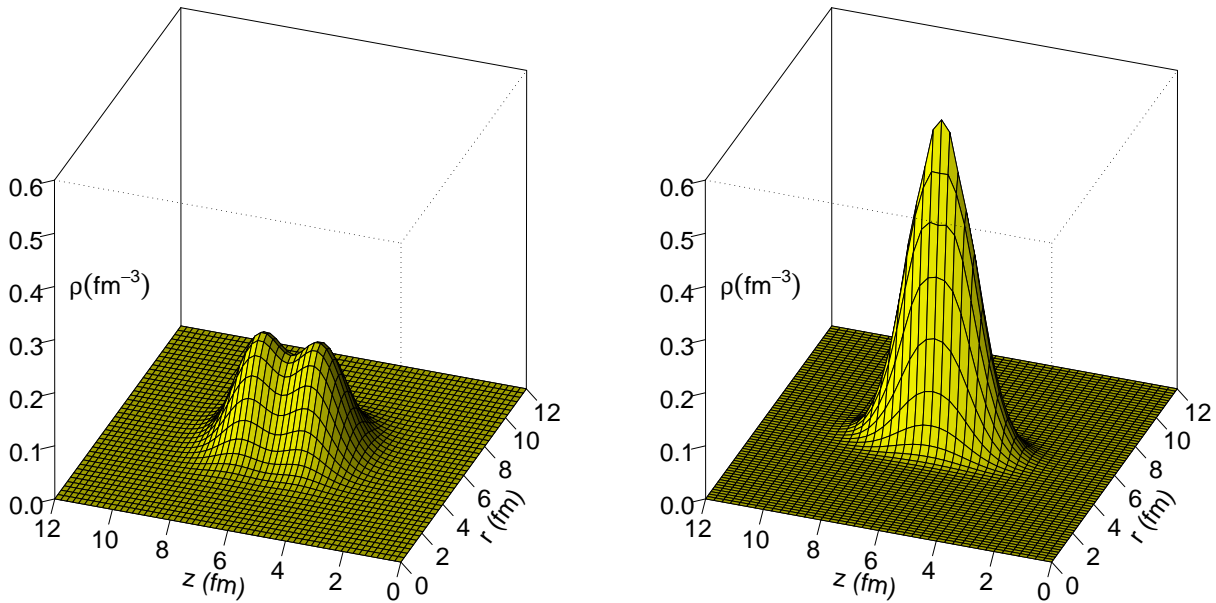}}
\caption{
3D plots of nucleon densities for the nucleus $^8$Be without (left) and
with (right) antiproton calculated within the NL3 model.
}
\label{fig11}
\end{figure*}

As a second example, we investigate the effect of a single antiproton
inserted into the~$^8$Be nucleus. In this case only the NL3
parametrization was used (the effect is similar for all three RMF
models). The normal $^8$Be  nucleus is not spherical, exhibiting a
clearly visible $\alpha-\alpha$~structure  with the deformation
$\beta_2\simeq 1.20$ in the ground state. As one can see from
Fig.~\ref{fig11}, inserting an antiproton gives rise to the compression
and change of nuclear shape, which results in a much less elongated
nucleus with $\beta_2\simeq 0.23$. Its maximum density increases by a
factor of three from $1.3\hsp\rho_0$ to $4.1\hsp\rho_0$\,. The cluster
structure of the ground state completely vanishes. A similar effect has
been predicted in Ref.~\cite{Aka02} for the $K^-$ bound state  in the
$^8$Be nucleus. In our case the binding energy increases  from
$52.9$~MeV (the experimental value is $56.5$~MeV) to about $700$~MeV.

\begin{figure*}[htb!]
\centerline{\includegraphics[width=12cm]{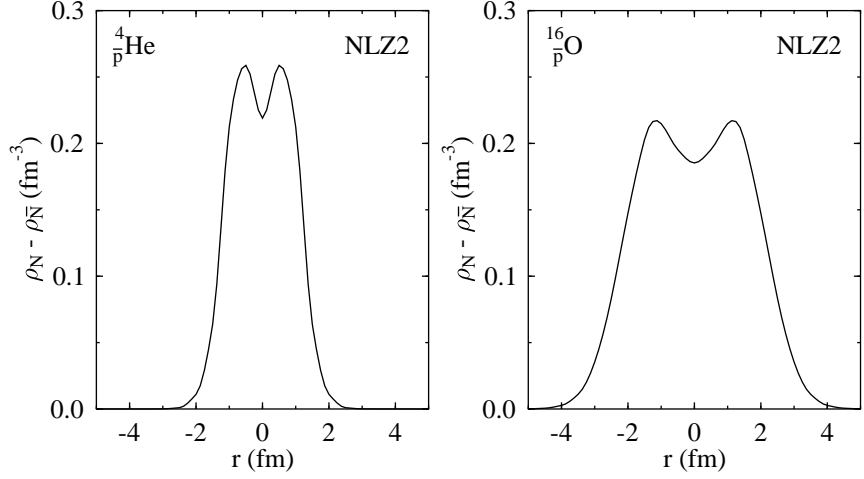}}
\caption{
The profiles of net baryon density, $\rho_N - \rho_{\ov{p}}$\,,
in the $\ov{p}\,+^{4}$He (left) and $\ov{p}\,+^{16}$O (right) systems,
calculated within the NLZ2 model.
}
\label{fig12}
\end{figure*}
Figure \ref{fig12} represents profiles of the net baryon
density i.e. the sum of the proton and neutron densities minus the
antiproton density. For both considered systems, there is a dip in the
center surrounded by the region with baryon density increased as
compared to normal nuclei.

\begin{table}[htb]
\caption{Total binding energies of $\ov{p}+^{16}$O system
with antiproton occupying different states.}
\vspace*{3mm}
\begin{ruledtabular}
\begin{tabular}{l|l|l|l|l}
$J_z^{\pi}$ & $1/2^+$ & $3/2^-$ & $1/2^-$ & $1/2^-$ \\\hline
NL3 & -1051 & -1008 & -920 &  -780   \\
NLZ2 &  -828 &  -938 &  -804 &  -876 \\
\end{tabular}
\end{ruledtabular}
\label{tab:states}
\end{table}
Now we address the question of wether the antiproton $1/2^+$ state is
energetically the most favorable configuration of the \Oap\, system or
not. To answer this questions, we put the antiproton into various
states and calculate the total binding energy. The results are shown in
Table \ref{tab:states}. In the case of NL3, the lowest $1/2^+$ state
indeed corresponds to the configuration with lowest energy. However,
within the NLZ2 calculation a state with the antiproton in the $3/2^-$
state corresponds to the ground state of the system.
\begin{figure*}[htb!]
\vspace*{-2cm}
\centerline{\includegraphics[width=11cm]{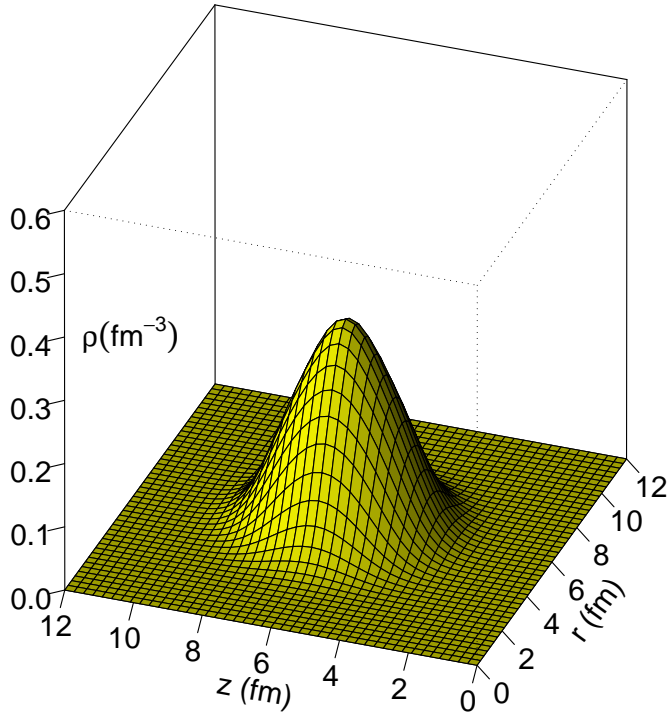}}
\vspace*{-3cm}
\caption{3D plot of nucleon density in the in the bound
$\ov{\Lambda}\,+^{16}$O system calculated with the NLZ model.}
\label{fig13}
\end{figure*}
The $3/2^-$ state has a different spatial distribution and thus, leads
to a different shape of the potential felt by nucleons. So, even though
the antiproton in the $3/2^-$ state is less bound than in the $1/2^+$
state, nucleons become more bound in this case and the
binding energy of the whole system increases.

\subsection{Light systems with antilambdas}

\begin{figure*}[thb!]
\centerline{\includegraphics[width=12cm]{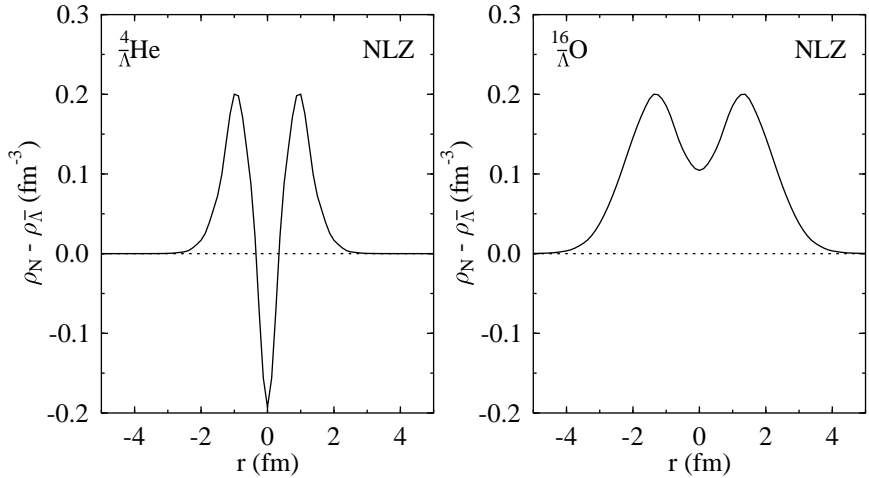}}
\caption{
Radial profiles of the net baryon density, $\rho_N -
\rho_{\ov{\Lambda}}$\,, in the $\ov{\Lambda}\,+^{4}$He (left) and
$\ov{\Lambda}\,+^{16}$O (right) systems calculated with the NLZ model.
}
\label{fig14}
\end{figure*}
Bound states of nuclei with antihyperons are especially interesting
because we expect longer life times in this case
(see Sect.~V).
We performed calculations of
the~$^{4}\hspace*{-6pt}$\raisebox{-5pt}{$\sst\ov{\Lambda}$}\hsp He and
\Oal\, systems within the NLZ model. The assumed $\ov{\Lambda}$ couplings
with the meson fields are given in Table~\ref{tab:parl}.

Figure~\ref{fig13} shows the 3D plot of the sum of proton and neutron
densities in the \Oal\, nucleus. Comparison with the results for the
\Oap\, systems (see Fig.~\ref{fig7}) shows noticeably lower nucleon
densities in this case. However, we still predict significant
compression of nuclei containing antihyperons. Figure~\ref{fig14}
gives profiles of the net baryon density in the
$^{4}\hspace*{-6pt}$\raisebox{-5pt}{$\sst\ov{\Lambda}$}\hsp He and
\Oal\, nuclei. It is interesting that the central density dip is even
more pronounced here as compared to nuclei with antiprotons (see
Fig.~\ref{fig12}). This effect can be explained by a stronger
localization of the antilambda due to its larger effective mass.

\begin{figure*}[htb!]
\centerline{\includegraphics[height=8cm]{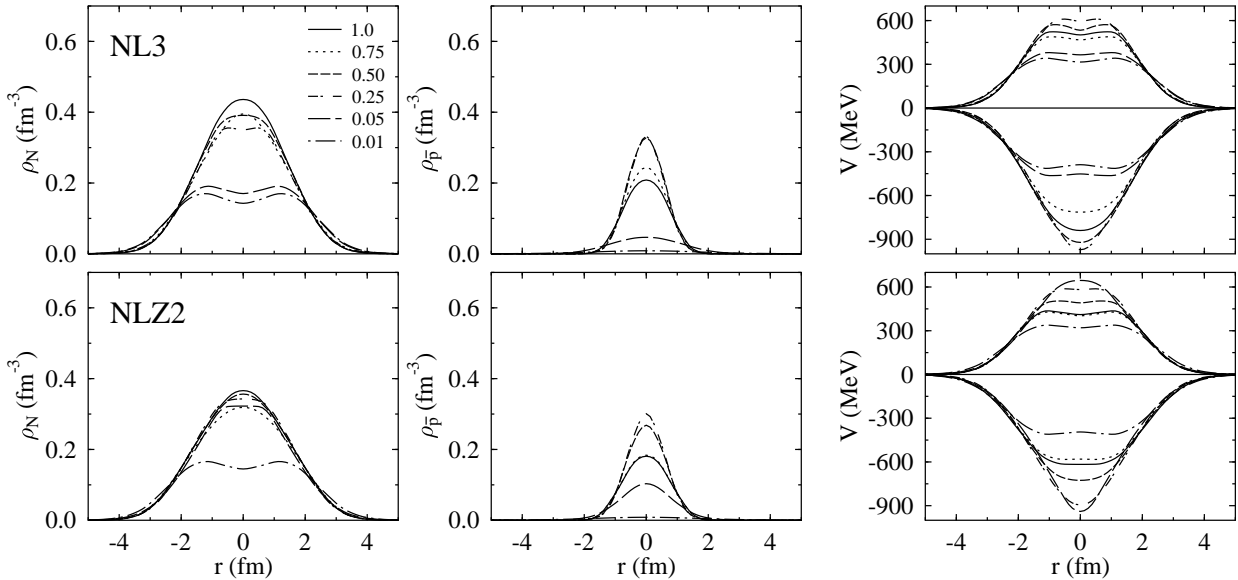}}
\caption{The profiles of densities and mean--field potentials in the
bound $\ov{p}\hsp+^{16}$O system calculated with reduced antiproton
coupling constants. The upper (lower) panel corresponds to the NL3
(NLZ2) model. Different curves correspond to different values of the
para\-meter~$\xi$\, defined in~\re{gpar1}. Shown are the nucleon
densities (left), antiproton densities (middle) and scalar and vector
potentials for nucleons (right). Note, that at $\xi<1$ the antiproton
potentials are smaller due to reduced $\ov{p}$ couplings.}
\label{fig15}
\end{figure*}

\subsection{Calculations with reduced antibaryon couplings}

As was pointed out earlier, the G--parity transformation may not work
on the mean--field level. Therefore, we have performed calculations of
bound $\ov{p}+A$ systems with reduced antiproton couplings to mean
meson fields. Figure~\ref{fig15} shows the results for
the~\Oap\, nucleus obtained for several values of the parameter $\xi$
introduced in~\re{gpar1}.

\begin{figure*}[htb!]
\centerline{\includegraphics[width=10cm]{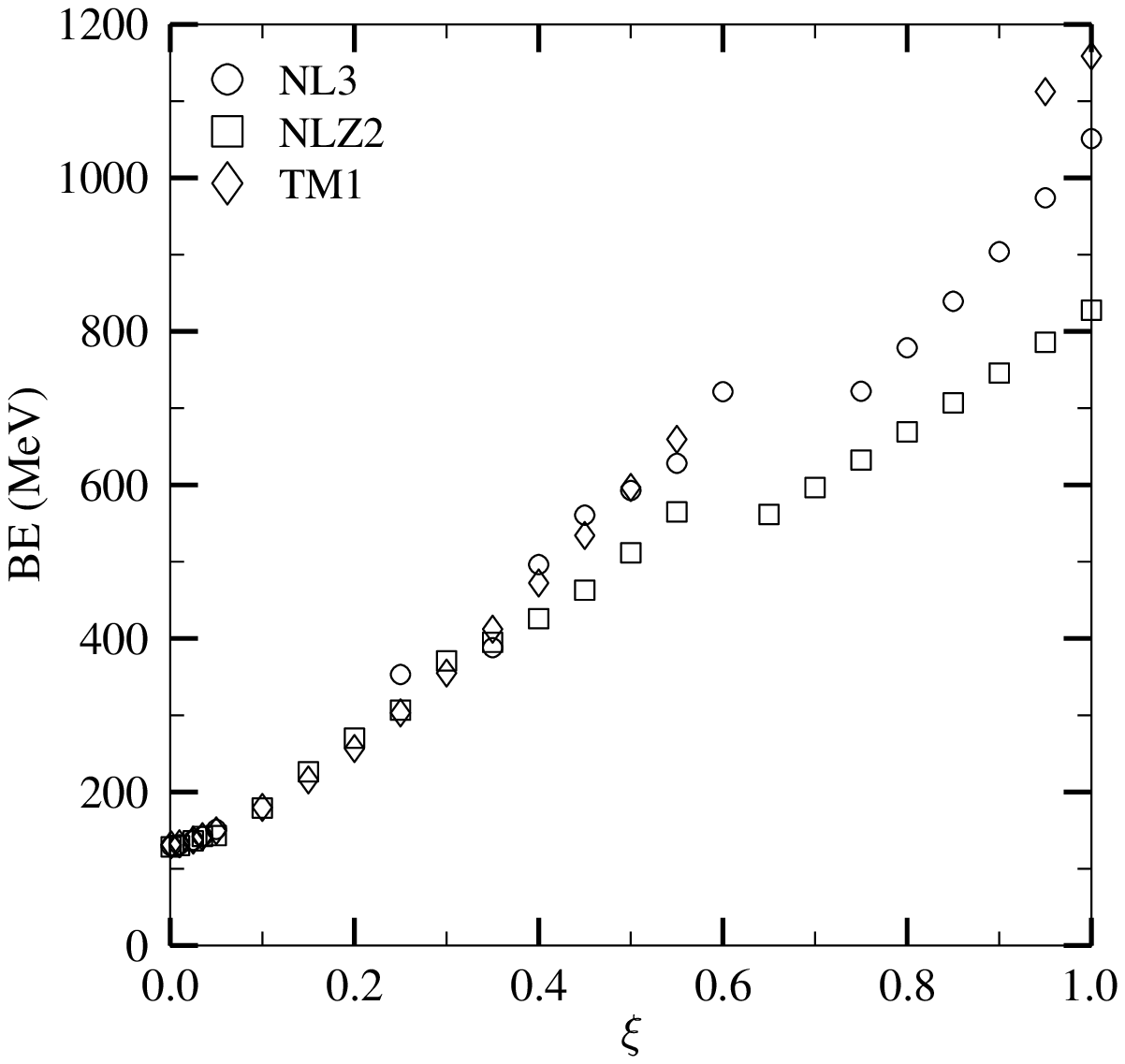}}
\vspace*{-2mm}
\caption{
Binding energy of $^{16}_{\hspace*{3pt}\ov{p}}$O calculated in
the TM1 and NLZ models as a function of the parameter $\xi$\,.}
\label{fig16}
\end{figure*}
One can see that at $\xi\gtrsim 0.25$, the compressed shape of the
nucleus is not very much affected. The largest compression, of course,
occurs for $\xi=1$, i.e. in the case of the exact G--parity.  At $\xi$
smaller than $0.05$ (NL3) or $0.01$ (NLZ2), the nucleon density
profiles practically coincide with the density distribution of the
normal $^{16}$O nucleus. On the other hand, the antiproton density
profile becomes rather flat in this case. Somewhere in between, there
exists a critical $\xi$ value separating these two regimes.  Detailed
calculations with different $\xi$ show that there is no smooth
transition but rather an abrupt jump. This is an indication that, to a
large extent, the shell effects control the structure of these systems.

\begin{figure*}[htb!]
\centerline{\includegraphics[height=8cm]{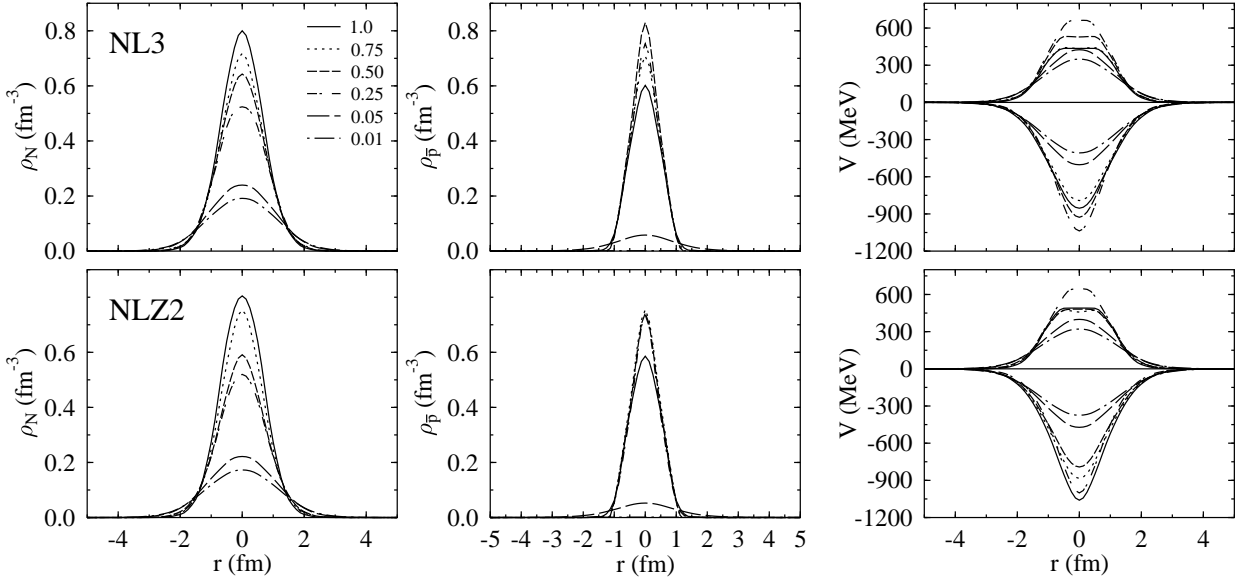}}
\caption{\label{fig17}Same as in Fig.~\ref{fig15} but for
$^{4}$He with an antiproton.
}
\end{figure*}
The results of these calculations show that the conclusions made in
Ref.~\cite{Bue02} and in the present work do not strongly depend on
the actual values of $g_{\sigma\ov{N}}$ and $g_{\omega\ov{N}}$. As
a~consistency check, we verified that at $\xi\to 0$ the binding energy
of \Oap\, goes to the value of normal~$^{16}$O~\footnote{
In the same limit the antiproton wave function
resembles the one for a free particle.
}.
This is demonstrated in Fig.~\ref{fig16} where one can see
strong increase of the binding energy for $\xi\gtrsim 0.1$\,.

Similar results take place for the $\ov{p}\hsp +^{4}$He system. They
are presented in Fig.~\ref{fig17}. Of course, this is system is
rather small and the mean-field approximation may be not accurate in
this case.  Nevertheless, we expect to get a qualitative picture even
for such a light system. At $\xi=1$ the nucleon density in the
$^4\hspace*{-5pt}$\raisebox{-4pt}{$\sst\ov{p}$}\hsp He nucleus reaches
a maximum value of about $6\hsp\rho_0$ which is noticeably larger than
in \Oap\hsp. As in the case of~\Oap\hsp, depending on the $\xi$ value,
there are two different configurations of the bound system, namely a
highly compressed one and the other resembling normal helium plus a
quasi free antiproton. For both RMF models, the reduction of antiproton
coupling constants up to~$\xi\simeq 0.25$ does not produce a strong
effect in the density distribution.

\subsection{Rearrangement of nuclear structure and Dirac sea}

\begin{figure*}[htb!]
\centerline{\includegraphics[width=15cm]{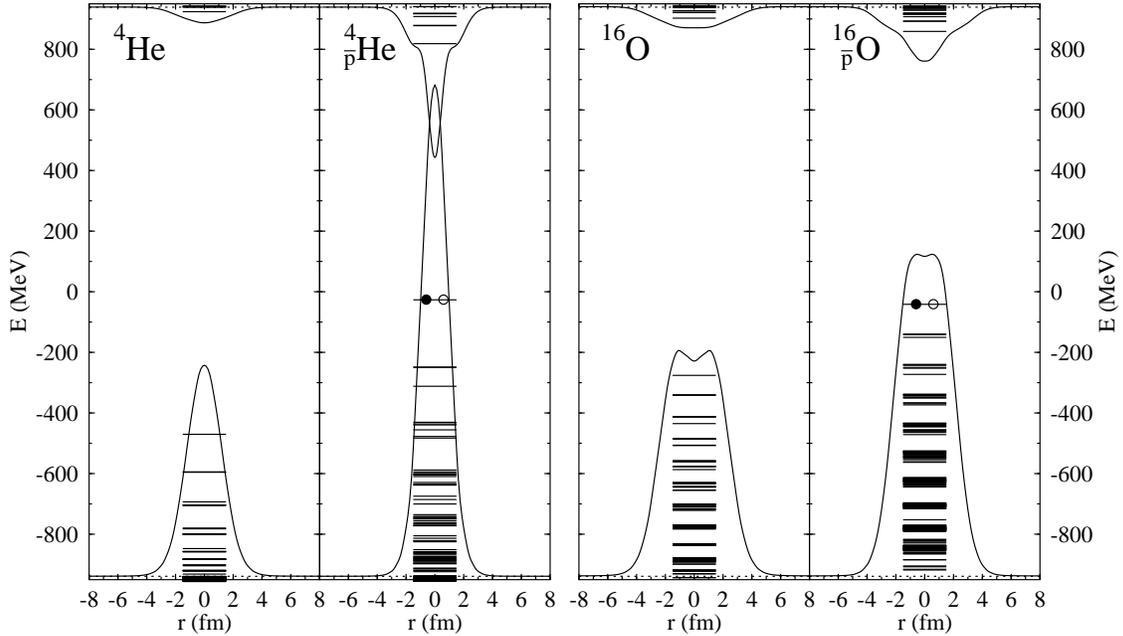}}
\caption{
Proton and antiproton potentials as well as
corresponding single--particle levels for the systems $^{4}$He,
$\ov{p}\,+^{4}$He (left side) and $^{16}$O, $\ov{p}\,+^{16}$O (right side)
calculated within the NLZ2 model. The full and open dots in the Dirac sea
show the most bound occupied and vacant states, respectively.
}
\label{fig18}
\end{figure*}
The presence of an antiproton within a light nucleus leads to a drastic
rearrangement of its structure and to the polarization of the Dirac
sea. We illustrate this effect in Fig.~\ref{fig18}, where the
effective potentials and single--particle levels are shown for helium
and oxygen, in each case with and without an antiproton. A hole at the
deepest bound level in the lower well corresponds to a positive energy
antiproton. After inserting an antiproton the nuclear structure is
strongly  rearranged. Both the nucleonic as well as the antiparticle
states become deeper which gives rise to increase of the binding energy
by several hundred~MeV. The lower well in the case of the
$\ov{p}\,+^{4}$He system is very narrow and its bottom overlaps with
the positive energy well. The uncertainty relation prevents, however,
the strongest bound antiparticle and particle states to be very close
in energy. The lower well in the case of oxygen also becomes narrower
as compared with normal $^{16}$O nucleus.

The rearrangement of the Dirac sea due to the presence of an antiproton
leads to increasing number of negative energy states. The spin--orbit
potential for antiparticles is very small, because, in contrast to
nucleons, scalar and vector potentials for antiprotons nearly cancel
each other~\footnote{
This leads to the so--called spin symmetry of the antinucleon
states discussed in Ref.~\cite{Zho03}.
}.
Still, there are gaps between the antiparticle levels of
the order of a hundred MeV (see Fig.~\ref{fig10}). This is a
consequence of a smaller width of the lower well as compared to the
negative energy well in the normal nucleus.

\subsection{Heavy nuclei containing antibaryons}

\begin{figure*}[htb!]
\centerline{\includegraphics[width=16cm]{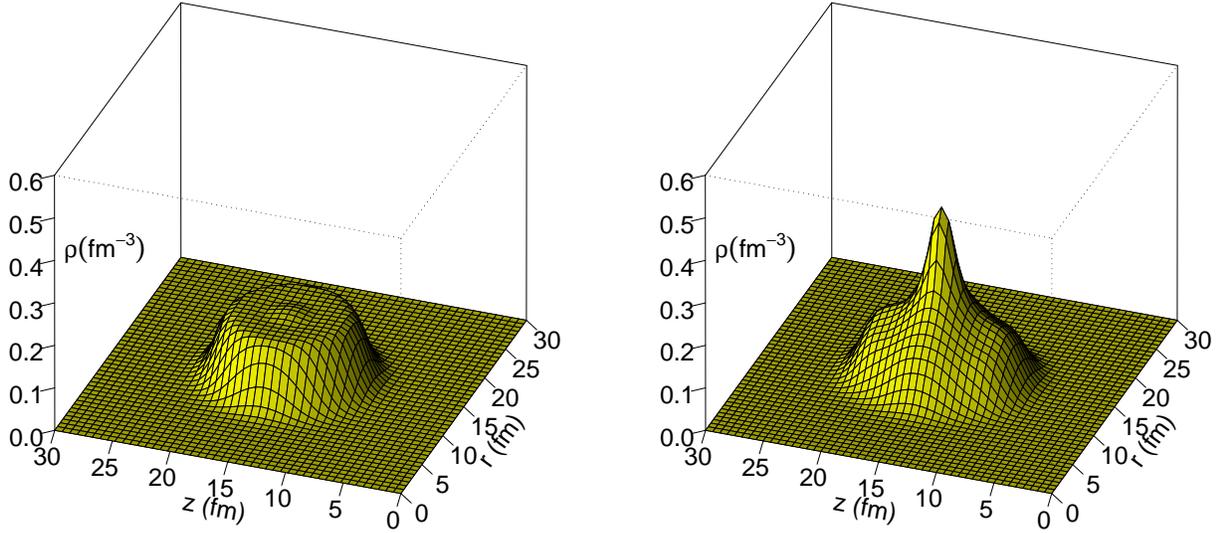}}
\caption{
3D plots of nucleon density in the $^{208}$Pb~nucleus (left) and in the
bound $\ov{p}\,+^{208}$Pb\, system calculated within the NL3 model.}
\label{fig19}
\end{figure*}

\begin{figure*}[htb!]
\vspace*{-3cm}
\centerline{\includegraphics[width=11cm]{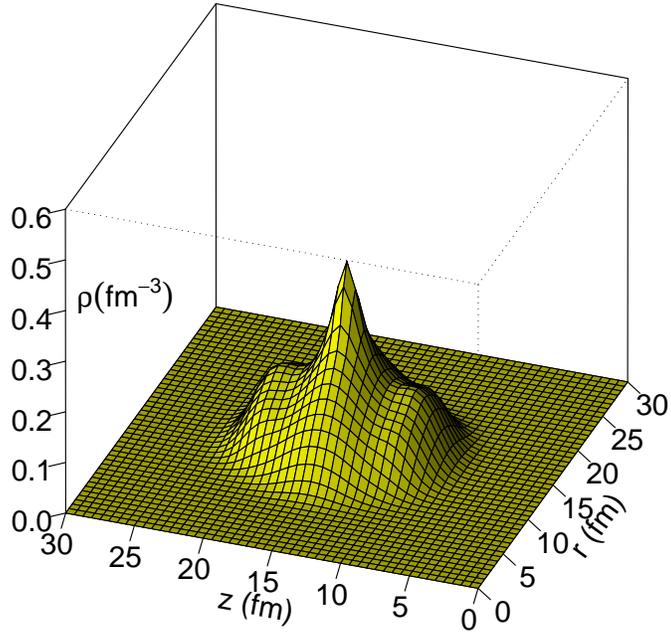}}
\vspace*{-3cm}
\caption{
3D plots of nucleon density in the bound $\ov{\Lambda}\,+^{208}$Pb
system calculated within the NLZ model.}
\label{fig20}
\end{figure*}

\begin{figure*}[htb!]
\vspace*{4mm}
\centerline{\includegraphics[height=6.5cm]{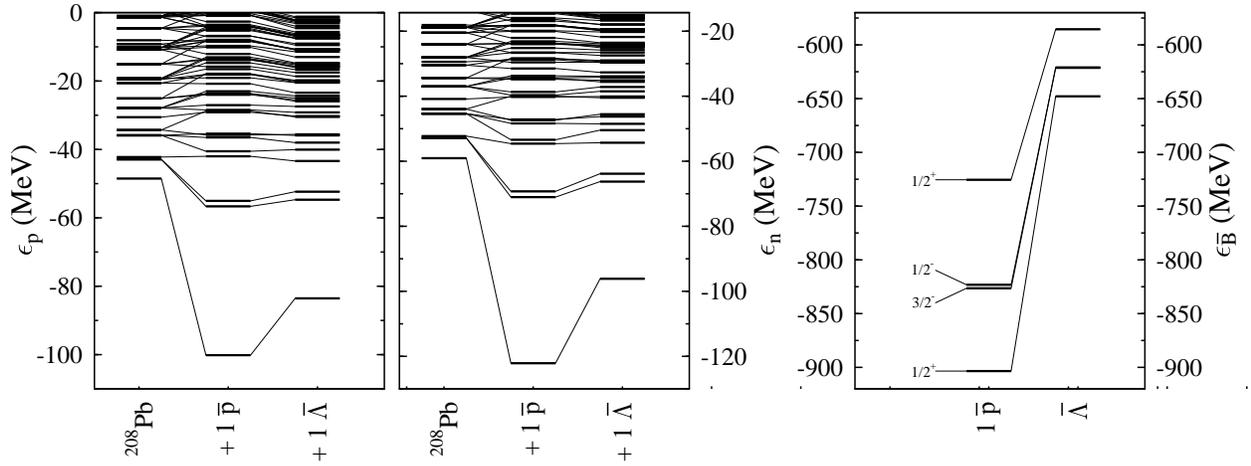}}
\vspace*{-4mm}
\caption{
Single--particle proton (left), neutron (middle) and antiproton  spectra
for the systems $^{208}$Pb, $\ov{p}\,+^{208}$Pb (NL3 forces) and
$\ov{\Lambda}\,+^{208}$Pb (NLZ2 set).}
\label{fig21}
\end{figure*}

In this section we investigate the effect of antiprotons and antilambdas
inserted into heavy nuclei. Because of larger radii of these objects
($R\gtrsim 7$~fm), one may expect appearance of a local compression
zone in the central region instead of a more homogeneously compression
in lighter systems.
This is exactly what follows from our calculation.
Fi\-gu\-re~\ref{fig19} shows the sum of proton and neutron densities, for
the case of doubly--magic lead nucleus with one deeply--bound
antiproton. Due to the presence of an antiproton, a small core of
highly compressed nuclear matter appears at in the center of the
\Pbap\, nucleus. In addition, the lead nucleus becomes deformed and
acquires a prolate shape.
As shown in Fig.~\ref{fig20} a~similar structural change occurs when
implementing the $\ov{\Lambda}$ particle into the $^{208}$Pb target.
Figure~\ref{fig21} shows single--particle spectra of protons, neutrons
and antiprotons in the normal lead as well as in the bound \Pbap\, and
$^{208}\hspace*{-8pt}$\raisebox{-5pt} {$\sst\ov{\Lambda}$} Pb nuclei.
One can see that implementing antiparticles results in a strong change
of shell structure. Due to the axial deformation, the system looses its
degeneracy and the shell structure becomes partly washed out.  Only the
$1/2+$ and $3/2-$ levels exhibit large binding. The
deepest antibaryon states have binding energies of more than 900 (600) MeV
in the case of the \mbox{\Pbap}
($^{208}\hspace*{-8pt}$\raisebox{-5pt} {$\sst\ov{\Lambda}$} Pb)
system. On the other hand, some of the single--particle states become even less
bound in the lead nucleus with antibaryons.

These results can be qualitatively understood from the fact that the
nucleon potential in the center of the bound system is very deep but
narrow. Two effects prevent deeper binding of single--particle levels.
First, a narrow potential gives rise to large uncertainties in particle
momenta. This in turn increases the kinetic energies of particles and
therefore, reduces their binding. Second, the single--particle states
with larger angular momenta are mainly localized at larger radii
and thus do not have much overlap with the deep central region of
the potential.
The total binding energy of the \Pbap\, nucleus predicted by the NL3 calculation
equals 2412 MeV or 11.5 MeV per particle. This is significantly smaller than
61.8 MeV per particle in the \Oap\, system. Furthermore, the total energy gain
in the binding energy after inserting an antibaryon into the $^{208}$Pb nucleus
(780 MeV) is smaller than in the $^{16}$O case (920 MeV).

\begin{figure*}[htb!]
\centerline{\includegraphics[width=16cm]{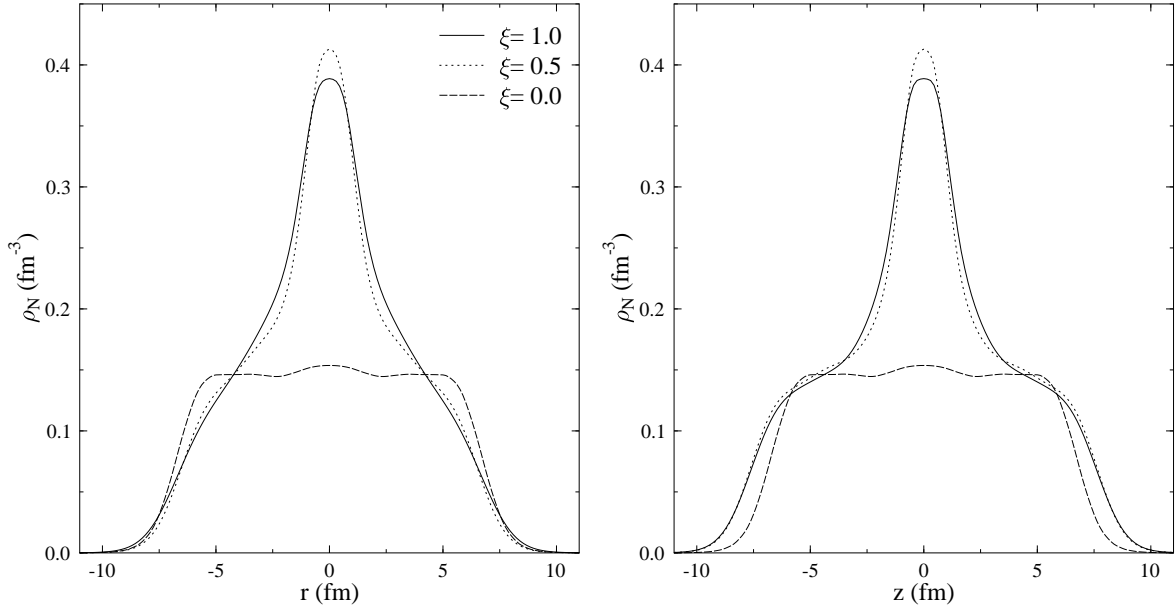}}
\caption{The profiles of nucleon density in the $^{208}_{\hspace*{6pt}\ov{p}}$\,Pb
system at different values of the paramater~$\xi$\,. The left and right panels
correspond, respectively, to the equatorial and coaxial planes.
}
\label{fig22}
\end{figure*}
Figure~\ref{fig22} shows the nucleon density profiles in the
\Pbap\, system calculated with $\xi=1,0.5$ and 0 (the normal Pb
nucleus). Because of the axial deformation we show separately the
profiles for radial (equatorial) and coaxial planes.
This figure demonstrates a very interesting behavior. Contrary
to a naive expectation, reducing $\xi$ from 1 to 0.5 leads to increasing
density of the central core~\footnote
{
In the case of the $\ov{p}$\,+O system this effect
is seen in the antiproton density profiles (see Fig.~\ref{fig15}).
}.
We believe that this can be explained
by the competition of two opposite trends taking place at decreasing
$\xi$\,. The first one is the reduction of the binding potential
while the second one is the increase of the antiproton effective mass.
The second effect leads to a stronger localization of the antiproton
wave function near the minimum of the effective potential. As a
consequence, the potential acting on surrounding nucleons
increases in the central regions.

\subsection{Systems containing several antibaryons}

It is amazing to consider nuclear systems with more than one
trapped antibaryon. For instance let us consider
doubly--magic oxygen containing different
numbers of antiprotons, namely $N_{\ov{p}}=2,4,6,8$\, and 10.
\begin{figure*}[htb!]
\centerline{\includegraphics[width=10cm]{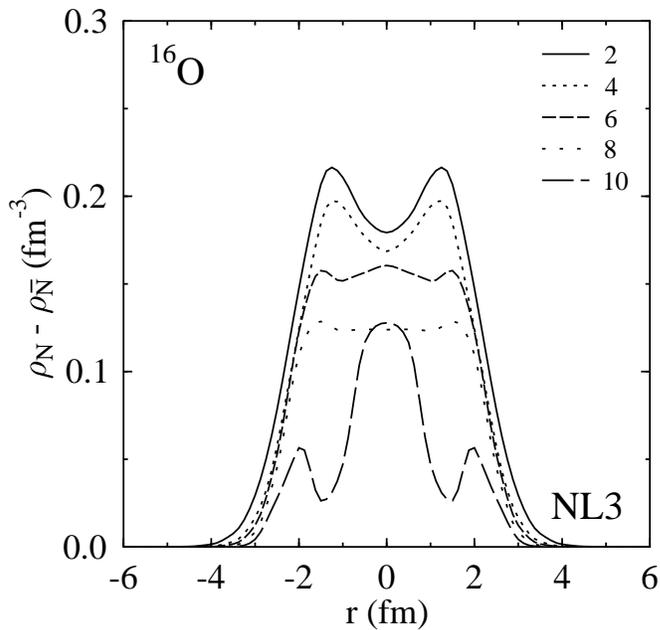}}
\caption{
Net baryon density, $\rho_N - \rho_{\ov{p}}$, for the $^{16}$O nucleus
with 2, 4, 6, 8 and 10 antiprotons, calculated within the
NL3 model.}
\label{fig23}
\end{figure*}
The corresponding baryon densities are shown in
Fig.~\ref{fig23}. The total binding energies for the
systems containing 2, 4, 6, 8 and 10 antiprotons are 1541, 2792, 3847,
5006, and 6300 MeV, respectively. As we can expect, with increasing
number of antiprotons, the region of reduced net baryon density becomes
broader. In the case $N_{\ov{p}}=10$ one can see a qualitative change
of nuclear structure, with appearance of a baryon density peak in the
center region and a~zone of reduced density at the nuclear periphery.

\begin{figure*}[htb!]
\centerline{\includegraphics[width=14cm]{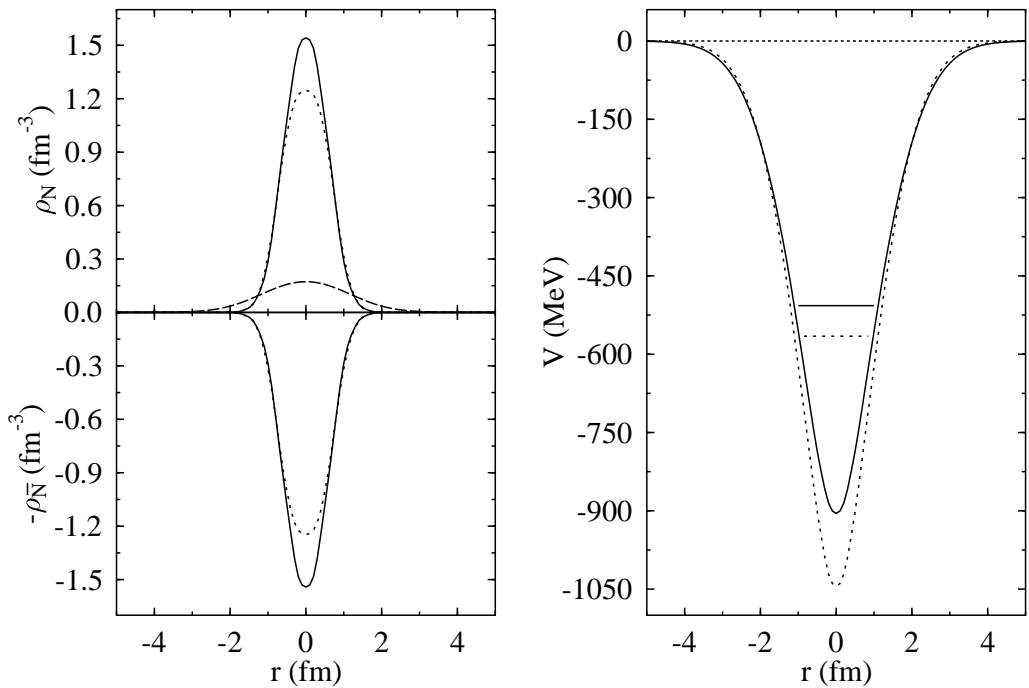}}
\caption{
The system $\alpha-\ov{\alpha}$ calculated within the NLZ2 (solid line) and NL3
(dotted line) models.
Shown are baryon and antibaryon densities (left) as well as
the scalar potential and energies of the lowest single--particle
states (right). The dashed line shows the nucleon density
of the normal~$\alpha$ particle calculated within the NLZ2 model.}
\label{fig24}
\end{figure*}
One can also think of most extreme case of finite systems with
equal numbers of baryons and antibaryons, i.e. about systems
with total baryon number $B=0$\,. These systems can be called
self--conjugate nuclei, since their charge conjugation
leads to the same object. Let us first consider the bound
$\alpha-\ov{\alpha}$ system. As seen in Fig.~\ref{fig24},
the (anti)particle densities in this system
reach about $10\hsp\rho_0$\,. Of course, one should expect
a breakdown of the RMF model with nucleonic degrees of freedom
at such high densities. Most probably the (anti)baryons will dissolve
into a deconfined state of cold and dense $q\ov{q}$ matter
(see Sect.~III).

\begin{figure*}[htb!]
\centerline{\includegraphics[width=14cm]{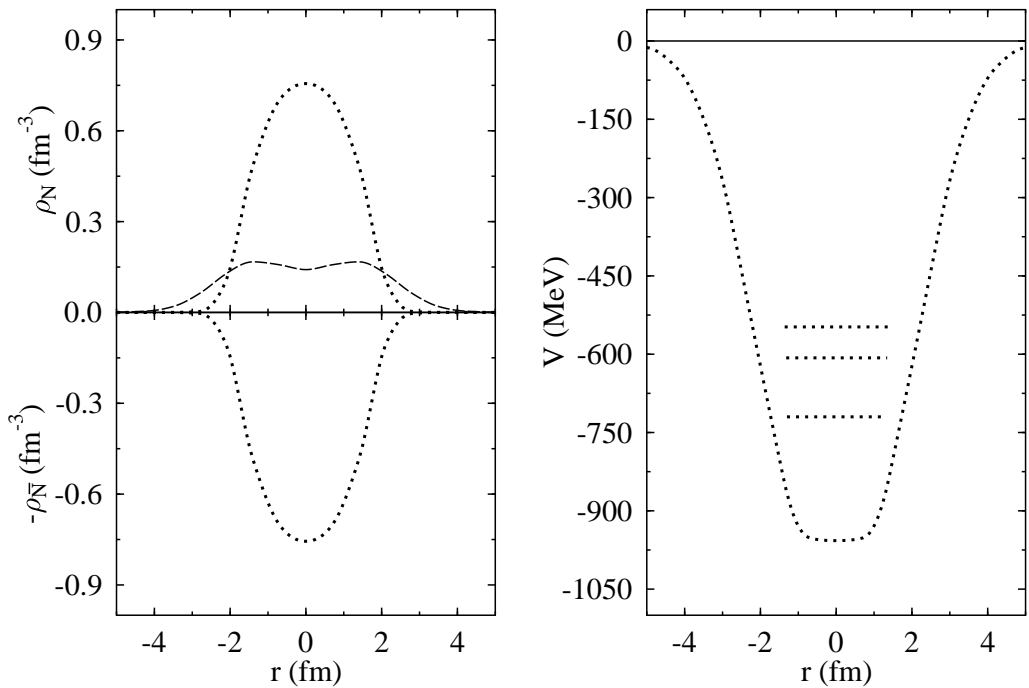}}
\caption{
The system  $^{16}$O--$^{16}\ov{{\rm O}}$  calculated within the NL3
model. The dotted lines show the baryon and antibaryon densities (left) as well as the
scalar potential and the lowest single--particle states (right). The dashed line
represents the nucleon density of the normal $^{16}$O nucleus.}
\label{fig25}
\end{figure*}
Due to the full symmetry between baryons and antibaryons (or quark and
antiquarks) such systems have both vanishing baryon and electric
charges. Accordingly, the vector fields, namely the $\omega$ meson and
Coulomb fields, vanish too.  Only the scalar density and with it the
scalar potential are nonzero in a symmetrical system.  Since nucleons
and antinucleons have the same scalar potential, they can occupy the
same single--particle states. In this case the spatial parts of the wave
functions of protons, neutrons, antiprotons and antineutrons  are
identical. Hence, the spatial overlap between particles and
antiparticles is maximal.  On the other hand, due to the very high
density of baryons and antibaryons, the effective masses are very small
in the core region.  The calculated binding energy of the
$\alpha-\ov{\alpha}$~system is very large, it is 2649 MeV for NL3 and
2235 MeV for NLZ2. This is about 100 times the value for a single
$\alpha$ particle and 40 times the binding energy of normal $^{8}$Be
nucleus.

As a second example we consider the $^{16}$O--$^{16}\ov{{\rm O}}$
system. The calculated profiles of (anti)nucleon density and scalar
potential are shown in Fig.~\ref{fig25}~\footnote{
Note that we were not able to achieve convergence of the iteration
procedure in the NLZ2 calculation.
}.
In this case we predict formation of a highly compressed system
with maximal nucleon and antinucleon densities of
about $5\rho_0$\,. One can see in Figs.~\ref{fig24}--\ref{fig25}
that radial profiles of the scalar potential in symmetric
$A+\ov{A}$\, systems are noticeably broader as compared to the (anti)nucleon
density distributions. Such a behavior follows from the finite range of $\sigma$
meson field (see \re{mess}).

\section{Annihilation of deeply bound antibaryons}

\subsection{$\ov{N}N$ annihilation in vacuum}

At low energies the total cross section of $\ov{p}p$
annihilation in vacuum can be well approximated as~\cite{Dov92}
\bel{ancr}
\sigma^{\rm\hsp ann}_{\ov{p}p}=C+\frac{D}{~v_{\rm rel}}\,,
\ee
where $v_{\rm rel}$ is the relative velocity of $\ov{p}$\,
with respect to the proton, $C=38$\,mb, $D=35\,{\rm mb}\cdot {\rm c}$\,.
In the following we neglect the isotopic and Coulomb effects
assuming that $\ov{p}p$ and $\ov{p}n$ annihilation cross
sections are approximately equal.
Using this approximation we discuss below the isotopically
averaged $\ov{N}N$ interactions.

When the $\ov{N}N$ c.m. energy $\sqrt{s}$ is close to the vacuum
threshold value, i.e. at \mbox{$\sqrt{s}\to 2m_N\simeq 1.88$\,GeV}, the
relative velocity becomes small and the $\ov{N}N$ annihilation cross
section is approximately proportional to~$1/v_{\rm rel}$~\footnote
{
At very low antiproton kinetic energies, below about 200 keV in the
lab frame, the Coulomb effects in $\ov{p}p$ annihilation become
important~\cite{Lan77}. In this case $\sigma^{\rm ann}_{\ov{p}p}$
deviates from the parametrization~(\ref{ancr}) increasing approximately
as $1/v_{\rm rel}^2$ at $v_{\rm rel}\to 0$.
}.
This limiting case of annihilation ''at rest'' ($v_{\rm rel}\to 0$)
has been extensively studied in Refs.~\cite{Bac83,Sed88,Obe97,
Cry94,Cry96,Ams98,Cry03}. Experimental data on many exclusive
annihilation channels $\ov{p}p\to c\to n_{\pi}\pi$  are also available.
Here intermediate states $c$ include direct pions as well as heavy
mesons $\eta\hsp (547), \rho\hsp (770), \omega\hsp (782) \ldots$

\begin{table}[p]
\vspace*{-1.5cm}
\caption{Exclusive channels of $\ov{p}p$ annihilation at rest in
vacuum}
\label{tab:anbr}
\vspace*{3mm}
\begin{ruledtabular}
\begin{tabular}{l|l|c|c|l}
$n_\pi$&   $c$  &$\sqrt{s}_{\rm thr} ({\rm GeV})$  & $B_c (\%)$ & Refs.\\
\colrule
2& $2\pi^0$~~& 0.27 & 0.07\hsp & \cite{Cry03}\\[-1.5mm]
~& $\pi^+\pi^-$~~& 0.28 & 0.31 & \cite{Ams98}\\
\hline
3& $(3\pi^0)_{\rm\hsp dir} $ & 0.41 & 0.7\hspace*{7pt}& \cite{Cry03}\\[-1.5mm]
~& $(\pi^+\pi^-\pi^0)_{\rm\hsp dir}$ & 0.42
& 1.8\hsp\footnote{~see~\re{bexa}}
&\cite{Ams98}
 \\[-1.5mm]
~& $\pi^0\rho^0$~~& 0.91 & 1.7\hspace*{7pt}  & \cite{Ams98}\\[-1.5mm]
~& $\pi^\pm\rho^\mp$~~& 0.91 & 3.4\hspace*{7pt}  & \cite{Ams98}\\
\hline
4& $(4\pi^0)_{\rm\hsp dir}$ & 0.54 &
0.5\hsp\footnote{~average between two values given in Ref.~\cite{Cry03}}
&\cite{Cry03}
\\[-1.5mm]
~& $(\pi^+\pi^-\hsp 2\pi^0)_{\rm\hsp dir}$ & 0.55 &
7.8\hsp\footnote{~calculated as
$B_{\hsp\ov{p}p\to\pi^+\pi^-2\pi^0}-B_{\hsp\ov{p}p\to\pi^0\omega}-
B_{\hsp\ov{p}p\to\rho^+\rho^-}$}
&\cite{Bac83,Sed88,Ams98}
\\[-1.5mm]
~& $(2\pi^+2\pi^-)_{\rm\hsp dir}$ & 0.56 &
4.2\hsp\footnote{~calculated as
$B_{\hsp\ov{p}p\to 2\pi^+2\pi^-}-B_{\hsp\ov{p}p\to\pi^+\pi^-\rho^0}$}
&\cite{Sed88,Obe97}
\\[-1.5mm]
~& $\pi^0\omega$ & 0.92 & 0.6\hspace*{5pt}  & \cite{Ams98}\\[-1.5mm]
~& $\pi^+\pi^-\rho^0$ & 1.05 &
3.6\hsp\footnote{~average between two values referred in
Ref.~\cite{Sed88}}
&\cite{Sed88}
\\[-1.5mm]
~& $\rho^+\rho^-$ & 1.54 & 0.9\hspace*{5pt} & \cite{Sed88}\\
\hline
5& $(5\pi^0)_{\rm\hsp dir}$ & 0.68 &
0.5\hsp\footnote{~calculated as
$B_{\hsp\ov{p}p\to 5\pi^0}-B_{\hsp\ov{p}p\to 2\pi^0\eta}
B_{\hsp\eta\to 3\pi^0}$}
&\cite{Cry94,Cry96}
\\[-1.5mm]
~& $(\pi^+\pi^-\hsp 3\pi^0)_{\rm\hsp dir}$ & 0.69 &
20.1\hsp\footnote{~calculated as
$B_{\hsp\ov{p}p\to\pi^+\pi^-3\pi^0}-
B_{\hsp\ov{p}p\to 2\pi^0\eta} B_{\hsp\eta\to\pi^+\pi^-\pi^0} -
B_{\hsp\ov{p}p\to\pi^+\pi^-\eta} B_{\hsp\eta\to 3\pi^0} -
B_{\hsp\ov{p}p\to 2\pi^0\omega}$}
&\cite{Bac83,Cry94,Cry03}
\\[-1.5mm]
~& $(2\pi^+2\pi^-\pi^0)_{\rm\hsp dir}$ & 0.70 &
10.4\hsp\footnote{~calculated as
$B_{\hsp\ov{p}p\to 2\pi^+2\pi^-\pi^0}-
B_{\hsp\ov{p}p\to\pi^+\pi^-\eta} B_{\hsp\eta\to\pi^+\pi^-\pi^0} -
B_{\hsp\ov{p}p\to\pi^+\pi^-\omega}-B_{\hsp\ov{p}p\to\rho^0\omega}$}
&\cite{Sed88,Cry94,Ams98}
\\[-1.5mm]
~& $2\pi^0\eta$ & 0.82 & 0.7  & \cite{Cry94}\\[-1.5mm]
~& $\pi^+\pi^-\eta$ & 0.83 & 1.3  & \cite{Cry94}\\[-1.5mm]
~& $2\pi^0\omega$ & 1.05 & 2.6 & \cite{Cry03}\\[-1.5mm]
~& $\pi^+\pi^-\omega$ & 1.06 & 6.6 & \cite{Ams98}\\[-1.5mm]
~& $\rho^0\omega$ & 1.55 & 2.3 & \cite{Ams98}\\
\hline
6& $(\pi^+\pi^-4\pi^0)_{\rm\hsp dir} $ & 0.82 &\hspace*{2pt}
1.9\hsp\footnote{~calculated as
$B_{\hsp\ov{p}p\to\pi^+\pi^-n\pi^0\,(n\geq 2)}-
B_{\hsp\ov{p}p\to\pi^+\pi^-2\pi^0}-B_{\hsp\ov{p}p\to\pi^+\pi^-3\pi^0}$
using data from Ref.~\cite{Sed88}}
&\cite{Bac83,Sed88}
\\[-1.5mm]
~& $(2\pi^+2\pi^-2\pi^0)_{\rm\hsp dir}$ & 0.83 &
13.3\hsp\hsp\footnote{~calculated as
$B_{\hsp\ov{p}p\to 2\pi^+2\pi^-2\pi^0}-
B_{\hsp\ov{p}p\to\omega\eta} B_{\hsp\eta\to\pi^+\pi^-\pi^0} -
B_{\hsp\ov{p}p\to 2\omega}$}
&\cite{Bac83,Cry03}
\\[-1.5mm]
~& $(3\pi^+3\pi^-)_{\rm\hsp dir}$&0.84&\hspace*{2pt}2.0&\cite{Sed88}\\[-1.5mm]
~& $\omega\eta$ & 1.32 & 1.5  & \cite{Cry03}\\[-1.5mm]
~& $2\omega$ & 1.54 & 3.0 & \cite{Cry03}\\
\hline
7& $(2\pi^+2\pi^-3\pi^0)_{\rm\hsp dir}~~$ & 0.97 &\hspace*{2pt}
4.0\hsp\footnote{~calculated as $B_{\hsp\ov{p}p\to 2\pi^+2\pi^-3\pi^0} -
B_{\hsp\ov{p}p\to\pi^0\omega\eta} B_{\hsp\eta\to\pi^+\pi^-\pi^0}$}
&\cite{Bac83,Cry03}
\\[-1.5mm]
~& $(3\pi^+3\pi^-\pi^0)_{\rm\hsp dir}$ & 0.98 &\hspace*{1pt} 1.9\hspace*{3pt}  &
\cite{Sed88}\\[-1.5mm]
~& $\pi^0\omega\eta$ & 1.47 &\hspace*{2pt} 1.0\hspace*{3pt}  &
\cite{Cry03}\\
\end{tabular}
\end{ruledtabular}
\end{table}
Various channels of the $\ov{p}p$ annihilation at rest
are listed in Table~\ref{tab:anbr}. They are sorted into groups
with different final pion multiplicities $n_\pi$\,.
The most important exclusive channels $c$ are given in the second column.
In the third column we give the corresponding
threshold energies in the c.m. frame (i.e. the total mass
of particles in the intermediate state). The fourth column shows
the observed branching ratio of a given channel $c$:
\bel{brra}
B_c=\sigma_{\ov{p}p\to c}/\sigma^{\rm\hsp ann}_{\ov{p}p}\,.
\ee
In the last column we give references to publications
where experimental information on branching ratios has been
found. The subscripts ''dir'' in the second column mean that
only nonresonant contributions are given in the
corresponding line. For example,
the branching ratio of the $(\pi^+\pi^-\pi^0)_{\rm\hsp dir}$
channel has been calculated using the formula
\bel{bexa}
B_{\ov{p}p\to(\pi^+\pi^-\pi^0)_{\rm\hsp dir}}=
B_{\hsp\ov{p}p\to\pi^+\pi^-\pi^0}-
B_{\hsp\ov{p}p\to\pi^0\rho^0}-B_{\hsp\ov{p}p\to\pi^\pm\rho^\mp}\,,
\ee
where the first term in the r.h.s. is the total branching ratio of the
reaction $\ov{p}p\to\pi^+\pi^-\pi^0$ and the two other terms give the
contributions of the $\pi\rho$ intermediate states.  Here we take into
account that $\rho\to 2\pi$ is the dominant channel of the $\rho$ meson
decay~\cite{Pdg02}. The channels $c=3\pi^0,\hsp
4\pi^0,\hsp\pi^+\pi^-4\pi^0,\hsp 3\pi^+3\pi^-$ and $3\pi^+3\pi^-\pi^0$
are classified as direct, because no resonance contributions have been
found there. In Table~\ref{tab:anbr} we include only main annihilation
channels with $B_c>0.5\hsp\%$ (for $n_{\pi}>2$). We do not separate
resonances in channels with total number of intermediate mesons
exceeding three.

\subsection{Kinetic approach to annihilation in medium}

Let us first consider annihilation of slow antibaryons, $\Bb$\,, in
homogeneous nucleonic matter with a small admixture of antibaryons.
Using again the mean--field approach we treat the $\Bb+N$ matter as a
mixture of ideal Fermi gases of nucleons and antibaryons interacting
with mean meson fields. Assuming that annihilation on multinucleon
clusters is relatively small (see below), we consider here only the
binary $\BbN$ annihilation. Then the local rate of $\Bb$ annihilation
per unit volume can be written as a sum over all $\BbN$ annihilation
channels with production of $n\geq 2$ mesons in the final state:
\bel{lra}
\frac{dN_{\rm ann}}{dtdV}=\sum_n
\frac{dN_{\BbN\to M_1\hsp\ldots\hsp M_n}}{dtdV}\,.
\ee
Here $M_l\hsp (l=1,\ldots, n)$ denote secondary mesons. In the following we
disregard in--medium modifications of these mesons, assuming that
their 4--momenta $k_l$ satisfy the mass--shell constraints,
$k_l^2=m_l^2$\,, with vacuum masses $m_l$~\footnote{
This choice requires a special comment. It is well--known from microscopic
calculations that light mesons $\pi, K, \rho \ldots$ acquire strong modifications
in nuclear medium (see e.g.~\cite{Mig74,Mig90,Wam03}). Generally speaking, these
modifications originate from two types of self--energy diagrams associated
with short-- and long--range processes. The short--range contribution is proportional
to the local density i.e. requires an additional nucleon in the annihilation zone.
Moreover, the $s$--wave contribution for pions and $\rho$--mesons is proportional
to the isovector density which vanishes in the isosymmetric matter. We classify such
processes as multi--nucleon annihilation and consider them in Sect.~VF.
The long--range contributions are generated by the nucleon--hole and resonance--hole
diagrams which are characterized by the energy scale of about 100 MeV.
Therefore, they can only affect propagation of mesons at large distances
(exceeding about 2 fm) or at long times. This is why we believe that these in--medium
effects are irrelevant for the annihilation process.
}.

In the quasiclassical approximation one can describe the phase--space
density of antibaryons ($i=\Bb$) and nucleons ($i=N$)  by
distribution functions $f_i(x,\bm{p})$\,, where \mbox{$x=(t,\bm{r})$}
is the space--time coordinate and $\bm{p}$ is the 3--momentum of the $i$--th
particle. Within the mean--field model,
the kinematic part of the single--particle energy is written as
\bel{drel}
E_i=\sqrt{m_i^{* 2}+\bm{p}^2}\,,
\ee
where $m_i^*$ is the effective mass of the $i$--th particles ($i=\Bb, N$).
Their vector and scalar densities are determined by integrals
over 3--momenta:
\begin{eqnarray}
&&\rho_i(x)=\int d^3p\hsp f_i(x,\bm{p})\,,\label{vden}\\
&&\rho_{Si}(x)=\int d^3p\hsp\frac{m_i^*}{E_i}
f_i(x,\bm{p})\,.\label{sden}
\end{eqnarray}
At zero temperature we use Fermi
distributions
$f_i=\frac{\ds\nu_i}{\ds (2\pi)^3}\hsp\Theta (p_{Fi}-|\bm{p}|)$\,,
where $\nu_i$ is the spin--isospin degeneracy factor of the
$i$--th particles, $p_{Fi}=(6\pi^2\rho_i/\nu_i)^{1/3}$ is their Fermi
momentum and $\Theta (x)\equiv (1+{\rm sgn}\hsp x)/2$~\footnote{
Here and below we consider the static $\Bb+N$ matter in
its rest frame.
}.

Within the kinetic approach (see \mbox{e.g.~Ref.~\cite{Cas02}}), the rate of
the reaction \mbox{$\BbN\to c$}~(\mbox{$c=M_1\hsp\ldots M_n$})
can be written as
\begin{eqnarray}
&&\frac{dN_{\BbN\to c}}{dtdV}=
\int\frac{d^3p_{\Bb}}{E_{\Bb}}f_{\Bb}\hsp (x,\bm{p}_{\Bb})
\int\frac{d^3p_N}{E_N}f_N(x,\bm{p}_N)\times\nonumber\\
&&\int\prod\limits_{l=1}^{n}\frac{d^3k_l}{E_l}\hsp
W_c\hsp (P=p_{\Bb}+p_N|\hsp k_1,\ldots,k_n)\hsp\delta^{(4)}
(\mbox{$P-k_1-\ldots -k_n$})\,.\label{anc1}
\end{eqnarray}
Here $W_c$ is the $\BbN\to c$ transition probability,
$p_i=(E_i,\bm{p}_i)$  and $k_l=(E_l,\bm{k}_l)$ are the 4--momenta
of the initial particles and
the final mesons.

Following the standard arguments of a statistical approach
we assume that the transition probabilities
do not depend sensitively on the particles' momenta so that $W_c$
can be replaced by constants,
\bel{wapp}
W_c\hsp (P|\hsp k_1,\ldots,k_n)\simeq W_c={\rm const}\,.
\ee
Then the contributions of different reaction channels
are determined mainly by their invariant phase space volumes
\bel{phsi}
R_c(s)=\int\prod\limits_{l=1}^{n}\frac{d^3k_l}
{E_l}\hsp\delta^{(4)}(P-\sum\limits_{l=1}^{n}k_l)\,,
\ee
where $s=P^2=(p_{\Bb}+p_N)^2$ is the c.m. energy squared available
for the reaction.

Now \re{anc1} can be written as
\bel{anc2}
\frac{dN_{\BbN\to c}}{dtdV}=\rho_{\Bb}\hsp\rho_N W_c\left<\frac{R_c(s)}
{E_{\Bb}\hsp E_N}\right>\equiv\rho_{\Bb}\hsp\Gamma_c\,,
\ee
where $\Gamma_c$ is the partial annihilation width for channel $c$\,.
The angular brackets denote averaging over the momentum distribution
of incoming particles,
\bel{aver}
\left<{\cal O}\right>\equiv\int
\prod\limits_{i=\Bb,N}\hspace*{-3pt}\left(d^3p_i\hsp\frac{f_i(x,\bm{p}_i)}
{\rho_i(x)}\right){\cal O}\,.
\ee
where ${\cal O}$ is an arbitrary function of 3--momenta
$\bm{p}_{\Bb},\bm{p}_N$\,.

Let us consider first the limit of dilute matter, i.e.
$\rho_{\Bb},\rho_N\to 0$. In this case, \mbox{$p_{Fi}\to 0$},
\mbox{$m_i^*\to m_i$} ($i=\Bb,N$) and the distribution functions
$f_i(x,\bm{p})$ can be formally replaced
by~$\rho_i\hsp\delta^{(3)}(\bm{p})$\,.  As a consequence,
at~$\rho_{\Bb}, \rho_N\to 0$ we get the following relation:
\bel{dill}
\frac{dN_{\BbN\to c}}{dtdV}\simeq\rho_{\Bb}\hsp\rho_N W_c\frac{R_c(s_0)}
{m_{\Bb}\hsp m_N}\,.
\ee
Here $s_0=(m_{\Bb}+m_N)^2$ corresponds to the energy
available for $\BbN$ annihilation at rest
in vacuum. On the other hand, in this limit the reaction rate can be
expressed~\cite{Cas02} through the
vacuum cross section of the $\BbN$ annihilation at $v_{\rm rel}\to 0$:
\bel{dill1}
\frac{dN_{\BbN\to c}}{dtdV}\simeq\rho_{\Bb}\hsp\rho_N\hsp
(\sigma_{\BbN\to c}v_{\rm rel})_{\hsp 0}\,,
\ee
where the subscript ''$0$'' indicates that the quantity in brackets is taken
at \mbox{$\bm{p}_{\Bb,N}\to 0$}\,. By comparing
Eqs.~(\ref{dill}) and (\ref{dill1}) one can express the
transition probabilities $W_c$
by experimental cross sections of the $\BbN$ annihilation:
\bel{trpe}
W_c=\frac{m_{\Bb}\hsp m_N}{R_c(s_0)}(\sigma_{\BbN\to c}
v_{\rm rel})_{\hsp 0}\,.
\ee

In a general case of finite densities, the partial annihilation width,
defined in~\re{anc2}, can be expressed as
\bel{paw}
\Gamma_c=\Gamma_{0c}\left<\lambda_{\hsp c}(s)\hsp\frac{m_{\Bb}\hsp m_N}
{E_{\Bb}\hsp E_N}\right>\,,
\ee
where
\bel{paw0}
\Gamma_{0c}=\rho_N\hsp (\sigma_{\BbN\to c}v_{\rm rel})_{\hsp 0}\,,
\ee
is the axillary width, corresponding to the vacuum cross section, and
\bel{psrf}
\lambda_{\hsp c}(s)=\frac{R_c(s)}{R_c(s_0)}
\ee
is the phase--space suppression factor, which plays a central
role in our estimates.

In nuclear medium the single--particle energies of antibaryons and nucleons
are modified by the scalar and vector potentials. The minimum energy
of a $\BbN$ pair available for annihilation, i.e. the reaction $Q$ value,
is
\bel{qval}
Q=\sqrt{s_*}=m_{\Bb}^*+V_{\Bb}+m_N^*+V_N\,,
\ee
where $m_i^*$ and $V_i$ are effective masses and vector potentials
of particles ($i=\Bb, N$). In the case of antinucleons with the G--parity
transformed potentials $m_{\Nb}^*=m_N^*$ and $V_{\Nb}=-V_N$ hence
$Q=2m_N^*$\,.

This value can be significantly reduced as compared to the minimal
energy in vacuum, $\sqrt{s_0}$\,, which~leads to strong suppression of
the available phase space for annihilation products. Actual c.m.
energies of $\BbN$ pairs have a certain spread due to the Fermi motion
of nucleons. It easy to show that even at $\rho_N\sim 2\rho_0$ the
variation of the $Q$ values does not exceed 10\%. Taking into account
that uncertainties in scalar and vector potentials are of the same
order, we simplify the general expression (\ref{paw}) by replacing
$\lambda_{\hsp c}(s)$ with $\lambda_{\hsp c}(s_*)$\,.  Then the partial
annihilation width in the medium can be represented in a simple form
\bel{paw1}
\Gamma_c\simeq\Gamma_{0c}\hsp\lambda_{\hsp c}(s_*)\hsp J_{\BbN}\,.
\ee
Here
\bel{ovl}
J_{\BbN}=\phi_{\Bb}(x)\phi_N(x)\,,~~~\phi_i(x)=
\left<\frac{m_i}{E_i}\right>=
\frac{m_i}{m_i^*}\hsp\frac{\rho_{Si}}{\rho_i}~~(i=\Bb,N)\,.
\ee
Returning to the initial expression (\ref{anc2}) we see that the
reaction rate is proportional to the product of the scalar densities of
annihilating particles. In fact, this is required by the Lorentz
invariance of \re{anc2}, because factors $W_c$ and $\lambda_c$ are
Lorentz invariants.

The direct calculation of the scalar densities
$\rho_{Si}$ using Fermi distributions in \re{sden} yields
\bel{phii}
\phi_i=\frac{3}{2}\frac{m_i}{p_{Fi}}\hsp\Phi\hsp (\frac{m_i^*}{p_{Fi}})\,,
\ee
where $\Phi\hsp (z)$ is the dimensionless function defined in~\re{phif}.
For systems with small admixture of antibaryons, i.e.
when $p_{F\Bb}\ll p_{FN}, m_{\Bb}^*$\,,
one can use the approximate formulae
\bel{phbb}
\Phi\hsp (\frac{m_{\Bb}^*}{p_{F\Bb}})\simeq\frac{2}{3}\hsp\frac{p_{F\Bb}}
{m_{\Bb}^*}\,,~~~
\phi_{\Bb}\simeq\frac{m_{\Bb}}{m_{\Bb}^*}\,.
\ee
Here $m_{\Bb}^*$ is determined for pure nucleonic
matter with $\rho_{\Bb}=0$\,.

\subsection{Phase space suppression factors}

Let us now consider in more detail the phase space suppression factors
$\lambda_{c}$ introduced by~\re{psrf}.
\begin{figure*}[htb!]
\vspace*{-8cm}
\centerline{\includegraphics[width=12cm]{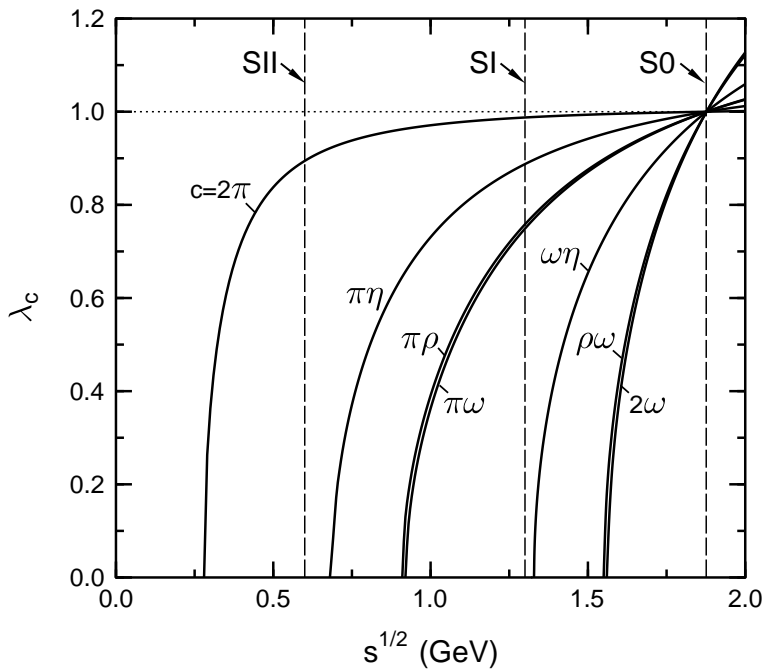}}
\caption
{Phase space suppression factors for $\NbN$ annihilation into two mesons
as functions of c.m. energy in the subthreshold region.
The vertical dashed lines indicate $\sqrt{s_*}$ values adopted
in the calculations
for infinite nuclear matter (see subsection D).}
\label{fig26}
\end{figure*}
In the particular case of the two--body annihilation channels
$\lambda_{c}$ can be calculated
analytically. Direct calculation using~\re{phsi} gives
\bel{phsi2}
R_{\ov{B}N\to M_1 M_2}(s)=2\pi\sqrt{\left[1-\frac{(m_1+m_2)^2}{s}\right]
\left[1-\frac{(m_1-m_2)^2}{s}\right]}\Theta (\sqrt{s}-m_1-m_2)\,,
\ee
where $m_1$ and $m_2$ are meson masses.
Figure~\ref{fig26} presents the results for the two--body channels
involving pions and the heavy mesons~$\eta, \rho, \omega$\,.
One can see a strong decrease
of $\lambda_c (s)$ in the subthreshold region, $s<s_0$,
which is getting more prominent for heavier final states.

\begin{figure*}[htb!]
\vspace*{-8cm}
\centerline{\includegraphics[width=12cm]{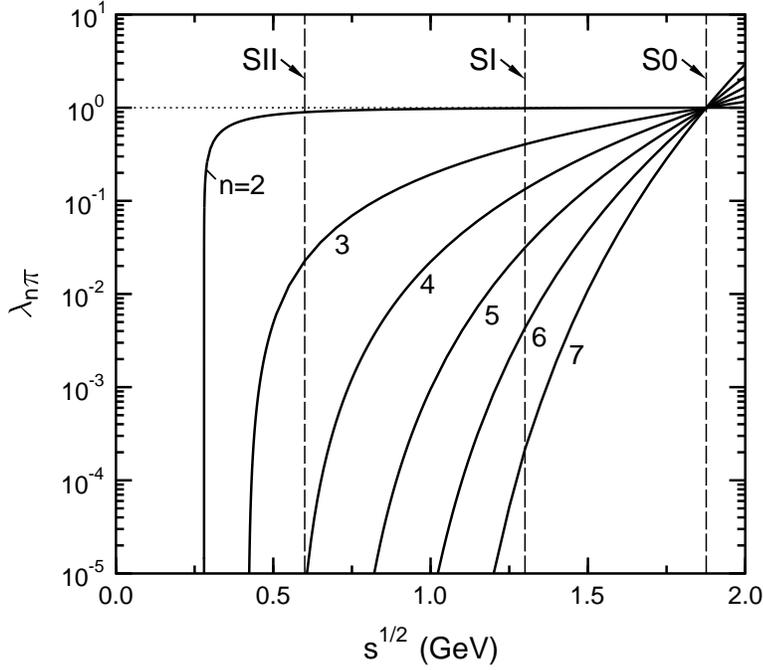}}
\caption
{Same as Fig.~\ref{fig26}, but for $\NbN$
annihilation into $n$ direct pions.
Note the change to a log scale on the vertical axis.}
\label{fig27}
\end{figure*}

For channels with more than two secondary mesons,
the multidimensional integrals in \re{phsi} are evaluated numerically
using the Monte Carlo method.
Figure~\ref{fig27} shows the energy dependence of factors
$\lambda_{\hsp c}$ for nonresonant channels
\mbox{$\ov{N}N\to (n\pi)_{\rm\hsp dir}$} with different pion
multiplicities $n$\,.
One can see that at given $s$ the reduction of phase space becomes more
and more important with increasing $n$\,. In the case
of the lightest annihilation channel,
$c=2\pi$\,, the phase space is only slightly reduced in the
subthreshold energy region. The noticeable deviation of
$\lambda_{2\pi}$ from unity takes place only in the vicinity of
the~$2\pi$ threshold, i.e. at $\sqrt{s}\to 2m_{\pi}$\,. On the
other hand, for $n>2$ factors $\lambda_{n\pi}(s)$  decrease very rapidly
with decreasing~$s$\,. For example, at $\sqrt{s}=1$\,GeV, the
phase space factors \mbox{$\lambda_{n\pi}\simeq 0.97,\, 0.19, 0.014$}
for \mbox{$n=2,3,4$}\,, respectively.

It is interesting to note that the trend in behavior of the phase space factors
above the threshold, $\sqrt{s}>2m_N$, changes to opposite, i.e. the multi--pion channels
become more and more important with increasing $\sqrt{s}$\,. This observation was used
earlier~\cite{Bra03} to explain fast chemical equilibration in nuclear collisions.
Thus, these two trends are complementary, but not inconsistent with each other.

\subsection{Annihilation widths in nuclear medium}

To calculate the factors $\Gamma_{0c}$ which enter
into the in--medium annihilation
widths (\ref{paw1}) we use vacuum branching ratios,
$B_c=\sigma_{\BbN\to c}/\sigma_{\BbN}^{\rm\hsp ann}$\,.
In the case of antinucleons ($\Bb=\ov{N}$) one has
\bel{anw01}
\Gamma_{0c}=B_c\rho_N\hsp (\sigma_{\ov{N}N}^{\rm\hsp ann}
\hsp v_{\rm rel})_{\hsp 0}=
\Gamma_0\hsp B_c\hsp\frac{\rho_N}
{\rho_0}~{\rm MeV}\,.
\ee
The numerical value $\Gamma_0=104$\,MeV is obtained when
the parametrization~(\ref{ancr}) is used for
the total $\NbN$ annihilation cross section.
In the calculations below we use the experimental $B_c$ values from
Table~\ref{tab:anbr}.

As an illustration, let us consider the
$\ov{N}N\to c$ partial widths for the following
two annihilation channels: $c=2\pi$ and $c=\pi\rho$\,.
\begin{figure*}[htb]
\vspace*{-7.5cm}
\centerline{\includegraphics[width=12cm]{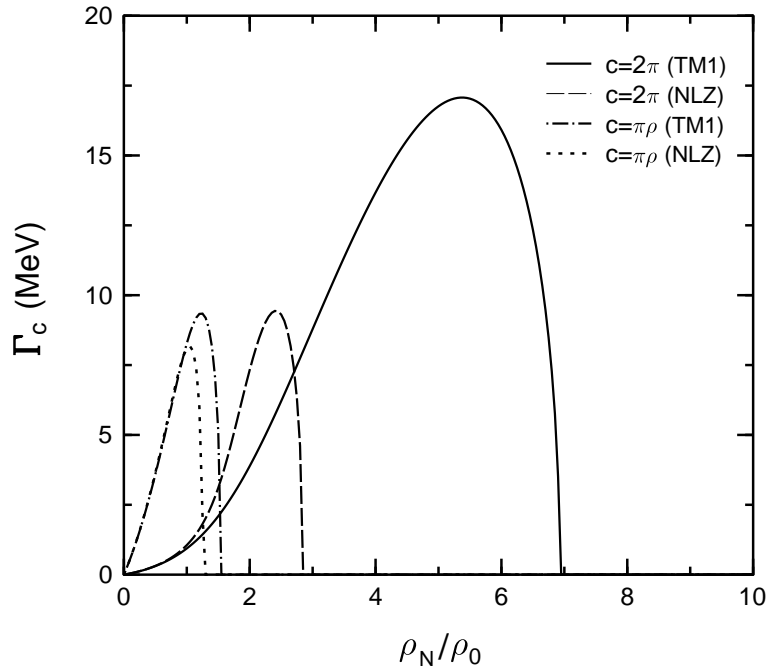}} \caption
{The $\ov{N}N\to c\,\,(c=2\pi, \pi\rho)$ partial widths in
nucleonic matter with density $\rho_N$\,.}
\label{fig28}
\end{figure*}
The in--medium widths are calculated from
\mbox{Eqs.~(\ref{paw1})--(\ref{anw01})}, in the limit
$\rho_{\ov{N}}\to 0$\,.  The corresponding branching rations
are obtained from the obvious relations
$B_{2\pi}=B_{2\pi^0}+B_{\pi^+\pi^-}$ and
$B_{\pi\rho}=B_{\pi^0\rho^0}+B_{\pi^{\pm}\rho^{\mp}}$\,, using data from
Table~\ref{tab:anbr}. To study the model dependence, the predictions of
two RMF models, NLZ~\cite{Ruf88} and TM1~\cite{Sug94}, have been
compared. In both cases, the antibaryon couplings are chosen according
to G--parity transformation, which leads to equal effective masses of
nucleons and antinucleons.

The widths $\Gamma_{2\pi}$ and $\Gamma_{\pi\rho}$ as functions
of nuclear density are shown in Fig.~\ref{fig28}.
At low nucleon densities, these widths deviate only slightly from the
linear dependence \mbox{$\Gamma_{0c}\propto\rho_N$}\,. However, at
\mbox{$\rho_N\gtrsim\rho_0$}, when the effective mass $m_N^*$ drops
significantly, the density dependence of the in--medium annihilation
widths becomes strongly nonlinear. Since the phase--space factors
$\lambda_{\hsp c} (s_*)$ vanish at \mbox{$\sqrt{s_*}=2\hsp
m_N^*<m_1+m_2$}\,, both channels become forbidden at large enough
densities.
\begin{figure*}[htb!]
\vspace*{-7.5cm}
\centerline{\includegraphics[width=12cm]{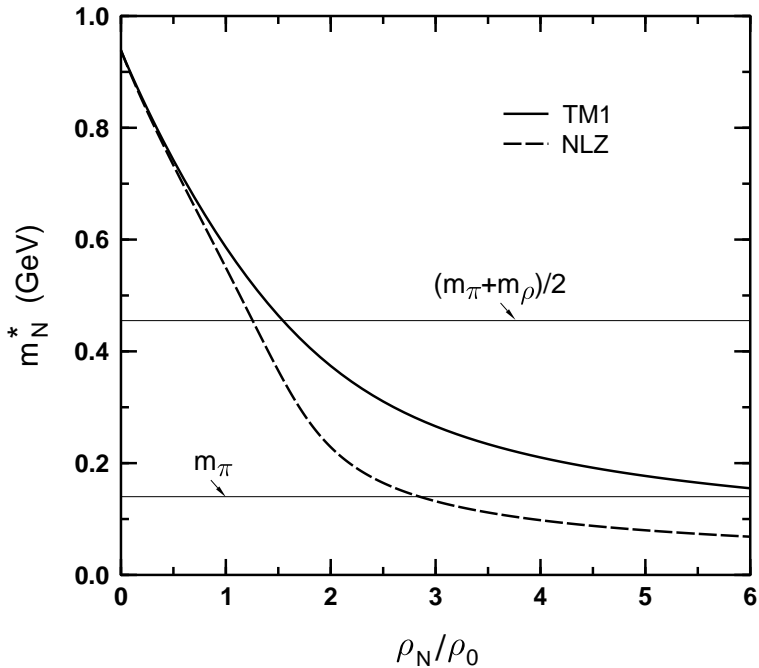}}
\caption
{Effective nucleon mass $m_N^*$ as a function of density
predicted by the NLZ and TM1 models.
Thin horizontal lines show threshold values of $m_N^*$
for $\ov{N}N$ annihilation
into $2\pi$ and $\pi\rho$ mesons.}
\label{fig29}
\end{figure*}
Figure~\ref{fig29} shows the density dependence of the nucleon
effective mass $m_N^* (\rho_N)$ for the same RMF models as in
Fig.~\ref{fig28}. One can see that the strong model dependence of
$\Gamma_{2\pi}$ is explained by a slower decrease of $m_N^*$ within the
TM1 model. It is interesting to note that all strong annihilation
channels would be closed when $m_N^*(\rho_N)<m_{\pi}$. In this case
only electromagnetic ($c=2\gamma $) or multi-nucleon (see subsection~F)
annihilation channels would limit the life time of an antinucleon in
nuclear medium.

It is instructive to calculate the annihilation widths
for the following three physically
interesting cases:
\begin{eqnarray}
{\rm ~S\hsp 0}:~~~~\rho_N=\rho_0\,,~~~~&&m_N^*=m_N\,,\label{ps0}\\
{\rm ~S\hsp I}:~~~~\rho_N=\rho_0\,,~~~~&&m_N^*=0.65\,{\rm GeV}\,,\label{ps1}\\
{\rm S\hsp II}:~~~\rho_N=2\rho_0\,,~~&&m_N^*=0.30\,{\rm GeV}\,.\label{ps2}
\end{eqnarray}
In the first parameter set, all in--medium
effects are disregarded, i.e.
\mbox{$s_*=s_0$}, \mbox{$\lambda_c=1$}, \mbox{$J_{\NbN}\simeq 1$}
and, therefore, $\Gamma_c\simeq \Gamma_{0c}$\,.
In the S\hsp I case we choose the density and effective mass which are
commonly accepted for equilibrium nuclear matter (without
antinucleons). Similar values are predicted by most RMF models. This
case is interesting for estimating the annihilation width in a
situation when the rearrangement effects due to the presence of
antibaryons are disregarded~\footnote
{
Formation of a deeply bound nucleus containing antibaryon is a
complicated dynamical process which takes a finite time. At the initial
stage of this process a target nucleus is not yet significantly
compressed. In this respect annihilation times calculated for the S\hsp
I case should be compared with the rearrangement time.
}.

For the case S\hsp II we take typical values for $\rho_N$ and $m_N^*$
as predicted by our calculations for the bound \Oap\, nucleus. Of
course, the main contribution to annihilation comes from the central
part of the nucleus where the antiproton is localized. As can be seen
in Fig.~\ref{fig9}, the antiproton wave function in the lowest
bound state of \Oap\, practically vanishes at
\mbox{$r\simeq R_{\ov{p}}=1.5$}\,fm. The values of density and effective
mass in the S\hsp II case correspond approximately to the values
obtained by averaging the $\rho_N$ and $m_N^*$ radial profiles in the
interval $r\leq R_{\ov{p}}$\,.

\begin{table}[ht]
\caption{Characteristics of $\ov{N}$ annihilation in cold nuclear
matter.}
\vspace*{3mm}
\label{tab:amed}
\begin{ruledtabular}
\begin{tabular}{l|c|c|c|c|c|c}
Parameter set&\multicolumn{2}{c|}{\mbox{S\hsp 0}}&
\multicolumn{2}{c|}{\mbox{S\hsp I}}&
\multicolumn{2}{c}{\mbox{S\hsp II}}\\
\cline{1-7}c &\hsp $B_c(\%)$\hsp &\hsp $\Gamma_{c}$ (MeV)
&$\lambda_c$&\hsp $\Gamma_{c}$ (MeV)
&$\lambda_c$&\hsp $\Gamma_{c}$ (MeV)\\
\colrule
$2\pi$& 0.38 & 0.39 & 0.99\,& 0.78 & 0.89 & 5.3\\
\colrule
$(3\pi)_{\rm\hsp dir}$& 2.5 & 2.6 & 0.40 & 2.1 & 0.023 & 0.91\\
$\pi\rho$& 5.1 & 5.3 & 0.76 & 8.0 & 0 & 0\\
\colrule
$(4\pi)_{\rm\hsp dir}$& 12.5 & 13.0 & 0.13 & 3.4 &
5.6$\cdot 10^{-6}$& 1.1$\cdot 10^{-3}$\\
$\pi\omega$& 0.6 & 0.62 & 0.75 & 0.93 & 0 & 0\\
$2\pi\rho$& 3.6 & 3.7 & 0.074\, & 0.55 & 0 & 0\\
$2\rho$& 0.9 & 0.93 & 0 & 0 & 0 & 0\\
\colrule
$(5\pi)_{\rm\hsp dir}$& 31.0 & 31.1 & 0.032 & 2.1 & 0 & 0\\
$2\pi\eta$& 2.0 & 2.1 & 0.16 & 0.66 & 0 & 0\\
$2\pi\omega$& 9.2 & 9.5 & 0.070 & 1.33 & 0 & 0\\
$\rho\hsp\omega$& 2.3 & 2.4 & 0 & 0 & 0 & 0\\
\colrule
$(6\pi)_{\rm\hsp dir}$& 17.2 & 17.8 & 4.2$\cdot 10^{-3}$ & 0.15 & 0 & 0\\
$\omega\eta$& 1.5 & 1.6 & 0 & 0 & 0 & 0\\
$2\hsp\omega$& 3.0 & 3.1 & 0 & 0 & 0 & 0\\
\colrule
$(7\pi)_{\rm\hsp dir}$& 5.9 & 6.1 & 1.9$\cdot 10^{-3}$ & 0.023 & 0 & 0\\
$\pi\omega\eta$& 1.0 & 1.0 & 0 & 0 & 0 & 0\\
\end{tabular}
\end{ruledtabular}
\end{table}
\begin{table}[ht]
\caption{
The widths of $\NbN\to n\pi$ channels as well as the
total width of $\ov{N}$ annihilation (all in MeV) for different
parameter sets introduced in Eqs.~(\ref{ps0})--(\ref{ps2}).
}
\vspace*{3mm}
\label{tab:awme}
\begin{ruledtabular}
\begin{tabular}{c|c|c|c}
$n$& S\hsp 0 & S\hsp I & S\hsp II\\
\hline
2 & 0.4 & 0.8 & 5.3 \\
3 & 7.9 & 10.1 & 0.9 \\
4 & 18.2\, & 4.8 & $10^{-3}$ \\
5 & 46.1 & 4.0 & 0 \\
6 & 22.5 & 0.2 & 0 \\
$\geq 7$ & 8.9\baselineskip 12pt\footnote{\,Obtained
as difference between
$\sum\limits_c\Gamma_{0c}\simeq 104$\,MeV (see \re{anw01})
and the total width
of $\NbN\to n\pi$ channels with $n\leq 6$\,.
}\baselineskip 24 pt
& 0.02\footnote{\,Only $n=7$ contribution included.} & 0 \\
\hline
{\rm total} & 104 & 19.9 & 6.2\\
\end{tabular}
\end{ruledtabular}
\end{table}
Table~\ref{tab:amed} presents numerical values of the \mbox{$\NbN\to c$}
widths $\Gamma_c$ calculated for the above parameter sets using
Eqs.~(\ref{paw1}), (\ref{anw01}).  In these calculations we use the
values \mbox{$J_{\NbN}\simeq 2.0$\,(S\hsp I)}\, and~7.6 (S\hsp II),
which follow from Eqs.~\mbox{(\ref{ovl})--(\ref{phbb})}.  In
Table~\ref{tab:amed}, annihilation channels are grouped according to the
multiplicity of pions in the final state $n$\,. As compared to the
vacuum extrapolated values (S\hsp 0), the in--medium annihilation
widths are significantly reduced for the channels with $n$ exceeding 3
( S\hsp I) and 2 (S\hsp II).  Many annihilation channels are strongly
suppressed or even completely closed due to reduced $Q$ values.  For
instance, the $(5\pi)_{\rm dir}$ channel, most important in the vacuum,
is suppressed by a factor $\sim 15$ in the case S\hsp I. All
channels with heavy mesons $\eta, \rho, \omega$ are closed in the case
S\hsp II.

Table~\ref{tab:awme} presents the $\NbN\to n\pi$ widths for
different total number of pions in the final state $n$\,. These widths are
calculated by summing partial widths of channels with the
same pion multiplicities
in Table~\ref{tab:amed}. The total annihilation width is obtained by summing
all partial contributions,
\bel{totw}
\Gamma=\sum\limits_{c}\Gamma_c\,.
\ee
Based on results presented in Table~\ref{tab:awme}, we conclude that
the $\Nb$ annihilation width in the medium can be suppressed by large
factors $\sim 5$ (S\hsp I) or even 15 (S\hsp II) as compared to the
naive estimate of the case S\hsp 0.  Corresponding life times are
\bel{ltme}
\tau\equiv\frac{\hbar}{\Gamma}\simeq
1.9~({\rm S\hsp 0}),~~9.9~({\rm S\hsp I})~,
~~32~({\rm S\hsp II})~{\rm fm/c}\,,
\ee
Life times in the range 10--30 fm/c open the possibility for
experimental studies of bound antinucleon--nucleus systems.

\begin{figure}[ht]
\vspace*{-7.5cm}
\centerline{\includegraphics[width=12cm]{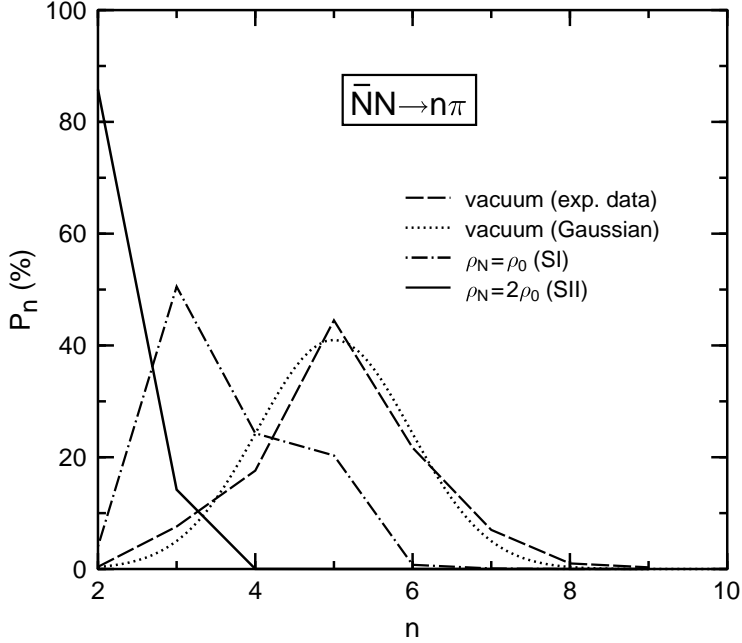}}
\caption{Probabilities of the $\ov{N}N\to n\pi$ annihilation as a
function of the pion multiplicity $n$. Dotted lines corresponds to
the Gaussian fit~\cite{Dov92} of experimental $\ov{p}p$ data.}
\label{fig30}
\end{figure}
Even larger suppression of annihilation may be expected for nuclei
containing antihyperons. Unfortunately, due to absence of detailed data
on $\ov{B}N$ annihilation, it is not possible to perform analogous
studies for $\ov{B}=\ov{\Lambda},\,\ov{\Sigma}$\,,\,\ldots Available
data~\cite{Gje72,Eis76} on $\ov{\Lambda}p$ annihilation show that
$\sigma^{\rm\hsp ann}_{\ov{\Lambda}p}\lesssim\sigma^{\rm\hsp
ann}_{\ov{p}p}$ at $p_{\rm\hsp lab}\sim 10\,{\rm GeV/c}$\,. On the
other hand, due to the appearance of a heavy kaon already in the
lowest--mass annihilation channel, $\ov{\Lambda}N\to\pi K$, one may
expect significant in--medium suppression of this channel, contrary to
the reaction $\NbN\to 2\pi$\,, which is enhanced as compared to the
vacuum (see Table~\ref{tab:awme}).

It is worth to mention another qualitative effect, which
accompanies annihilation of antibaryons in nuclei: the strong
modification of the distribution $P_n$ in the number of secondary
pions,~$n$\,, as compared to the $\BbN$ annihilation in vacuum.
These distributions are calculated as $P_n=\Gamma_{\BbN\to
n\pi}/\Gamma$ and presented in Fig.~\ref{fig30} for
$\ov{B}=\ov{N}$. The dashed curve corresponds to the S\hsp 0 data
from Table~\ref{tab:awme}~\footnote{ To obtain $P_n$ at $\rho_N\to
0$\,, one can use the S\hsp 0 results from this Table since in
this case $\rho_N$ cancels in the definition of $P_n$\,. }. In
this case the maximum is at $n=5$ and the shape is not far from
Gaussian (dotted line). It is clearly seen that the maximum of
$P_n$ shifts to smaller values, $n=3$ and 2, in matter with
densities~$\rho_0$ (S\hsp I) and $2\rho_0$ (S\hsp II). This effect
may be used to select events with formation of superbound nuclei
containing antinucleons, e.g. by rejecting events with more than
two soft pions in the final state.

\subsection{Annihilation of antibaryons in finite nuclei}

In the end of this section we estimate the life times of bound
antibaryon--nucleus systems with respect to annihilation. Since the
antibaryon is localized in a central core of the nucleus, our former
assumption of a homogeneous antibaryon density is not valid.  The
minimal energy available for annihilation is now given by
\bel{avle1}
\sqrt{s_*}=Q={\rm min}(E_{\Bb}+E_N)=m_{\Bb}+m_N-B_{\Bb}-B_N\,,
\ee
where $B_{\Bb}$\,\,and $B_N$ are the binding energies of the annihilating
partners.  The annihilation rates can be calculated by averaging the
local expression (\ref{anc2}) over the volume where the antibaryon wave
function $\psi_{\Bb}\hsp (x)$ is essentially nonzero. Taking into account
that the antibaryon density distribution $\rho_{\Bb}$ is normalized to
unity, after a simple calculation we get
\bel{fwid}
\Gamma=\sum\limits_{c}\Gamma_c,~~~~\Gamma_c=\Gamma_0 B_c
\lambda_{\hsp c}(s_*)
\widetilde{J}_{\BbN}\,.
\ee
Here we have introduced the overlap integral appropriate for localized density
distributions,
\bel{ovli}
\widetilde{J}_{\BbN}=\frac{m_{\Bb}\hsp m_N}{\rho_0}\int dV
\frac{\rho_{S\Bb}}{m_{\Bb}^*}\,\frac{\rho_{SN}}{m_N^*}\,.
\ee

\begin{table}[htb]
\caption{ Characteristics of antiproton annihilation in the
$^{16}_{\hspace*{3pt}\ov{p}}$O system calculated within the NLZ2
and NL3 models.The third and fifth columns shows the results of
calculation with reduced antiproton couplings ($\xi=0.5$).}
\vspace*{3mm}
\label{tab:ao16}
\begin{ruledtabular}
\begin{tabular}{l|c|c|c|c}
&\multicolumn{2}{c|}{\mbox{NLZ2}} &\multicolumn{2}{c}{\mbox{NL3}}\\
\cline{1-5}
$\xi$&1 & 0.5 &1 & 0.5 \\
\colrule
$B_N$\,(MeV)\,\footnote{
~Calculated as average between single--particle binding
energies $B_p$ and $B_n$\,
(see Fig.~\ref{fig10}).}
 & 88 & 83 & 113 & 102 \\
$B_{\ov{p}}$~(MeV)
& 897 & 500 & 1079 & 583 \\
$Q$\,(MeV)
& 892 & 1294 & 685 & 1192\\
$\lambda_{\hsp 2\pi}$ & 0.96 & 0.99 & 0.92 & 0.98\\
$\lambda_{\hsp 3\pi}$ & 0.13 & 0.40 & 0.046 & 0.32\\
$\lambda_{\hsp 4\pi}$ & $8.4\cdot 10^{-3}$ & 0.13 & $2.9\cdot 10^{-4}$
& 0.078 \\
$\sum\limits_cB_c\lambda_{\hsp c}$\,(\%) & 0.81 & 9.4 & 0.47 & 6.9 \\
$\widetilde{J}_{\ov{p} N}$ & 10.1 & 8.5 & 46.3 & 33.1 \\
\colrule
$\Gamma$\,(MeV) & 8.5 & 83 & 23 & 236\\
$\tau$\,(fm/c) & 23 & 2.4 & 8.7 & 0.84\\
\end{tabular}
\end{ruledtabular}
\end{table}

We have performed numerical calculations for the \Oap\, system
using the NLZ2 and NL3 parameter sets with two choices of antiproton
coupling constants corresponding to $\xi=1$ and $\xi=0.5$\,.
The results are summarized in Table~\ref{tab:ao16}. One can notice
strong model dependence of $\widetilde{J}_{\ov{p} N}$
which is mainly caused by its high sensitivity to the in--medium effective
masses~\footnote{It is instructive to compare
the values of overlap integrals for $\xi=1$ with the value
$J_{\NbN}\rho_N/\rho_0=15.2$ obtained from the previous estimates
for infinite matter in the case~S\hsp II.
}.
As one can see from Fig.~\ref{fig9}, the (anti)nucleon effective mass
in the~\Oap\, system becomes rather small in
the central region, especially for the NL3 parameter set. This explains
large value of $\widetilde{J}_{\ov{p} N}$ in this case.

The appearance of effective masses in
the overlap integral~(\ref{ovli}) is an artifact of quasiclassical approximation
used in our study of the in--medium annihilation.
For its validity the effective potential should vary slowly
over a Compton wave length of corresponding particles,
$\lambdabar=1/m^*_j\,(j=\ov{B},N)$\,.
The approximation breaks down when $m^*_j$ become small. Therefore,
one can use \re{ovli} only as a  rough estimate~\footnote{
In fact, the overlap integral is proportional
to $I=\left<\frac{\ds 1}{\ds E_{\ov{B}}E_N}\right>$ where $E_j=\sqrt{m_j^{*\hsp 2}(r)+p^2}$
is the kinetic energy of the j--th particle and angular brackets
denote averaging over single--particle wave functions of nucleons and an antibaryon.
Due to a finite momentum spread of the wave functions $I$ is always finite
even if $m_j^*(r)$ vanishes. This is why we think that the quasiclassical
expression~(\ref{ovli}) which contains $m_i^*$ in the denominator overestimates
the overlap integral and thus, the annihilation width of a bound antibaryon.
}.

The estimated total annihilation widths for the \Oap\, system are
also presented in Table~\ref{tab:ao16}. In the case of pure G--parity
($\xi=1$) they are in the range 9--23 MeV, depending on the RMF
parametrization. The corresponding life times are 9 -- 23 fm/c, in
agreement with the values presented in~\re{ltme}.
One can indeed talk about the delayed annihilation
due to in--medium effects. These results are very promising from the
viewpoint of experimental observation of deeply bound
antibaryon--nuclear systems.

However, much larger annihilation widths are
predicted by calculations with reduced $\ov{B}N$ couplings ($\xi=0.5$).
This is a consequence of increased $Q$ values and larger number of open
annihilation channels as compared to the case $\xi=1$\,.
The corresponding life times of about 1--2 fm/c are perhaps too short
for pronounced observable effects. We present here these results to demonstrate
uncertainties in our present knowledge of the in--medium annihilation.

\subsection{Multi--nucleon annihilation}

In addition to the in--medium effects considered above, there are
several other processes which can influence the antibaryon annihilation
inside the nucleus. Basically, one should consider two types of
processes.  Processes of the first type are of the long--range nature.
They include rescattering and absorption of primary mesons on the
intranuclear nucleons. Since primary pions produced in the annihilation
have energies in the $\Delta$--resonance region, they can effectively
be absorbed in the chain of reactions:  $\pi+N\rightarrow
\Delta,~\Delta+N\rightarrow N+N$\,. As a result of meson--nucleon and
nucleon--nucleon rescatterings, multiple particle-hole excitations will
be produced. The nucleus will eventually heat up and emit a few
nucleons. These processes can be well described by the intranuclear
cascade model (see Ref.~\cite{Bon95}). We expect that they cannot
change noticeably the life time of the trapped antibaryon.

Processes of the second type may, in principle, significantly affect
the annihilation probability.  They include new annihilation channels
which are not possible in vacuum, for instance, the emission and
absorption of a virtual meson. Since a virtual particle can propagate
only within a Compton wave length ($1/m_{\pi}\simeq$ 1.4 fm for
pions), this process should be of the short-range nature. In other
words, the recoiling nucleon should be very close to the annihilation
zone which is characterized by the radius of about 1 fm \cite{Kle02}.
Thus, in addition to the annihilation on a single nucleon, considered
so far, there exist new annihilation channels involving two and more
nucleons. It is customary to classify these annihilation channels by
the net baryon number of a combined system, i.e. $B=0$, 1, 2,...
Obviously, the channels with $B\neq 0$ must contain not only mesons but
also baryons in the final state. Probably, the most famous process of
this kind was proposed by Pontecorvo many years ago \cite{Pon56}:
\begin{equation}
\overline{p}+d\rightarrow p+\pi^-\,.
\ee
It has only one pion and one nucleon in the final state ($B=1$).
Despite of a seemingly simple final state (two charged particles), up
to now this process has not been convincingly identified
experimentally. One may expect that an analogous process with $B=1$,
\begin{equation}
\overline{B}+(NN)\rightarrow M+N\,,
\ee
can also occur on a correlated 2-nucleon pair in heavier nuclei.  The
next most interesting reaction of this kind would be the annihilation
on a 3-nucleon fluctuation:
\begin{equation}
\overline{B}+(NNN)\rightarrow N+N\,.
\ee

The relative probability of multi-nucleon annihilation can be estimated
on the basis of a simple geometrical consideration. Let us assume that
the annihilation proceeds through an intermediate stage when an
"annihilation fireball" \cite{Raf80} is formed. It is widely accepted
(see e.g. Ref.~\cite{Kle02}) that within this fireball baryons and
antibaryons are dissolved into their quark--antiquark--gluon
constituents.  The radius of this fireball, $R_{\rm ann}$, can be found
from the cross section of the inverse reaction $p+p\rightarrow
p+p+N+\overline{N}$\,.  This cross section is certainly only a fraction
of the inelastic $pp$ cross section, $\sigma_{in}^{pp}$, which is
about~30~mb in the vicinity of the $\ov{p}p$ production threshold.
Assuming that $\sigma_{in}^{pp}\simeq 2\pi R_{\rm ann}^2$ we get
$R_{\rm ann}\simeq 0.8$~fm, which is in agreement with other estimates
\cite{Kle02,Wei93}.

Now we can estimate the average number of nucleons,
which are present in the annihilation zone around an antibaryon
\begin{equation}
\ov{n}=\ov{\rho_N(r)}\cdot\frac{4\pi}{3}R_{\rm ann}^3\,.
\ee
Taking
$\ov{\rho_N(r)}\simeq 2\rho_0\mbox{$=0.3$}$~fm$^{-3}$, one obtains
$\ov{n}\simeq 0.6$. Finally, we assume that the actual number of
nucleons present in the annihilation fireball is distributed according
to the Poisson law, $P_n=\ov{n}^{\hsp\hsp n}\exp{(-\ov{n})}/n!$. This
gives the probabilities of different channels,
\begin{equation}
P_0=0.55,~~P_1=0.33,~~P_2=0.10,~~P_3=0.02,~~P_4=0.004,...
\ee
The relative probability of multi--nucleon annihilation channels is now
given by
\begin{equation}
\frac{\sum\limits_{n\geq 2}P_n}{P_1}=\frac{0.10+0.02+...}{0.33}<0.4\,.
\ee
Therefore, we expect that the channels with
$B>0$ may lead to~40\% reduction of the antibaryon life times in
nuclei, as compared to estimates including only the $B=0$ channel.

Experimental information on multi--nucleon annihilation channels is
very scarce.  We mention here a recent paper by the OBELIX
collaboration \cite{Obe02}, where the $\ov{p}$ annihilation at rest in
$^4$He was studied. They have analyzed the annihilation channels with
2$\pi^+$ and 2$\pi^-$, with and without a fast proton
($p_{\rm lab}>$\hsp 300 MeV) in the final state. The events where such a fast
proton was present were associated with the annihilation on a
2--nucleon  fluctuation i.e. with the $B=1$ channel.  The corresponding
branching ratios are
\begin{eqnarray}
&&B(\ov{p}+^4{\rm He}\rightarrow 2\pi^+2\pi^-)=(1.42\pm 0.19)\%\,,\\
&&B(\ov{p}+^4{\rm He}
\rightarrow 2\pi^+2\pi^-\hspace*{-2pt}~p_{\hsp\rm fast})=(0.098\pm 0.02)\%\,.
\end{eqnarray}
Thus, these data indicate that the relative contribution of the $B=1$
channel is less than~10\%. Certainly, more exclusive data are needed to
assess the importance of multi--nucleon annihilation channels. These
processes, in particular the Pontecorvo--like reactions, are
interesting by themselves, irrespective of their contribution to the
total annihilation rate. They may bring valuable information about the
physics of annihilation in nuclei.

\section{Formation in $\bm{\ov{p}A}$ reactions}

In this section we present estimates of formation probability of deeply
bound antibaryon-nuclear systems in $\ov{p}A$ reactions.  It is well
known from $\ov{p}p$ experiments that the annihilation cross section
$\sigma_{\rm ann}(\sqrt{s})$ is very large at low energies \cite{Pdg02}
(see its low-energy parametrization given by Eq. (\ref{ancr})).
By this reason slow antiprotons in $\ov{p}A$
interactions  are absorbed at far periphery of nuclear
density distribution. This situation can be avoided by using high-energy
antiprotons, whose annihilation cross section is strongly reduced so that
they can penetrate deeply into the nucleus.

As most promising from the experimental point of view we consider the
following reactions
\begin{eqnarray}
&&\ov{p}+N\rightarrow \ov{N}_{\rm slow}+N+\pi\,,
~~E_{\rm th}=0.3~{\rm GeV}\,, \label{1}\\
&&\ov{p}+N\rightarrow \ov{\Lambda}_{\rm slow}+\Lambda+\pi\,,
~~E_{\rm th}=1.13~{\rm GeV}\,,\\
&&\ov{p}+N\rightarrow \ov{\Lambda}_{\rm slow}+N+\ov{K}\,,
~~E_{\rm th}=1.58~{\rm GeV}\,, \label{3}\\
&&\ov{p}+N\rightarrow \ov{N}_{\rm slow}+N+\ov{N}_{\rm slow}+N\,,
~~E_{\rm th}=5.63~{\rm GeV}\,, \label{4}\\
&&\ov{p}+N\rightarrow \ov{N}_{\rm slow}+N+
\ov{\Lambda}_{\rm slow}+\Lambda\,,~~E_{\rm th}=7.11~{\rm GeV}\,,
\end{eqnarray}
where $N$ stands for nucleons $(p,n)$,~$E_{\rm th}$ is the threshold
energy for a corresponding channel.  Here slow antibaryons have a
chance to be trapped by a target nucleus leading to the formation of a
deeply bound antibaryon-nuclear system. The fast reaction
products ($N,~\pi,~\Lambda,~\ov{K}$) can be used as triggers.

One may wonder, why do we need antiproton beams? Wouldn't it be easier
to use for the same purpose much cheaper and widely available proton
beams?  The reason is that the threshold energy for the
baryon-antibaryon pair production in $pp$ collisions is very high, at
least 5.6 GeV for producing a $\ov{p}p$ pair. This means that the
reaction products are fast in the lab frame where the target nucleus is
at rest. In this situation the capture of an antibaryon by the target
nucleus is strongly suppressed. In the case of the antiproton beam the
corresponding threshold energies are quite low, as in reaction listed
above, so that the antibaryon capture becomes in principle possible.
It is interesting to note that antiproton beams open unique possibility
to produce simultaneously two antibaryons (the last two reactions above)
which can then form a bound state like an antideuteron.

In general the formation probability of a deeply--bound
antibaryon--nuclear state can be expressed as
\begin{equation} \label{Pf}
P_{\rm form}=w_{\rm cent}\cdot w_{\rm surv}\cdot
w_{\rm stop}\cdot w_{\rm capt}\,,
\end{equation}
where $w_{\rm cent}$ is a fraction of central events selected in a
given experiment (typically \mbox{$w_{\rm cent}\simeq 10\%$}), $w_{\rm
surv}$ is the antiproton survival probability. The last two factors in
Eq.~(\ref{Pf}) give, respectively, the stopping and capture
probabilities.  For estimates we assume that an antiproton penetrates
to the center of a nucleus, i.e. traverses without annihilation a
distance of about $R$ (the nuclear radius) in target matter of normal
density,
\begin{equation}
w_{\rm surv}={\rm exp}\left(-\rho_0\sigma_{\rm ann}R\right)\,.
\end{equation}

To be captured into a deep bound state an incident antiproton must
change its energy and momentum in, at least, one inelastic collision
inside the nucleus.  This can be achieved, for instance, by producing a
fast pion or kaon carrying away excessive energy and momentum. The
probability of such an event is denoted by $w_{\rm stop}$ in Eq.
(\ref{Pf}). Obviously, the energy loss should be equal to the energy
difference of the initial and final nuclei. The final $\ov{B}$ momentum
should be comparable with the momentum spread of its bound-state wave
function. One can estimate this momentum spread as
$\Delta p\simeq \pi/R_{\ov{p}}$, where $R_{\ov{p}}$ is the rms
radius of the antibaryon
density distribution.  From numerical calculations presented in
Sect.~IV we find $R_{\ov{p}}\simeq$1.5 fm so that $\Delta p\simeq$ 0.4 GeV.

The probability of a single inelastic
$\ov{p}N$ collision can be estimated by assuming Poissonian distribution
in the number of such collisions, $w_n=\ov{n}^{\hsp n}\exp{(-\ov{n})}/n!$,
so that \mbox{$w_1=\ov{n}\exp{(-\ov{n})}$}.
The mean number of inelastic collisions on a distance $r=R$ is
\begin{equation}
\ov{n}= \frac{R}{\lambda_{\rm in}}=\rho_0\sigma_{\rm in}R\,,
\end{equation}
where $\lambda_{\rm in}$ is the mean free path between inelastic collisions
and $\sigma_{\rm in}$ is the total $\ov{p}N$ inelastic cross section.

In fact only a small fraction of inelastic collision leads to a desired
energy-momentum loss. This fraction can be calculated from the
differential inelastic cross section,
$\frac{\displaystyle{\rm d}\sigma_{\rm in}}{\displaystyle{\rm d}^3p}$, for
the reaction $\ov{p}N\rightarrow \ov{B}X$. Explicitly we define
$w_{\rm stop}=w_1\cdot w_{\rm loss}$ with
\begin{equation}
w_{\rm loss}\equiv w(p_{\ov{B}}^{\rm\hsp lab}<\Delta p)=
\frac{1}{\sigma_{\rm in}}\int\limits_{p_{\ov{B}}^{\rm\hsp lab}<\Delta p}
d\sigma_{\ov{p}N\rightarrow \ov{B}X}~.
\end{equation}
Experimental data on the reaction $\ov{p}N\rightarrow \ov{B}X$ in a GeV
energy region are quite scarce. We have found only one paper
\cite{Ban79} where detailed results from the bubble chamber experiment
at CERN were reported. The differential cross sections for the reaction
$\ov{p}p\rightarrow \ov{\Lambda}X$ were measured for several $\ov{p}$
momenta from 3.6 to 10 GeV/c. Using these data we estimate
$w(p_{\ov{\Lambda}}^{\rm\hsp lab}<\Delta p)$ in the range  $10^{-4}\div
10^{-5}$ for $\ov{p}$ incident momenta 3.6$\div$12 GeV/c. The
integrated cross section for the reaction
$\ov{p}p\rightarrow\ov{\Lambda}X$ at 3.6 GeV/c is approximately 0.5~mb
as compared to about 50 mb for the total inelastic $\ov{p}p$ cross
section. The ratio of these two values, $\sim 10^{-2}$, gives the
probability of the $\ov{p}\rightarrow\ov{\Lambda}$ conversion, which is
small in accordance with the Zweig rule. In considered region of
bombarding energies we can obtain $w(p_{\ov{p}}^{\rm\hsp lab}<\Delta
p)$ simply rescaling $w(p_{\ov{\Lambda}}^{\rm\hsp lab}<\Delta p)$ by a
factor $\sim 10^2$.

\begin{table}[ht]
\caption{Formation probabilities and reaction rates for producing
antibaryon--nuclear systems in several typical reactions. The values of
$P_{\rm\hsp form}$ are obtained by multiplying probabilities from the
first three columns with $w_{\rm\hsp cent}w_{\rm\hsp capt}$\,, where
\mbox{$w_{\rm\hsp cent}=w_{\rm\hsp capt}=0.1$}\,. The reaction rates
are obtained as $Y=P_{\rm form}\cdot L\cdot \sigma_{\rm geom}$ with
$L=2\cdot 10^{32}\,{\rm cm}^{-2}\hsp {\rm s}^{-1},\,
\sigma_{\rm geom}=0.27$\, and 1.1 b for $^{17}$O and $^{209}$Bi,
respectively.}
\vspace*{3mm}
\label{tab:forp}
\begin{ruledtabular}
\begin{tabular}{c|l|c|c|c|c|r}
$p_{\rm\hsp lab}$\,(GeV/c)&~~~~~~~~~reaction & $w_{\rm\hsp surv}$ & $w_1$
&$w_{\rm\hsp loss}$&$P_{\rm\hsp form}$& Y (s$^{-1}$)\\
\colrule
0.8&
\mbox{$\ov{p}+^{17}{\rm\hspace*{-3pt} O}\to$\hsp\Oap$\,
+\hsp p\hspace*{-1pt}+\hspace*{-1pt}\pi^{-}$}
& 0.03 &0.10\footnote
{
~In this estimate we use the value $\sigma_{\rm in}\simeq 3$\,mb motivated
by measurements for the reactions \mbox{$p+^{14}{\rm N}\to\pi$} at
$E_p^{\rm kin}=300$\, MeV
reported in Ref.~\cite{Jak97}.
}
&0.1&$3\cdot 10^{-6}$& 160\\
3.6&
\mbox{$\ov{p}+^{17}{\rm\hspace*{-3pt} O}\to$\hsp\Oap
$\hsp+\hsp p\hspace*{-1pt}+\hspace*{-1pt}\pi^{-}$}
& 0.26 &0.36&$10^{-2}$&$9\cdot 10^{-6}$& 510\\
3.6&
\mbox{$\ov{p}+^{17}{\rm\hspace*{-3pt} O}\to$\hsp\Oal
$\hsp+\hsp p\hspace*{-1pt}+\hspace*{-1pt}K^-$}
& 0.26 &0.36&$10^{-4}$&$9\cdot 10^{-8}$& 5\\
3.6&
\mbox{$\ov{p}+^{209}{\rm\hspace*{-3pt} Bi}\to$\hsp\Pbap
$\hsp+\hsp n\hspace*{-1pt}+\hspace*{-1pt}\pi^+$}
& 0.07 &0.29&$10^{-2}$&$2\cdot 10^{-6}$& 440\\
3.6&
\mbox{$\ov{p}+^{209}{\rm\hspace*{-3pt} Bi}\to$\hsp\Pbal
$\hsp+\hsp p\hspace*{-1pt}+\hspace*{-1pt}\ov{K}^{\hsp 0}$}
& 0.07 &0.29&$10^{-4}$&$2\cdot 10^{-8}$& 4\\
10&
\mbox{$\ov{p}+^{17}{\rm\hspace*{-3pt} O}\to$\hsp\Oap
$\hsp+\hsp p\hspace*{-1pt}+\hspace*{-1pt}\pi^-$}
& 0.53 &0.37&$10^{-3}$&$2\cdot 10^{-6}$& 110\\
10&
\mbox{$\ov{p}+^{17}{\rm\hspace*{-3pt} O}\to$\hsp\Oal
$\hsp+\hsp p\hspace*{-1pt}+\hspace*{-1pt}K^-$}
& 0.53 &0.37&$10^{-5}$&$2\cdot 10^{-8}$& 1\\
10&
\mbox{$\ov{p}+^{209}{\rm\hspace*{-3pt} Bi}\to$\hsp\Pbap
$\hsp+\hsp n\hspace*{-1pt}+\hspace*{-1pt}\pi^+$}
& 0.22 &0.21&$10^{-3}$&$5\cdot 10^{-7}$& 110\\
10&
\mbox{$\ov{p}+^{209}{\rm\hspace*{-3pt} Bi}\to$\hsp\Pbal
$\hsp+\hsp p\hspace*{-1pt}+\hspace*{-1pt}\ov{K}^{\hsp 0}$}
& 0.22 &0.21&$10^{-5}$&$5\cdot 10^{-9}$& 1\\
\end{tabular}
\end{ruledtabular}
\end{table}
The last factor in Eq. (\ref{Pf}), $w_{\rm capt}$, is the most uncertain
quantity. It is determined by the matrix element between the initial,
plane wave state and the final, localized $\ov{B}$ state. Moreover,
because of the nuclear polarization effects, in particular, the reduced
effective mass and finite width, the bound antibaryon is off the vacuum
mass shell. The realistic calculation of $w_{\rm capt}$ will
require a special effort which is out of the scope of the present
paper. For our estimates below we take a fixed value $w_{\rm capt}=0.1$.

In Table \ref{tab:forp} we present resulting formation probabilities
for several typical reactions. One can see that $P_{\rm form}$ is on
the level of $10^{-7}\div 10^{-5}$ for bound $\ov{p}+A$ systems.  In
this Table we also give reaction rates calculated for the parameters of
antiproton beams planned at the future GSI facility~\cite{GSI} (the
beam luminosity is $L=2\cdot 10^{32}$ cm$^{-2}\cdot$\hsp s$^{-1}$,
antiproton energies from 30 MeV to 15 GeV). We see that expected
reaction rates are in the range from tens to hundreds desired events
per second. Such rates seem to be well within the present detection
possibilities.

Let us consider now two examples which clarify the reaction kinematics.
For simplicity we use plane-wave functions labeled by the
particle momentum ${\bf p}$.  We are interested in a reaction which
leads to the final $\ov{B}$ trapped in a deeply bound state with energy
$E_j=\sqrt{(m_j-S_j)^2+{\bf p}_j^2} +V_j$, where
$j=\ov{N},\ov{\Lambda},\ldots$ For instance, consider the reaction
(\ref{1}) where an antiproton with initial energy $E_{\rm beam}$ is
captured in a target nucleus $(A,Z)$ after colliding with a bound
nucleon $N$ with energy $E_N$.  If the recoiling nucleon, say a
neutron, leaves the nucleus the energy of the final nucleus
$_{\ov{p}}\hsp (A-1,Z)$ is changed by
\begin{equation}
\Delta E=E_{\ov{p}}^{\prime}-E_n\simeq
\left(\sqrt{(m_p-S_{\ov{p}})^2+\bm{p}_{\ov{p}}^{\hsp\prime\hsp 2}}+
V_{\ov{p}}\right)-\left(\sqrt{(m_n-S_n)^2+{\bf p}_n^2}+V_n\right)\,,
\end{equation}
where $\bm{p}_n$ is the initial neutron momentum and
$\bm{p}_{\ov{p}}^{\hsp\prime}$ is the final antiproton momentum, both
defined inside the nucleus. Assuming that $\bm{p}_{\ov{p}}^{\hsp\prime}
\simeq \bm{p}_n$ and $S_{\ov{p}}\simeq S_n$, $V_{\ov{p}}=-V_p$
(G-parity) we find that the binding energy of the final nucleus is
changed by \mbox{$\Delta B=-\Delta E\simeq (V_p+V_n)$}.  Neglecting
small recoil effects, we now estimate the energy of the emitted pion as
\mbox{$E_{\pi}\simeq E_{\rm beam}+\Delta B$}. So, the observation of a
pion with energy exceeding the incident antiproton energy would provide
a strong evidence in favor of the $\ov{p}$--nuclear bound state
formation. Such pions can be well separated from the annihilation pions
which are produced at a later stage of the reaction. Due to the
expected large energy gain such non-annihilation pions can be produced
even at subthreshold antiproton energies.

Analogous consideration can be done for the reaction (\ref{3}) leading
to the formation of an $\ov{\Lambda}$-nuclear bound system,
$_{\ov{\Lambda}}\hsp (A-1,Z)$. The energy difference between the initial and
final nuclei is in this case
\begin{equation}
\Delta E\simeq (m_{\Lambda}-m_N)-(S_{\Lambda}-S_N)-(V_{\Lambda}+V_N)\,.
\ee
With $S_{\Lambda}\sim S_N$ and $V_{\Lambda}\sim V_N$ this energy
difference is negative. The energy balance requires that the energy
$E_{\rm beam}-\Delta E$ is carried away by an emitted antikaon which,
therefore, should be fast in the lab frame. We propose to use fast
pions and antikaons as triggers for selecting desired events.

It is necessary to emphasize that the antibaryon
capture reactions discussed above are relatively fast as compared
to the characteristic time of target nucleus rearrangement,
$\tau_{\rm nuc}$. Therefore, initially antibaryons will occupy
''non--modified'' levels which are predicted for ordinary nuclei.
Rearrangement of nuclear structure, in particular, the creation of
a local compression, as discussed in Sect.~IV, will take a longer
time. One can estimate this time~as
\begin{equation}
\tau_{\rm nuc}= R_{\rm core}/c_s\,,
\end{equation}
where $R_{\rm core}$ is the radius of a compressed core and $c_s$ is
the sound velocity. Taking $R_{\rm core}\simeq 1.5$ fm and $c_s=0.2\,c$
we get $\tau_{\rm nuc}\simeq 7.5$ fm/c. In the light of estimates
presented in Sect.~V, this time is of the same order as the predicted
life time of antibaryon-doped nuclei. Better understanding of the
rearrangement process can be achieved only by carrying out dynamical
simulations on the basis of the TDHF or molecular dynamics approaches.

Finally we estimate the temperature of an antibaryon-nuclear system which
might be expected after the rearrangement of its structure.
As one can conclude from Fig.~\ref{fig18} after the rearrangement the antiproton
binding energy increases by about 220 MeV in \Oap\, and 400 MeV in
$^4\hspace*{-4pt}$\raisebox{-4pt}{$\sst\ov{p}$}\hsp He. Assuming that
this binding energy gain $\Delta E$ is transferred into heat, we can estimate
the temperature of the final system using the Fermi gas formula $\Delta E=aT^2$\,,
where $a\simeq A/10$\,MeV is the nuclear level density parameter. For \Oap\,this
gives $T\simeq 12$\,MeV which is much lower than the temperatures $\sim 100$\,MeV
associated with high--energy nuclear collisions. Therefore, we can indeed speculate
about cold compression of nuclei induced by antibaryons as an alternative to the
shock--wave compression in heavy--ion collisions~\cite{Sch74}. In the latter case,
as follows from the Hugoniot adiabate, compression of nuclear matter is always
accompanied by its strong heating.

\section{Observable signatures of bound antibaryon--nuclear systems}

If bound antibaryon--nuclear systems exist and live long enough
they can manifest themselves in several ways. In this section we
discuss their possible signatures.

\subsection{Transitions from atomic to nuclear states}

If a slow antiproton is first captured on a Coulomb orbit forming an
antiproton atom, it can later on make transition on a deep nuclear
bound state. This process was first proposed in Ref.~\cite{Won84} where
antiproton--nuclear bound states were studied within the
nonrelativistic optical model. Such a transition will be accompanied by
the emission of a monoenergetic photon, pion or kaon, depending on the
antibaryon type. According to our calculations, many discrete states
may exist in the relativistic antiproton potential. Their binding
energies and thus, the transition energies may vary from a few tens to
several hundred MeV. The probability of such transitions is determined
by the matrix element of the corresponding transition operator between
the Coulomb and nuclear antiproton wave functions. As shown in
Ref.~\cite{Won84}, the partial width associated with such transitions
could be as small as $10^{-4}-10^{-5}$ of the total width of the atomic
level. The total width is obviously determined by the annihilation on
the nuclear surface. Therefore, a special effort should be made to find
this signal in the huge background of direct annihilation events.

\subsection{Super--transitions from upper to lower well}

\begin{figure*}[htb!]
\vspace*{-5mm}
\centerline{\includegraphics[width=9cm]{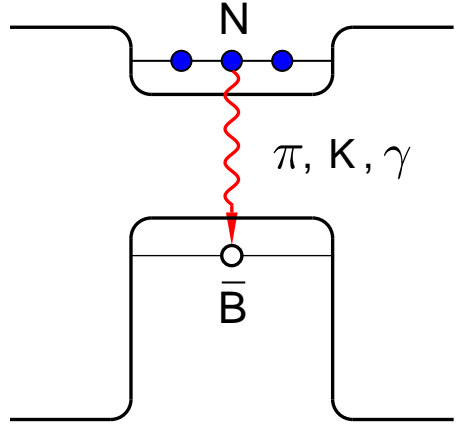}}
\vspace*{-1.5cm}
\caption{
Annihilation of antibaryon--nuclear system as a ''super--transition''
between the Fermi and Dirac seas.}
\label{fig31}
\end{figure*}
As pointed out earlier, high--energy antiprotons have a better chance
to penetrate deep into the nucleus.  Under certain conditions discussed
earlier they can be trapped into a deep bound state.
We have demonstrated in Sect.~V that the annihilation can be delayed in
this case due to the reduction of energy released in this process.  An
antibaryon sitting in the nucleus can be viewed as a hole in otherwise
filled Dirac sea.  This state represents a strong excitation of the
nucleus which soon or later will decay. Among other decay modes the most
interesting is a {\it super-transition} when a nucleon from a discrete
level of the Fermi sea jumps into the hole state in the Dirac sea. This
can be achieved by emitting a single photon, pion or kaon with energy
of about $Q=E_N+E_{\ov{B}}$ and isotropic angular distribution in the
nucleus rest frame. Such process is forbidden in vacuum by the
energy--momentum conservation laws. But in the considered case the
recoil momentum can be carried away by the residual nucleus. This is
analogous to the Pontecorvo--like reactions discussed in Sect.~V.  This
process is illustrated in Fig.~\ref{fig31}. The appearance of
relatively narrow lines with energies of about \mbox{$0.5-1$~GeV} and
width $\Delta E\simeq 20\div 50$~MeV in the spectrum of secondary
photons or mesons would be a direct signature of the deeply--bound
antibaryon--nuclear states.

\subsection{Explosive multifragmentation}

\begin{figure*}[tb!]
\vspace*{-1cm}
\centerline{\includegraphics[width=14cm]{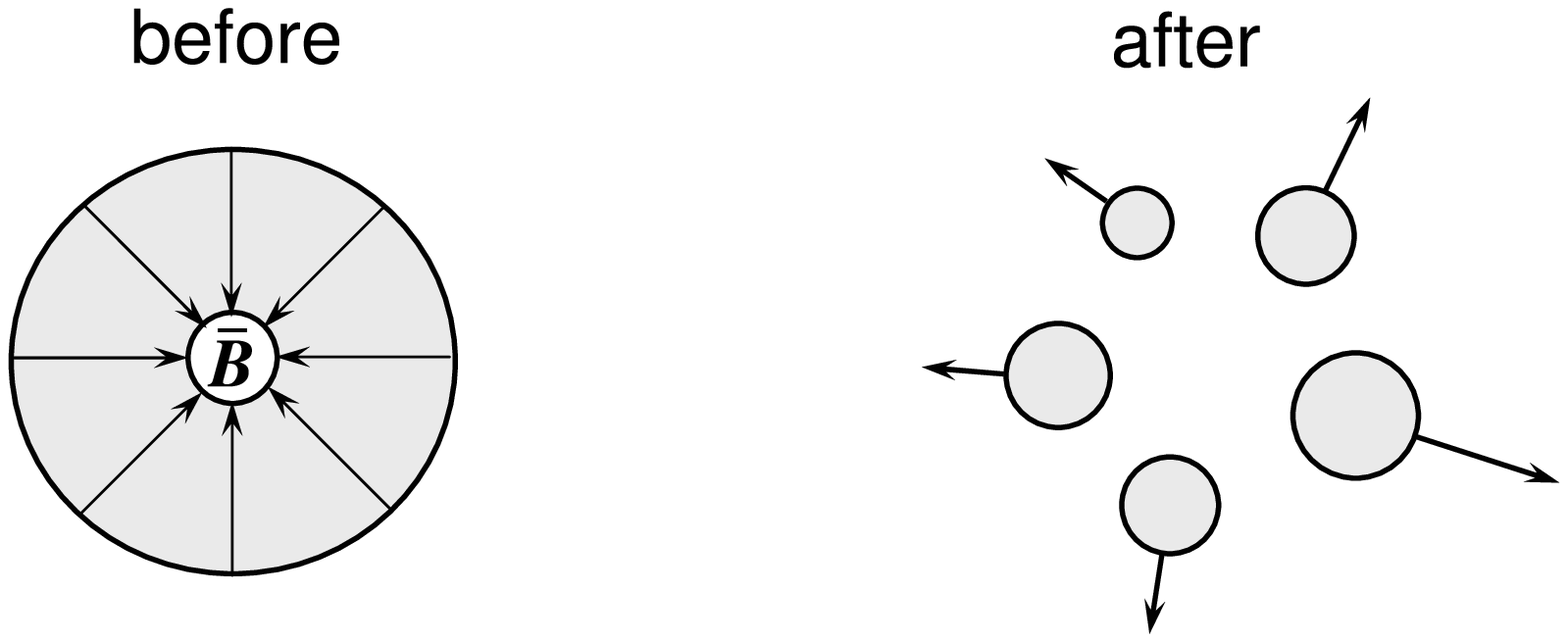}}
\caption{
Multifragmentation (right) of a nuclear
remnant formed after the antibaryon annihilation in a bound
$\ov{B}$--nuclear system (left).}
\label{fig32}
\end{figure*}
Another signal may come from the explosive disintegration of a
compressed nucleus after the antibaryon annihilation. When the
driving force for the compression disappears, the nucleus will
expand as a compressed spring.
As a result of the collective expansion the nucleus will be torn
apart into fragments, as illustrated in Fig.~\ref{fig32}. This process
can be observed by measuring collective velocities of these fragments.
Multifragmentation of nuclei induced by high--energy antiprotons has been studied
earlier~\cite{Bea99}, but with limited statistics. So far only minimum
bias events have been analyzed. In this case the distribution of
fragments follows closely predictions of the statistical
multifragmentation model~\cite{Bon95} where no collective effects
are included. We emphasize again that a dedicated study with
proper triggering is needed to find explosive events.
Modern experimental technics allow to unambiguously distinguish
between purely thermal and flow--driven multifragmentation.

\subsection{Multi--quark--antiquark clusters}

It is interesting to look at the antibaryon--nuclear systems from a
somewhat different point of view. An antibaryon implanted into a
nucleus acts as a strong attracting center for nearby nucleons.
Due to uncompensated attractive force these nucleons acquire
acceleration towards the center. As the result of this inward
collective motion the nucleons will pile up and produce a
compression zone around the antibaryon. If such a process were
completely elastic, it would look like a monopole--type
oscillations around the equilibrium configuration found by solving
static equations (see Sect.~IV). In this dynamical process even
stronger compression can be reached as compared with the one
predicted by this static configuration. The maximum compression is
achieved when the initial potential energy generated by the
antibaryon is transformed into the compressional energy. Simple
estimates show that local baryon densities up to $5\rho_0$
may be obtained in this way.
\begin{figure*}[tb!]
\centerline{\includegraphics[width=16cm]{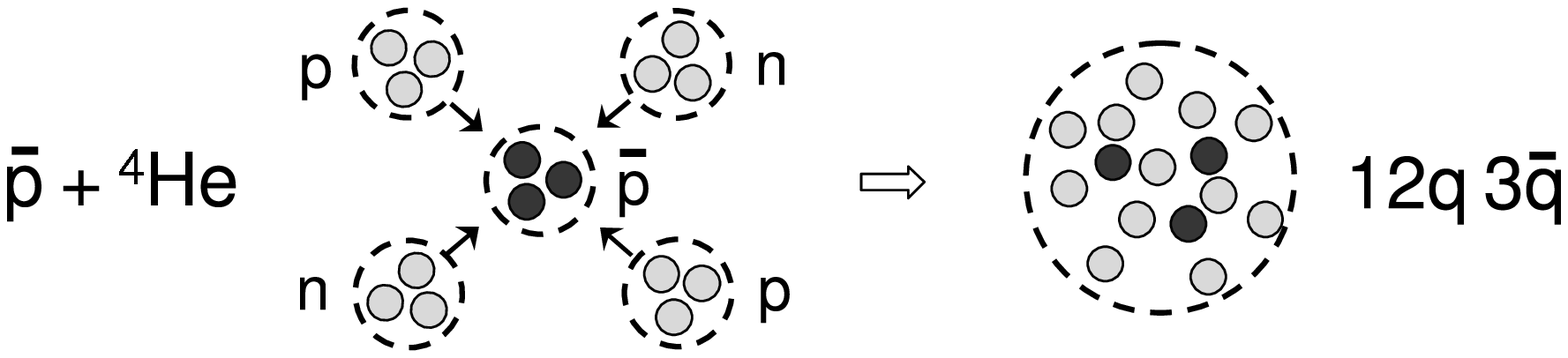}}
\vspace*{5mm}
\caption{
Schematic picture of deconfined
quark--antiquark droplet formed in interaction of an antiproton
with the $^4$He nucleus.}
\label{fig33}
\end{figure*}
It is most likely that the deconfinement transition will occur at this
stage and a high--density cloud containing an antibaryon and a few
nucleons will appear in the form of a multi--quark--antiquark cluster.
One may speculate that the whole $^4$He or even $^{16}$O nucleus can be
transformed into the quark phase by this mechanism. This process is
illustrated in~Fig.~\ref{fig33}.

As shown in Refs.~\cite{Mis99,Mis00} and shortly discussed in
Sect.~III (see Fig.~\ref{fig4}), an admixture of
antiquarks to cold quark matter is energetically favorable. The
problem of annihilation is now transferred to the quark level.
But the argument concerning the reduction of available phase space
due to the entrance--channel nuclear effects should work in this
case too. Thus one may hope to produce relatively cold droplets of
the quark phase by the inertial compression of nuclear matter
initiated by an antibaryon. A similar mechanism for producing
large quark bags was proposed in Ref.~\cite{Raf80}. There the idea
was that the annihilation fireball moving through the nucleus
can grow by absorbing nucleons on its way.

Thus, the annihilation process in nuclear environment may serve as a
breeder for creating new multi--quark--antiquark structures.
We believe that among the decay modes of multi--quark--antiquark
systems, like the one depicted in Fig.~\ref{fig33}, not only usual hadrons,
but also exotic states such as pentaquarks ($uudd\ov{d},uudd\ov{s},\ldots$)
or heptaquarks ($uuu\ov{u}dd\ov{d},uu\ov{u}dd\ov{d},\ldots$) can be present.
Moreover, more exotic states like
baryonium ($3q\hsp 3\ov{q}$), deutronium~($6q\hsp 3\ov{q}$), tritonium
($9q\hsp 3\ov{q}$) and even helionium ($12q\hsp 3\ov{q}$\,,~see
Fig.~\ref{fig33}) can also be searched for. Discovery of such states would significantly
extend our knowledge about the quark--gluon structure of matter.  Such
a study will be complimentary to the active search for exotic
multi--quark systems stimulated by the discovery of the
pentaquark~\cite{Nak03}.

\section{Summary and discussion}

In this paper we have demonstrated that antibaryons are of significant
interest for nuclear physics. They can provide new and important
information about the nature of strong interactions and structure of
the vacuum. As well known, the relativistic nuclear models predict deep
antibaryon potentials in nuclei generated by coherent action of mean
scalar and vector fields. To verify this picture we study a real
antibaryon bound in the nucleus instead of considering
filled states of the Dirac sea. On the basis of the RMF model we
performed detailed calculations of such bound antibaryon--nuclear
systems taking into account the rearrangement of nuclear structure due
to the presence of antibaryons. The self--consistent calculations
lead to stronger bindings as compared to previous studies  where
the rearrangement effects were ignored.  What is even more important,
our calculations predict strong local compression of nuclei induced
by the antibaryon. This opens a principal possibility of producing
relatively cold and compressed nuclear matter in the laboratory. In
contrast, in high-energy nuclear collisions the compression is always
accompanied by a strong heating of nuclear matter.

Our second goal in this paper was to deceive a common delusion that the
antibaryon annihilation in nuclei is so strong that it will shadow all
other processes. We have performed detailed calculations assuming that
annihilation rates into different exclusive channels are proportional
to the available phase space. Due to reduced effective masses and
(partial) cancelation of vector potentials, the energy available for
annihilation of slow antibaryons (the reaction Q--value) is
significantly reduced as compared to the minimum vacuum value,
$m_N+m_{\ov{B}}$. Then many channels of in--medium annihilations are
simply closed. This may lead to a dramatic reduction, by factors
5$\div$20, of the total annihilation rate. We estimate life times of
deeply-bound antibaryon-nuclear systems on the level of 2$\div$20~fm/c
that makes their observation feasible. This large margin in the life
times is mainly caused by uncertainties in  antibaryon coupling constants
as well as the overlap integrals
between the antibaryon and nucleon scalar densities. We have also analyzed
multi--nucleon annihilation channels (Pontecorvo-like reactions) and
found their contribution to be less than 40\% of the single--nucleon
annihilation.

We believe that bound antibaryon-nuclear systems can be produced by
using antiproton beams of multi--GeV energy, e.g. at the future GSI
facility.  Since the annihilation cross section drops significantly
with energy, a high--energy antiproton can penetrate deeper into the
nuclear interior.  Then it can be stopped there in an inelastic
collision with a nucleon, e.g. via the reaction
$A\hspace{1pt}(\ov{p},N\pi)\hspace{1pt}_{\ov{p}}\hsp A'$, leading to the
formation of a $\ov{p}$--doped nucleus. Reactions like
$A\hspace{1pt}(\ov{p},N\hsp\ov{K})\hspace{1pt}_{\ov{\Lambda}}A'$\, and
$A\hspace{1pt}(\ov{p},\Lambda\pi)\hspace{1pt}_{\ov{\Lambda}}A'$ can be
used to produce a $\ov{\Lambda}$--doped nucleus.  Fast mesons, nucleons
or lambdas can be used for triggering such events.  Our estimates of
the formation probability in a central $\ov{p}A$ collision give the
values \mbox{$10^{-5}-10^{-6}$}. With the antiproton beam luminosity of
10$^{32}$ cm$^{-2}\,$s$^{-1}$ planned at GSI this will correspond to
the reaction rate from tens to hundreds of desired events per second.

We have proposed several observable signatures which can be
used for detection of antibaryon--nuclear bound states.
They include exotic annihilation channels with emission of a single photon,
pion or kaon; explosive multifragmentation of nuclei, and formation of
multi--quark--antiquark clusters. The possibility of using antibaryon
annihilation in nuclei to produce droplets of relatively cold quark matter
is interesting by its own. This mechanism will work irrespective to
the existence of antibaryon--nuclear bound states.

There remain many problems to be studied. The list of most
urgent questions includes:
\vspace*{-1cm}
\begin{tabbing}
~\=$\bullet$~\=how G--parity works in a many--body system and what
are the antibaryon couplings\\
\>\>to the meson fields?\\
\>$\bullet$\>what is the role of exchange terms and
dispersive corrections?\\
\>$\bullet$\>how large is the unphysical self--interaction of antibaryons
within the RMF approach?\\
\>$\bullet$\>how to extend the RMF approach to include
self--consistently imaginary contribution\\
\>\>to the antibaryon self energy?\\
\>$\bullet$\>how accurate are quasiclassical estimates of life times
of bound antibaryon--nuclear\\
\>\>systems?\\
\>$\bullet$\>how these life times are affected by in--medium
modifications of secondary mesons?
\end{tabbing}
\vspace*{-2mm}
In addition, the nuclear rearrangement dynamics after the antibaryon
capture should be studied on the basis of a dynamical approach like the
TDHF model. Transport calculations of $\ov{p}A$ reactions are needed
to check our simple estimates of the formation probabilities and
annihilation rates.  We are  planning to address these questions in the
future work.

Our most general conclusion is that the antibaryon--nuclear physics is
a broad field of research which is not yet explored sufficiently well.
It may bring about new interesting phenomena ranging from unusual
annihilation channels to exotic antibaryon--nuclear or multi-quark-antiquark
bound states. This
is a nonperturbative domain of strong interactions and therefore the
progress in this field can be achieved only by close cooperation of
theory and experiment.

\begin{acknowledgments}
The authors thank P. Braun--Munzinger, E. Friedman, A. Gal,
C. Greiner, J. Knoll, J.-M. Richard, S. Schramm, and T. Yamazaki for useful
discussions.  This work has been supported by the GSI (Germany),
the RFBR Grant 03--02--04007 and the MIS Grant NSH--1885.2003.2.
\end{acknowledgments}

\end{document}